\documentclass[journal,comsoc]{IEEEtran}
\usepackage[utf8]{inputenc}
\usepackage{datatool}
\usepackage{ifthen}
\usepackage{xstring}

\usepackage[nocompress,sort]{cite} %
\usepackage[pdftex]{graphicx}
\graphicspath{{src/fig/}}
\DeclareGraphicsExtensions{.pdf,.jpeg,.png}
\usepackage[caption=false,font=normalsize,labelfont=sf,textfont=sf]{subfig} %
\usepackage{comment}

\usepackage{acronym}

\usepackage[hyphens]{url}
\usepackage[hidelinks]{hyperref} %
\usepackage{xspace}
\usepackage{amsmath}
\usepackage{wasysym}
\usepackage{amssymb}
\usepackage[group-separator={,}]{siunitx}
\usepackage{paralist}

\newcommand{\slightlybold}{\fontseries{b}\selectfont}

\makeatletter
\newcounter{subparagraph}[paragraph] %
\def\subparagraph{\@startsection{subparagraph}{5}{2.5\parindent}{0ex plus 0.1ex minus 0.1ex}{0ex}{\normalfont\normalsize\itshape\normalfont\normalsize\itshape\slightlybold}}%
\makeatother

\makeatletter
\def\subsubsection{\@startsection{subsubsection}{3}{\parindent}{0ex plus 0.1ex minus 0.1ex}{0ex}{\normalfont\normalsize\itshape\slightlybold}}%
\def\paragraph{\@startsection{paragraph}{4}{2\parindent}{0ex plus 0.1ex minus 0.1ex}{0ex}{\normalfont\normalsize\itshape\normalfont\normalsize\itshape\slightlybold}}%
\makeatother

\newcommand{\ie}{i.\@\,e.\@\xspace}
\newcommand{\eg}{e.\@\,g.\@\xspace}

\newcommand{\etal}{et~al.\@\xspace}

\newcommand{\nburl}[1]{\mbox{\url{#1}}} %
\makeatletter
\DeclareRobustCommand*\textsubscript[1]{%
  \@textsubscript{\selectfont#1}}
\def\@textsubscript#1{%
  {\m@th\ensuremath{_{\mbox{\fontsize\sf@size\z@#1}}}}}
\makeatother

\newcommand{\super}[1]{\textsuperscript{#1}}

\usepackage[table]{xcolor}
\usepackage{rotating}
\usepackage{array}
\usepackage{makecell}
\usepackage{multirow}
\usepackage{xcolor}
\usepackage{mdframed} %
\usepackage{arydshln} %

\renewcommand\theadfont{\bfseries}
\newcolumntype{L}{>{\raggedright\arraybackslash}X}

\usepackage{pifont}

\usepackage{tikz}

\newcommand{\DEBUG}[1]{}

\usepackage{morewrites} 
\usepackage{mastertable}

\addAttackerCapability{airpassive}{Radio Passive}
\addAttackerCapability{airactive}{Radio Active}
\addAttackerCapability{airuser}{User Traffic}
\addAttackerCapability{ss7}{SS7 Interface}
\addAttackerCapability{pstn}{PSTN Interface}
\addAttackerCapability{inet}{Internet Traffic}
\addAttackerCapability{physical}{Nondestructive Physical}

\addRootCause{RCspec}{Specification Issue}
	\addCause{pat}{Unsecured Pre-Authentication Traffic}{RCspec}
	\addCause{nonmutal}{Non-Existing Mutual Authentication}{RCspec}
	\addCause{weakcrypto}{Weak Cryptography}{RCspec}
	\addCause{resource}{Resource Usage Asymmetry}{RCspec}
	\addCause{insecSS7}{Insecure Inter-Network Protocol}{RCspec}

\addRootCause{RCimpl}{Implementation Issue}
	\addCause{insecureimpl}{Insecure Implementation}{RCimpl}
	\addCause{leakyimpl}{Leaky Implementation}{RCimpl}

\addRootCause{RCnsci}{Protocol Context Discrepancy}
	\addCause{crosslayer}{Cross-Layer Information Loss}{RCnsci}
	\addCause{accounting}{Accounting Policy Inconsistency}{RCnsci}

\addRootCause{RCshared}{Wireless Channel}
	\addCause{channel}{Channel Characteristics}{RCshared}

\addComponent{sim}{SIM}
\addComponent{ue}{User Equipment}
\addComponent{enb}{Base Station, (e)NodeB}
\addComponent{core}{Core Network}
\addComponent{hss}{HLR, HSS}
\addComponent{pgw}{Packet Gateway}
\addComponent{sgw}{PSTN/Voice Gateway}
\addComponent{inter}{Interconnection Gateway}

\begin{document}
\title{On Security Research Towards \\Future Mobile Network Generations}

\author{David Rupprecht*,~
        Adrian Dabrowski*,~
        Thorsten Holz,~
        Edgar Weippl,~
        and Christina Pöpper\\
        \thanks{* These authors contributed equally to this work.}
        \thanks{David Rupprecht and Thorsten Holz are with Horst Görtz Institute for IT-Security (HGI) at Ruhr-Universität Bochum, Bochum, Germany (email: david.rupprecht@rub.de; thorsten.holz@rub.de).}        \thanks{Adrian Dabrowski and Edgar Weippl are with SBA Research, Vienna, Austria (email: adabrowski@sba-research.org; eweippl@sba-research.org).}
        \thanks{Christina Pöpper is with New York University Abu Dhabi,
Abu Dhabi, United Arab Emirates (email: christina.poepper@nyu.edu).}
	\thanks{Revised Version 2018-03-01}
}

\markboth{}{On Security Research Towards Future Mobile Network Generations}
\maketitle

\begin{abstract}

Over the last decades, numerous security and privacy issues in all three active mobile network generations have been revealed that threaten users as well as network providers. In view of the newest generation (5G) currently under development, we now have the unique opportunity to identify research directions for the next generation based on existing security and privacy issues as well as already proposed defenses. This paper aims to unify security knowledge on mobile phone networks into a comprehensive overview and to derive pressing open research questions. To achieve this systematically, we develop a methodology that categorizes known attacks by their aim, proposed defenses, underlying causes, and root causes. Further, we assess the impact and the efficacy of each attack and defense. We then apply this methodology to existing literature on attacks and defenses in all three network generations. By doing so, we identify ten causes and four root causes for attacks. 

Mapping the attacks to proposed defenses and suggestions for the 5G specification enables us to uncover open research questions and challenges for the development of next-generation mobile networks. The problems of unsecured pre-authentication traffic and jamming attacks exist across all three mobile generations. They should be addressed in the future, in particular to wipe out the class of downgrade attacks and, thereby, strengthen the users' privacy. Further advances are needed in the areas of inter-operator protocols as well as secure baseband implementations. 
Additionally, mitigations against %
denial-of-service attacks by smart protocol design represent an open research question.

\end{abstract}


\IEEEpeerreviewmaketitle

\section{Introduction}
\IEEEPARstart{O}{ver} the past decades, mobile communication has become an integral part of our daily life. For instance, in 2016 the mobile network comprised 4.61 billion users~\cite{StatNo2016} and the revenue of all mobile network operators totaled 1,331 billion USD~\cite{StatRe2016}. In many markets, the number of mobile Internet subscribers has outnumbered the stationary ones. A vast and diverse mobile communication and application ecosystem has emerged. These applications include private as well as business communication, and even critical infrastructures. For example, payment services, energy infrastructure, and emergency services (\eg, FirstNet \cite{First2016}) highly depend on mobile networks. As a consequence, the reliability and security of mobile networks have become a substantial aspect of our daily lives.

However, over the last years, a large body of literature has revealed numerous security and privacy issues in mobile networks. There is a broad set of attacks \cite{Shaik2015, Kune2012,TS24.301,Arapinis2012,Paget2010,Mjolsnes2017} that affect the users' privacy and data secrecy, the mobile network operators' revenue, and the availability of the infrastructure. Various countermeasures against these attacks have been proposed, some of which have become security features of new mobile generations. Besides the academic community, the non-academic community also substantially contributed to the comprehension of mobile network security. Unfortunately, attacks and countermeasures were mostly considered in an isolated manner and the research efforts have not been systematized or categorized into a big picture. However, these insights are necessary to develop generic countermeasures instead of separate fixes or mitigations. For example, messages being exchanged before the authentication and key agreement is the cause of multiple attacks \cite{Shaik2015, Forsberg2007, Golde2016}. Considering the attacks separately, one might not assume that this is a broader problem present in all three mobile generations.

As network standards tend to stay in use for decades, structural or backwards-incompatible changes are only possible for new network generations. We would like to use the window of opportunity with regards to 5G for the development of future mobile security specifications in order to eliminate insecure legacies. While considering the next mobile network generation, we systematized the research efforts of the last decades to improve and provide a basis for future security research and specifications. 
Since the contributions in mobile network security research are fragmented, we develop a methodology to categorize attacks and their countermeasures and thus provide an abstract overview on the topic.
We project the design errors and attacks across the network generations to illustrate the specifications' development.
Furthermore, we give an outlook on future developments in mobile communication and map the extracted issues to them. 
Finally, we identify open research questions regarding mobile network security and point out challenges for future specifications. 

The scope of our survey is on the technical side of mobile networks (Figure~\ref{fig:generic_architecture}) to fill a blank space between highly researched topics. 
There are surveys on mobile applications such as secure messaging systems \cite{Unger2015} and mobile operating systems, \eg, Android \cite{Acar2016}. On the other hand, there are generic Internet security surveys \cite{Ullrich2014} and telephony security contributions focusing on fraud attacks~\cite{Tu2016,Gupta2015,Sahin2017}. Recently, Jover \cite{Jover2017} pointed out 5G security challenges, but without systematizing prior work.

In summary, the main contributions of this article are as follows:
\begin{compactitem}

\item We develop a \textbf{systematization methodology} for attacks and defenses in mobile networks. Starting from security requirements, we classify the attacks by their aims. We use attack characteristics for estimating the attack impact, \eg, different attacker capabilities. Defenses characteristics help us to describe the advancement for defenses. To gain an abstract overview on the topic, we logically group technical attacks and defenses into \textit{causes} and high-level \textit{root causes}. 

\item We \textbf{categorize and systematize} attacks and defenses on mobile networks using our systematization methodology to obtain a comprehensive picture of research in this field. To this end, we incorporate publications from the academic as well as non-academic communities to represent the big picture.

\item We derive \textbf{open research questions and challenges} building upon our systematization for further studies in both offensive and defensive work. We do this to shape the future research in the field of mobile network security. In order to achieve this, we investigate the shortcomings of existing work, the implications of future technologies, and the concrete challenges of defenses. We underline the challenges of future technologies by mapping implications of 5G technologies to open research questions.

\end{compactitem}

\section{Mobile Network Background}

In the following, we briefly describe the technical background of mobile networks, including an overview of the currently active generations and a generic overview of the network architecture.

\subsection{Generations}
\label{sec:Generations}
Over the years, the requirements for mobile networks have shifted from rather single-purpose networks (voice service) to multi-purpose networks (data). In the following, we introduce the currently active network generations.
\begin{compactitem}

\item \textbf{\acs{GSM} (2G, \acl{GSM})} is the first digital mobile communication system and was designed for voice transmissions. It uses circuit-switched scheduling in which fixed slots are allocated for transmissions over the air and on network components along the transmission path. The \ac{GPRS} is a packet-switched extension on top of the circuit-switched architecture. 

\item \textbf{\acs{UMTS} (3G, Universal Mobile Telecommunications System)}: 
In order to meet the increasing demand for data transmissions, the next generation was optimized for data transmission on the radio layer. Additionally, UMTS added new security features such as mutual authentication and new encryption algorithms. 
Although the network is packet-switched in its core, voice and SMS transmissions are still offered as distinct network services.

\item \textbf{\acs{LTE} (4G, Long Term Evolution)} uses a completely redesigned radio layer and a strict IP-based packet-switched architecture with guaranteed \ac{QoS} classes. 
In contrast to its predecessors, voice and SMS transmissions are no longer network services, but offered as IP-based services (SIP, VoIP) on top of a general-purpose IP data network. However, fallback options exists for phones or operators which do not support \ac{VoLTE} \cite[Sec. 8.2]{TS23.272}.

\end{compactitem}

Names and abbreviations for equivalent network components and concepts vary between the different network generations. Overall, we try to stay agnostic to the generations and access technologies. 
If we need to specially differentiate between different terms, we denote them with a \super{2G} for GSM, \super{3G} for UMTS, and \super{4G} for LTE.

\subsection{Network Architecture} 
\label{sec:architecture}

\begin{figure}[!t]
\hspace{-1mm}\includegraphics[width=1.03\columnwidth]{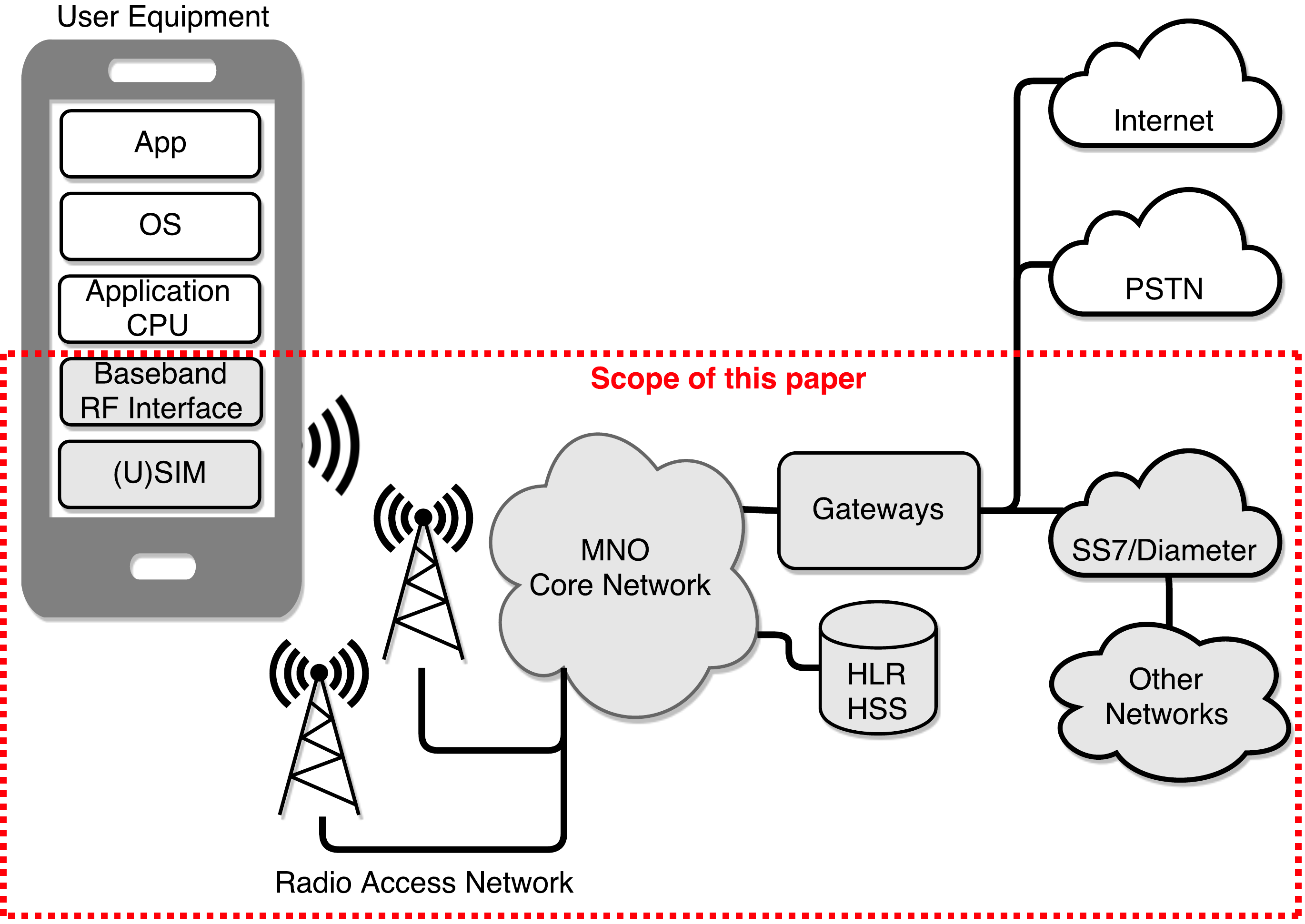}
\caption{Generic mobile network architecture and the scope of this paper.}
\label{fig:generic_architecture}
\end{figure}

Figure \ref{fig:generic_architecture} shows a generic network architecture including the scope of this paper. In general, the architecture consists of the following components: 

\subsubsection{User Equipment}
\label{sec:architecture_ue}
\label{sec:backgroundTMSI}
\label{sec:architecture_TMSI}
The \ac{UE}\super{4G} (\eg, smartphone) is the device utilized to communicate with the network and consume its services. It comprises different components, such as the application processor that runs the mobile operating system, the graphical user interface, and all its locally installed applications.
The baseband processor implements the mobile protocol stacks for multiple network generations and thereby establishes the communication with the network. The SIM\super{2G}/USIM\super{3G,4G} (\acl{USIM}) directly identifies a customer and stores the authentication information as a pair of the permanent identity (\acs{IMSI}, \acl{IMSI}) and the secret long-term symmetric key used for encryption and authentication. From outside, a user is identifiable and thus callable via a public phone number (called \ac{MSISDN}). Besides permanent identities, for privacy reasons temporary identities are dynamically allocated to the \ac{UE}, such as a \ac{TMSI} that is used for paging and core network communication.

\subsubsection{Radio Access Network}
\label{sec:architecture_ran}
The purpose of the \ac{RAN} is to transmit data between the \ac{UE} and the core network that provides service to the user. 
Therefore, the mobile phone establishes a radio connection to the base station (BTS\super{2G}, nodeB\super{3G}, eNodeB\super{4G}) that acts as a network access point. For mobility management, base stations are organized into cells which are in turn grouped for circuit-switched services into \acp{LA}\super{2G,3G}, and for packet-switched services into \acp{RA}\super{2G,3G} and \acp{TA}\super{4G}. 

\subsubsection{Core Network}
\label{sec:architecture_core}
The core network's task is to manage the connection mobility and to deliver the services, \eg, phone calls and Internet connection. For this mobility management, several core network elements are utilized. A central database, the \ac{HLR}\super{2G} or \ac{HSS}\super{3G,4G}, stores the authentication, mapping, and other information about the users. Its security functionality is often referred as \ac{AuC}. Core network elements manage the mobility, connection, and security establishment. \ac{SS7} is used within \acs{GSM} and \acs{UMTS} networks for signaling purposes such as mobility management and call setup as well as externally for roaming. \ac{SS7} was developed in the mid-1970s for landline networks and was later extended for mobile telephony networks. Unfortunately, the protocol only provides limited security mechanisms. Today, SS7 is mostly used as an SS7-over-IP adaptation. \acs{LTE} introduced new IP-based protocols for the core-network infrastructure, \eg, the SIP-based \ac{IMS} handles voice, video, and text messages.

\subsubsection{Inter Network}
\label{sec:architecture_inter}
\label{sec:architecture_ss7}
Many services require a connection to other communication networks such as the \ac{PSTN} or the Internet leading to the introduction of subsystems and gateways. 

Mobile networks are connected to each other via \ac{SS7} or its successor, the Diameter protocol for global inter-network operator roaming, text messages, and call forwarding. Diameter inherited most of the \ac{SS7} semantics, but offers improved authentication and confidentiality through the use of IPsec and \ac{TLS}.

\subsubsection{Radio Channel}
\label{sec:architecture_radiochannel}

The radio layer shares some common design choices between GSM, UMTS, and LTE, whereas other characteristics like frequencies, modulation, or access technologies are highly individual. All generations incorporate three main types of logical channels into the physical radio channel:~(i)~Broadcast control channels carry information about the base station, its neighbors, and the network configuration.~(ii)~Paging channels are used to call out for specific UEs when the network wants to transmit data to them.~(iii)~Dedicated channels are used for traffic to and from each single device. These are the only channels that can be encrypted and integrity protected, if initiated by the network.

\subsubsection{Pre-Authentication Traffic and Security Establishment} 
\label{sec:architecture_security}
Unless initiated by the network, the traffic is unencrypted, not integrity protected and, thus, not authenticated. This means that only dedicated traffic to and from a specific device is secured. Thus paging, other broadcasts, most of the radio resource allocations, and low-level signaling traffic are always unprotected. All traffic that happens before the setup of an authenticated session is defined as \emph{pre-authentication traffic}.
 
\phantomsection
\label{sec:architecture_enc}
The authenticated session is established via an \ac{AKA} protocol which is a challenge-response protocol, that authenticates the partner(s) and derives a session key for the communication. While \acs{GSM} only establishes user authentication, newer generations (\acs{UMTS}, \acs{LTE}) establish mutual authentication. The session keys are derived from a common long-term shared secret stored in the (U)SIM. 
The particular used \ac{AKA} depend on the deployed SIM, the operators \ac{AuC}, and the access technology. On GSM, the example algorithm COMP128 became the de-facto standard \cite[Ch. 16]{Walker2001}, albeit operators could issue SIMs with a custom algorithm. In later generation key derivations are split between the \ac{UE} and the \ac{USIM} where the publicly reviewed Milenage (and TUAK) algorithms are used. Operators are still able to customize AuC's and USIM's algorithms.

\subsubsection{Mobility Management and Paging}
\label{sec:architecture_mm}
When no active data transmission or phone call is ongoing, the phone goes into the idle state. In this state, the network only knows the coarse Location Area where the subscriber is located. The phone listens to the paging channel as an incoming phone call, message, or data triggers a paging message of the subscriber in the Location Area. Upon receiving a paging message, the phone contacts the network and requests a dedicated (logical) channel for further communication. Thus, only if the phone switches to another Location Area (circuit switched), it has to inform the network using a \textit{Location Update Request}.  Additionally, the phone sends \textit{periodic} location updates at a low interval (typically every 24h) to reassure the network of its continued presence. Analogue semantics exist for Routing Areas and Tracking Areas in packet-switched context. Additionally, each cell broadcasts a list of neighbor cells (\eg, their frequencies) to help the phone find these cells faster.

\subsubsection{RAN Sessions and Data Tunnels}
\label{sec:architecture_gtp}\label{sec:architecture_bearers}
As most data services need stable addresses, tunnels are used between the UE and an IP endpoint. These tunnels hide the mobile network's mobility management  and also allow to offer multiple connections to different IP networks, such as Internet access or private networks. Tunnels terminate at the packet gateways (\acl{P-GW}, \acs{P-GW}\super{4G}). If necessary, \ac{NAT} middleboxes separate the mobile network from the Internet by translating the private IP address and port to a public IP address and port. Tunnels aim at guaranteeing certain \ac{QoS} parameters, such as latency or bandwidth.

\begin{figure*}[t!]
\centering
\includegraphics[width=1\textwidth]{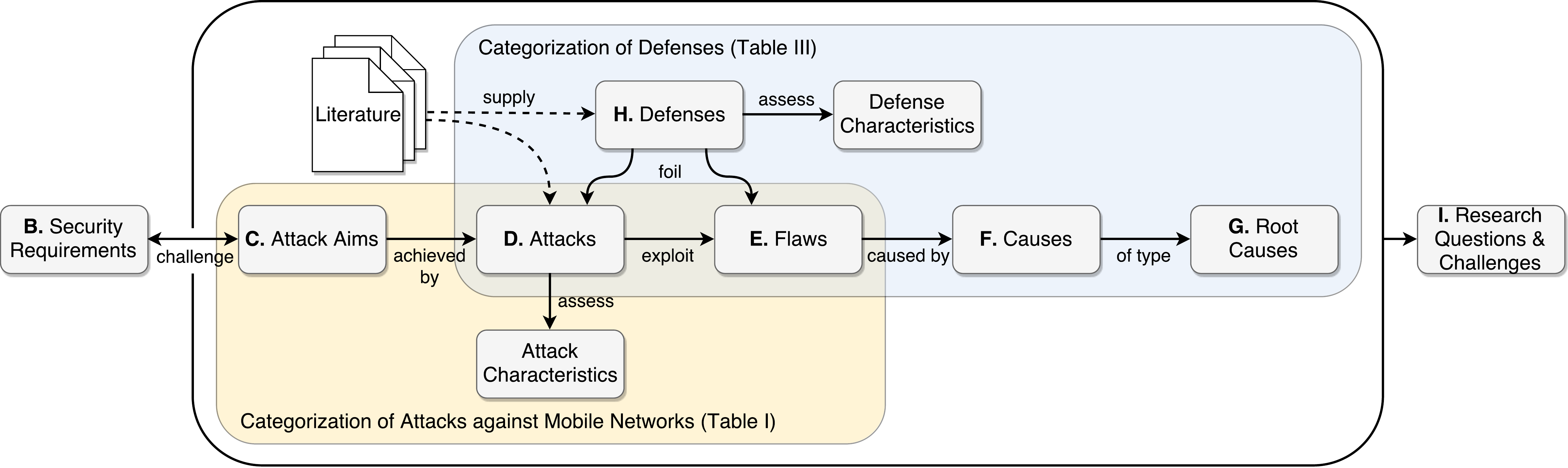}
\caption{Systematization methodology applied throughout this work. The letters correspond to the subsections of Section~\ref{sec:systematizationMethodology} where each aspect is described.}
\label{fig:methodology}
\end{figure*}

\section{Methodology of Systematization}
\label{sec:systematizationMethodology}
In this section, we introduce our systematization methodology that we apply to categorize attacks on mobile networks and their defenses. In the course of this article, we recognize and formulate research questions and challenges for future research based on this systematization.

\subsection{Methodology}

We structure the systematization process according to the flow depicted in Figure~\ref{fig:methodology}, beginning with the selection of particular security requirements, continuing with the assessment of recent attacks and defenses, and resulting in a set of research questions and challenges to shape future research in mobile network security. This process allows us to incorporate multiple aspects of offensive and defensive previous work resulting in a high-level perspective on essential root causes of security issues.

\textbf{Security requirements} define concrete features to protect mobile networks and their users. Such requirements are challenged by \textbf{attack aims}, \ie, the major interests of an attacker (Table~\ref{tab:results}). Accordingly, \textbf{attacks} are instantiations of these attack aims and exploit existing vulnerabilities in the definitions and implementations of systems or system components. In order to assess the impact of attacks on the mobile network, we use a set of \textbf{attack characteristics} that give a precise definition of an adversary's technical and organizational capabilities, \eg, the preliminaries for an attack.  We assess the scope of existing \textbf{defenses} in relation to known attack vectors (Table~\ref{tab:countermeasures}) and aggregate \textbf{defense characteristics} to assess their research and deployment status.
However, multiple flaws are often manifestations of a broader problem that we define as a \textbf{cause}. Such causes facilitate the differentiation of attacks into distinct classes, thus allowing to derive open research questions and challenges in relation to the current state of the art. All causes are grouped into four fundamental \textbf{root causes}, which form the logical structure of our systematization.

Below, we give a short example to illustrate the application of this method.

\paragraph*{Example} \hangindent=2em \textit{Radio Measurement Reports} are used for the maintenance of radio access networks and can be requested by a base station without authentication. This flaw can be exploited via the \textit{Radio Measurement Report Request Attack} that allows an attacker with \textit{active radio capabilities} to pinpoint a victim \cite{Shaik2015}. Hence, this attacks the user's privacy (\textit{attack aim}) as it breaks the location confidentiality \textit{requirement}. The attack can be executed in the non-authenticated state of UEs, allowing an active attacker to fake measurement report requests. Therefore, the \textit{flaw} is that requests for Radio Measurement Reports are part of the unsecured pre-authentication traffic (\textit{cause}). As this is a legitimate request according to the specification, the \textit{root cause} lies in the specification. A proposed \textit{mitigation} is to require this specific request to be authenticated. However, this does not fix the underlying general problem of insecure pre-authentication traffic. A \textit{generic solution} would be to eliminate pre-authenticated traffic completely, as it is the source of many other vulnerabilities as well. An \textit{open research question} is how to develop a privacy-preserving specification while keeping the maintainability of mobile networks.

\subsubsection*{Coverage of Literature}
We focus on academic research, \ie, scientifically peer-reviewed publications, as a foundation for the assessment. In addition, we use non-academic research (including publications, presentations, and demonstrations at hacker venues) and  white papers from industry. Even though these publications are not peer-reviewed, they complement the academic body of the systematization with a comprehensive picture of mobile network security. In particular, the hacker community has contributed a lot to the understanding of mobile protocols~\cite{Wagenknecht2016} and has provided tools that academic researchers have built upon.

We require that the literature must present security- or privacy-related attacks and defenses and need to be unique to mobile networks and not focus solely on applications. We specifically exclude mobile operating systems security and common challenges of the public phone network. Preferentially, the literature should have a high impact, \eg, it affects many users, is operable from a large distance, or produces considerable damage.

\subsubsection*{Structure} We base the structure of our systematization (Sections \ref{sec:rc_si}--\ref{sec:sys_wc}) on the root causes, and each section is further grouped into causes. Thus, attacks that are evoked by the same cause and root cause are logically grouped together. Mitigations and solutions are discussed directly within the respective subsections. Regarding the structure of this document, we traverse the systematization process~(Figure~\ref{fig:methodology}) backwards from root causes, to causes, to attacks. Research questions originate from causes, root causes, and the implications of 5G technologies.

\subsection{Security Requirements}
\label{sec:security_req}
Security requirements describe the demands that need to be met by the system in order to protect the interests of its stakeholders. For our systematization, we aim to establish generic and long-lasting security requirements spanning all three mobile network generations. However, the standardization bodies, \eg, \ac{3GPP}, have issued diverging requirements over time, which is why, they do not allow us to provide a holistic view and might not fit modern security concepts. In order to define generic and long-lasting security requirements, we therefore base our work on the publication of Avizienis~\etal~\cite{Avizienis2004} who define a taxonomy of dependable and secure computing and general security requirements that we underpin with some more concrete requirements published by the 3GPP~\cite{TS22.278,TS33.401,TS33.102}.

\subsubsection{Confidentiality} Avizienis~\etal~\cite{Avizienis2004} define confidentiality as the ``\textit{absence of unauthorized disclosure of information}''. This statement is substantiated by the 3GPP with the following requirement: ``\textit{the network shall provide several appropriate levels of user privacy including communication confidentiality, location privacy, and identity protection}'' \cite[p.~33]{TS22.278}. In detail, this means~\cite{TS22.278}:

\begin{compactitem}
\item Communication confidentiality: ``[\dots] \textit{contents, origin, and destination of a particular communication shall be protected from disclosure to unauthorized parties}''.

\item Identity protection: The network shall ``\textit{hide the identities of users from unauthorized third parties}''. 

\item Location privacy: The network shall ``\textit{hide the user location from unauthorized parties}''. 

\end{compactitem}

\subsubsection{Availability} Availability denotes the readiness and the continuity of correct services~\cite{Avizienis2004}. 
With the pervasiveness of mobile communications in our everyday lives, availability becomes a crucial factor for customers as well as part of critical infrastructures~\cite{Thompson2015,First2016}. 

\subsubsection{System Integrity} 
In contrast to data or transmission integrity, system integrity focuses on the hard- and software of the network components. 
Integrity is defined as the absence of unauthorized system alterations~\cite{Avizienis2004}. System integrity is an essential security requirement as it is crucial for the proper operation and trustworthiness of the system. 

\subsubsection{Unauthorized Service Access and Correct Charging} 
The service should only be accessible to authorized parties~\cite{TS22.278}. This requirement includes correct recording and offsetting call data records and other chargeable items~\cite{TS22.115}. In other words, a proper authorization and charging system should only allow subscribed services to be consumed and it should charge the \textit{right} user for the \textit{correct} volume~\cite{Peng2014}.

Subsequently, we will use these high-level security requirements to assign an attack aim to each identified attack that challenges one or more of those requirements.

\subsection{Attack Aims}
Each attack has a clear primary aim that challenges one of the identified security requirements. An attacker might also pursue a secondary attack aim. For example, using side-channels, an attacker can obtain the shared key on the SIM card that undermines primarily the secrecy aim. However, the attacker might also clone the SIM card for free calls and thereby commit fraud attacks (secondary). We define five distinct attack aims: 
\begin{compactitem}
\item \textbf{Attacks on Privacy:} This aim covers all attacks that undermine the privacy of the user, including the \textit{identity protection} and the \textit{location privacy}.

\item \textbf{Attacks on Secrecy:} This category includes attacks on \textit{communication confidentiality}, \eg, the content of the transmission.

\item \textbf{Denial of Service:} This attack aim contains all the  objectives that impact the \textit{availability} of services, or parts of them. Thus, \textit{downgrade attacks}, such as disabling encryption or stepping back to less secure protocols belong here.

\item \textbf{Attacks on Integrity:} This category comprises all the attempts which undermine the requirements for system integrity.

\item \textbf{Fraud Attacks:} This aim covers attacks that aim towards directly or indirectly targeting financial benefits for the attacker or financial losses for others.
\textit{Direct under-billing} attacks dodge service charges at the expense of the operator, whereas \textit{direct over-billing} produces financial loss to customers. \textit{Indirect fraud} includes scams or spam via telephone.

\end{compactitem}

\subsection{Attacks}
Attacks exploit system flaws under the defined attack aims. We use the following characteristics for an assessment of the attack impact (see Table~\ref{tab:results}). In general, as for Table~\ref{tab:results}, \sokCircle{2}~denotes a fully applicable attack for the characteristic, \sokCircle{1}~refers to limitations, and \sokCircle{0}~characterizes attacks that are not applicable.

\subsubsection{Attacker Capabilities}
\label{sec:attacker_model} 
\label{sec:attackerCapabilities}
For mobile radio attacks, an attacker often combines several capabilities to perform an attack (\eg, retrieving session keys over SS7 and passively monitoring and decoding traffic); thus, 
we describe the attacker model as a set of distinct capabilities (\ie, building blocks). We assume that the attacker is a-priori not in possession of any \textit{private} information (secret keys) of the victim, but might be in possession of \textit{public} identifiers such as the phone number (\ac{MSISDN}).
\begin{compactitem}

\item \textbf{Passive Radio}: An attacker with passive radio capabilities is able to capture radio transmissions, decode signals, and read raw messages. The recent developments of \acp{SDR} and re-purposed hardware render this type of attack quite affordable~\cite{gomez2016,osmocombb}. 

\label{sec:attackerCapabilityRadioPassive}

\item \textbf{Active Radio}: An attacker with this capability has full control over radio transmissions and is therefore able to put arbitrary messages on the radio channel. This enables an attacker to setup a own base station or a fully controllable phone stack using an \acs{SDR} \cite{gomez2016,osmocombb,openlte,openbts,Nikaein2014,Mjolsnes2017}.
\label{sec:attackerCapabilityRadioTrasnsmit}

\item \textbf{User Traffic}: The attacker is able to control or initiate traffic on a commodity mobile phone. The phone performs normal radio emissions, but the attacker accesses the higher (user-land controlled) network layers (\eg, IP) or dedicated network services (\eg, SMS). In most cases, this ability does not require a rooted or jail-broken phone.

\label{sec:attackerCapabilityUserTraffic}

\item \textbf{SS7/Diameter Interface}: An attacker with access to SS7/Diameter is able to send and receive \acl{SS7} or Diameter messages to and from other networks. Some network providers even sell these services \cite{Timberg2014}.
\label{sec:attackerCapabilitySS7}

\item \textbf{Nondestructive Physical}: A nondestructive-physical attacker temporarily has physical access to the victim's device, but neither destroys nor modifies hardware or software. Thus, the attack leaves no visible or measurable trace. We exclude the destructive attacker, because these visible traces would raise doubts by the users.
\label{sec:attackerCapabilityPhysical}

\item \textbf{PSTN Interface}: An attacker has access to voice or text services of the \acf{PSTN}. 
\label{sec:attackerCapabilityPSTN}

\item \textbf{Internet Traffic}: An attacker with the ability to access the Internet in a way that can specifically contact the victim's phone. That can be achieved by knowing the phone's public IP address and the TCP/UDP port mapping on the operator's packet gateway. Other possibilities include identifiers of chat services, instant messaging apps, or cloud messaging services (such as \ac{GCM}\cite{gcm2017} or Apple's Push Notifications~\cite{apn2017}), and the ability to transfer such messages.
\label{sec:attackerCapabilityInet}

\end{compactitem}

\subsubsection{Limitations of Attacker Capabilities}
For our systematization, we assume that the operator's authorized personnel is trusted and thus exclude such attacker capabilities from systematization. However, such attacks have occurred in the past and are a threat to the mobile user's data secrecy and privacy \cite{Prevelakis2007,Scahill2015}. For instance, in the 2005 Vodafone Greece incident \cite{Prevelakis2007}, a staff technician was suspected to have planted a backdoor in mobile switches that allowed copying traffic on government phones. 
In the Gemalto SIM key material theft~\cite{Scahill2015}, secret key material for the SIM cards was transferred by the use of unprotected means. However, such attacker capabilities are beyond the scope of this paper, as the attackers had the permissions in the first place and deliberately misused them.

\subsubsection{Target} 
The target category depicts who is harmed by the attack, and if there is a relation to other categories, \eg, privacy attacks predominantly target the user. We focus on the primary goal and disregard secondary effects such as bad publicity due to data breaches.

\subsubsection{Technology} 
\label{sec:technology}
This category maps the applicability of an attack to the three major access technology generations and assesses if there has been a security development, \eg, if defenses have been introduced in later access technologies, or if new technologies open new attack vectors. The former does not necessarily prevent attacks, as multiple downgrade attacks are known.

For example, only \acs{GSM} lacks mutual authentication (\textit{cause}), hence it is prone to \ac{MitM} attacks. However, UMTS and LTE are open to various downgrade attacks; therefore the problem will not be resolved until phones stop to (unconditionally) support GSM. Downgrade attacks that trick or force a specific party to fall back on older and less secure access technology must be kept in mind when discussing fixes or mitigations for new access technology generations. We filed downgrade attacks as part of \ac{DoS} attacks, as they deserve a separate review.

\subsubsection{Range}
The range of an attack is an indirect indicator of impact and cost. 
A higher range (\eg, a globally performable attack) increases its impact and versatility and might justify higher costs for an attacker. In contrast, an attack that requires more physical vicinity increases involvement of the attacker and reduces the set of victims. In Table \ref{tab:results}, we classify the range by technical boundaries: Physical access (\textbf{Phy}), same radio cell including simulated ones (\textbf{Cell}), same location area (\textbf{LA}), same network (\textbf{Net}), and globally executable attacks~(\textbf{Glo}).

\subsection{Flaw}
For our systematization, we define a flaw as a specific and distinct vulnerability that is exploited by a particular attack. We coalesced attacks that exploit the same technical flaw or are otherwise very similar in their technical or operational principle. This leads to a 1:1 relationship between flaws and attacks.

\begin{table*}[!pht]
	\sokRenderTableIfUpdate
	\caption{Categorization of mobile security attacks by their aim}
	\label{tab:results}
	\centering \resizebox{\textwidth}{!}{
		\renewcommand{\arraystretch}{1.3}
		\sokDisplaytable 	%
	} %
	\begin{minipage}{\textwidth}
		\raggedright
		\vspace{2mm}
		\hspace{3mm}
		\sokCircle{2} yes, applicable, needed for attack 
		\hspace{2em} \sokCircle{1} partially/supportive/optional 
		\hspace{2em} \sokCircle{0} no, not applicable, or does not apply
		\hspace{2em} \textbf{?} property unknown
	\end{minipage}	
\end{table*}

\subsection{Cause}
We group flaws that have similar technical or organizational reasons via a common cause. A cause is a broader technical reason summarizing multiple individual flaws and, if dealt with appropriately, would foil an entire class of attacks.

\subsection{Root cause}
Root causes are the underlying reason for certain classes of attacks; they are defined on an abstract level and independent from technology. Each root cause summarizes particular flaws and vulnerabilities according to their structural or causal dependency (Table~\ref{tab:root_cause_overview}) and is completely disjoint from other root causes. We use this abstract structure as the foundation for our systematization.

\subsection{Defenses}
In order to systematize and assess the defenses presented in the literature, we relate them to attacks, causes, and root causes. We show the coverage of the suggested defenses in Table~\ref{tab:countermeasures}. If a defense encounters all the attacks of a cause, it can be considered a \textit{generic} defense, otherwise it is a \textit{specialized} defense. We differentiate between two kinds of defenses, namely, \textbf{detections} (\LEFTcircle) and \textbf{mitigations} (\CIRCLE). While the detection of an attack is an important step to impede losses or disadvantages, it does not ultimately prevent the attack. Mitigations foil attacks by fixing the underlying flaw or by making attacks very unlikely to succeed.
 
Additionally, we evaluated the defenses according to a set of quality characteristics that help to assess the realization effort, the sustainability, and the current research status of a defense.

\subsubsection{Realization Method}
The realization method (Table~\ref{tab:countermeasures}) specifies \textit{how} a defense is achieved.
A \textbf{specification defense}~(S) needs to pass the \ac{3GPP} change-request process, such as protocol changes, to guarantee the interoperability. It has a higher realization effort time-wise, as it needs to pass the specification process. However, it will potentially reach more people in the long run, as new products are likely to implement such measures. In contrast, \textbf{implementation defenses}~(I) do not depend on a specification and can be implemented directly into network components. While this task can be accomplished by a vendor, only users of an updated product benefit from it.

The realization method is an indicator of the \textit{effort}, the \textit{swiftness} with which it can be rolled out, and the \textit{reachability}.

\subsubsection{Affected Components}
On the right hand side of Table~\ref{tab:countermeasures}, we denote which network components need modifications to implement a particular countermeasure. On the one hand, it indicates who needs to take action to roll-out a defense. On the other hand, as a rough estimate, the less components are affected by a defense measure, the easier it can be implemented in practice, and vice-versa. However, particular updates on the UE are hard to roll out due to the high number of units and manufacturers involved.

\subsubsection{Deployment Status}
The deployment status has more a practical than a scientific value: It helps to evaluate the feasibility of a defense, if information about its deployment is available. A defense which has been deployed by at least one operator can naturally be considered feasible. The following notation is used:
\begin{compactitem}
\item \textbf{No information} (?). No information about deployment.
\item \textbf{Not deployed} (\Circle). Not deployed by any operator.
\item \textbf{Partially deployed} (\LEFTcircle). At least one operator or phone that partially implements the defense. 
\item \textbf{Fully deployed} (\CIRCLE). At least one operator or phone that fully implements the countermeasure.
\end{compactitem}

\subsubsection{Research Status}
In the literature, defenses are discussed at vastly different levels of detail. Some papers about attacks conclude with rather vague defense proposals, while other works focus on the concrete realization and evaluation of a defense. The research status of a defense helps us to disclose possible shortcomings of the existing work. For this, we estimated the research status by examining the detail level and evaluation degree of a defense proposal and denote three status levels in Table \ref{tab:countermeasures}:
\begin{compactitem}
\item \textbf{Vague Proposal} (\Circle). A vague proposal mentions a defense without any precise scheme or architecture. 
\item \textbf{Concrete Proposal} (\LEFTcircle). A concrete proposal is a detailed scheme for a defense. However, the scheme is not evaluated.
\item \textbf{Evaluated Proposal} (\CIRCLE). The security of the proposal was comprehensively evaluated---preferably by more then one literature source. The evaluation can be done either theoretically and/or practically.
\end{compactitem}

\subsection{Research Questions and Challenges}
From the assessment of attacks and countermeasures, we derived open questions and challenges that should shape future security research in context of 5G mobile security. Following the systematization structure, we defined three leading questions or challenges for each cause:

\textbf{Shortcomings of Existing Work}. 
The limitations of prior research lead to individual sets of shortcomings for the different causes that we identified through our systematization. We define concrete starting points to address these shortcomings in future offensive and defensive work.

\textbf{Concrete Challenge of Defenses}. As we systematize the defenses in the context of causes, we emphasize the problems and challenges that all proposed defenses are trying to solve. We evaluate if the defenses are successful and might encounter multiple attacks in the cause. If they are not already solved successfully or just covering a small amount of attacks in a cause, we give starting points using literature of other related disciplines in which similar problems were solved. %

\textbf{Security Implications of 5G}. When possible, we match these research questions onto the current 5G specification development. As the 5G is currently under development, we briefly refer to discussions and suggestions made within technical reports \cite{TR33.899}. This matching assists us in finding the difference between current research and the specification process. From today's perspective, we point out security challenges of 5G technologies that should be addressed by future security research. To do this, we introduce the new technologies as well as the associated 5G challenges.

\begin{table}[!t]
	\centering
	\caption{Root causes related to causes}
	\label{tab:root_cause_overview}	
	\renderRootCauseOverview %
	\resizebox{\linewidth}{!}{\	
		\showRootCauseOverview
	}
\end{table}
\begin{table*}[!pht]
	\noindent %
	\caption{Categorization of defenses}%
	\label{tab:countermeasures}%
	\vspace{-2mm}
	\noindent\resizebox{!}{0.44\textheight}{\noindent	
	\renewcommand{\arraystretch}{1.4} %
	\setlength{\tabcolsep}{1.8pt} %
	\newcommand{\counterwidth}{85mm} %
	\hspace{-6mm}	
	\showCounterTable
	} %
	\scalebox{0.7}{\parbox{1.3\linewidth}{
		\raggedright
		\vspace{3mm}
		Countermeasures: \hspace{1em}
			\sokCircle{2} effective mitigation/solution 	
			\hspace{2em} \sokCircle{1} detection, partial mitigation, probabilistic mitigation
			\hspace{2em} \sokCircle{0} not applicable
		\newline
		Components: \hspace{1em}
			\sokCircle{2} implementation requires changes of that component
			\hspace{2em} \sokCircle{1} optional or minor changes
			\hspace{2em} \sokCircle{0} no changes necessary 
		\newline
		Realization Method: \hspace{1em} \textbf{S} by specification change/amendment/extension 
			\hspace{2em}\textbf{I} by implementation (update) 
		\newline
		Research Status: \hspace{1em} \sokCircle{0} vague proposal
		\hspace{2em} \sokCircle{1} concrete proposal
		\hspace{2em} \sokCircle{2} evaluated proposal 
		\newline
		Deployment Status: \hspace{1em} \textbf{?} unknown
			\hspace{2em} \sokCircle{0} not deployed or deployable
			\hspace{2em} \sokCircle{1} partially deployed
			\hspace{2em} \sokCircle{2} deployed and known to work from at least one Carrier
	}}
\end{table*}

\section{Systematization Overview}
In the following, we introduce the abstract \textit{root causes} and their individual \textit{causes} as used in this systematization 
(see Table~\ref{tab:root_cause_overview}). Later, we present each cause in detail in relation to attacks, defenses, and open research questions, following the reverse process of Figure~\ref{fig:methodology}. The results regarding attacks are condensed in Table \ref{tab:results} and as graph in Figure \ref{fig:graphviz_diagram_all} (see Appendix). Likewise, defenses are aggregated in Table \ref{tab:countermeasures}.

\textit{Specification Issues} originate from incomplete, inaccurate, or faulty definitions of system behavior and comprise five individual causes: \textit{Unsecured Pre-Authentication Traffic} allows to send messages to the phone or network prior to the key agreement and ciphering setup has taken place. 
\textit{Non-Existing Mutual Authentication} relates to an attack vector exclusive to \acs{GSM} networks, but is still an issue in recent technologies due to downgrade attacks. 
The use of \textit{Weak Cryptography} significantly reduces the required effort for attacks on encrypted data, while \textit{Insecure Inter-Network Protocols} undermine the users' privacy and confidentiality by poor protocol design choices. \textit{Resource Usage Asymmetry} enables an attacker to perform cheap requests that result in intensive operations on the network side and hence can lead to \ac{DoS}.

\textit{Implementation Issues} are either caused by \textit{Insecure Implementations} that open attack vectors in components of the mobile network, which can impair the system's integrity, or by \textit{Leaky Implementations}, which means that sensitive information can be accessed through unintended side channels.

The root cause \textit{Protocol Context Discrepancy} summarizes a class of security issues which use a protocol differently or in another environment than it was originally designed for.
\textit{Cross-Layer Information Loss} occurs at the interface of different layers in the network stack, \eg, when necessary, trustworthy security information is lost between network layers. \textit{Accounting Policy Inconsistencies} result from different accounting schemes that can be played against each other, \eg, voice data is charged in minutes, whereas other data is charged by volume.
  
The \textit{Wireless Channel} and its characteristics is essential for the transmission of information in mobile communication and comes with several physical limitations that impact the security. 

Following the structure of root causes, we discuss offensive and defensive characteristics of specific causes and derive open research questions and challenges for future mobile network technologies.

\soktablestart %

\section{Root Cause: Specification Issue}
\label{sec:rc_si}
Specifications ensure the interoperability between implementations by specifying protocols, state machines, and interfaces. However, there may exists issues in the specification that might lead to flaws that can be exploited by attackers. In the specification-related root cause, we collate all flaws that are based on specification issues. The reasons for these problems range from technical trade-offs to political motivation.

\subsection{Cause: Unsecured Pre-Authentication Traffic}
\label{sec:UnsecuredPreAuthTraffic}
The signaling traffic prior the security establishment with the \ac{AKA} protocol is unprotected: it is neither encrypted nor integrity-protected and thus unauthenticated\footnote{In LTE some uplink data is integrity protected but not encrypted.}. This leads to implicit trust between the phone and the network. In this unauthenticated state, the phone fully obeys the network, even if the latter is not genuine. A malicious usage of messages in this unauthenticated state can serve for downgrade, track, or locate a specific user or handset.

\subsubsection{Attacks}
\label{sec:IMSICatcher}
One prominent example for attacks based on unsecured pre-authentication is the deployment of \textit{fake base stations}. Fake base stations (also known as \textit{rogue} or \textit{fraudulent base stations}, \textit{IMSI Catchers}, \textit{cell-site simulators}, a \textit{DRT-Box}, or by product names such as \textit{Stingray}) are active devices simulating a genuine base station to the phone by broadcasting genuine network identifiers. These fake base stations exploit the fact that mobile phones cannot verify the authenticity of the network prior to the \ac{AKA} protocol. 

In the unauthenticated state, the attacker is allowed to ask for the permanent identity, such as the \acs{IMSI} or \acs{IMEI}, and can thus undermine the user's identity and location privacy \cite{ Nohl2014, Malette2015, AIMSICD, Paget2010, Mjolsnes2017, TS24.301, Borgaonkar2017}. 
Besides obvious requests such as the \textit{identity request}, an attacker can also use more subtle ways to determine the vicinity of a victim, \eg, with the \ac{AKA} linkability attack \cite{Arapinis2012}. 
Additionally, an attacker can repeatedly page the victim's IMSI~\cite{Arapinis2012} and, thereby, determine if a user is in radio range. Moreover, an attacker can retrieve a more precise location by requesting measurement reports from the victim's handset~\cite{Shaik2015, Forsberg2007} enabling an attacker to track a victim or to request the identity of people within radio range.

Furthermore, unsecured pre-authentication traffic allows \textit{downgrade attacks} to a less secure access technology by denying service using the \textit{tracking area update reject} or a combination of other messages  \cite{TS24.301,Zhang2016,Shaik2015}. This serves as a stepping stone for further attacks such as the \acs{GSM} \ac{MitM} attack (Section~\ref{sec:MITMimsicatcher}). Additionally, a fake base station can disable the location service on some phones from the late 2000s by sending out the country code of Egypt, because GPS receivers used to be forbidden in Egypt at that time~\cite[p.\,28]{Welte2008}, \cite{Farivar2008,nokiagps}.
Furthermore, Golde \etal~\cite{Golde2013} showed that unsecured pre-authentication uplink traffic in \acs{GSM} can be misused for a \ac{DoS} attack dropping calls in the entire location area by winning the race answering paging requests. This is a problem of the GSM state-machine specification, as it can not recover once it proceeds to the ciphering setup.

The missing protection of broadcast and paging messages also enables attacks that retrieve the temporary identity of a victim by triggering the paging process multiple times and statically analyzing the paged \ac{TMSI}s~\cite{Shaik2015, Kune2012}. 
An attacker can trigger the paging procedure in multiple ways: for example, with targeted Internet traffic, a short phone call and immediate hang-up, \eg, before the ring starts, or with a \textit{Silent SMS} that is a text message which is silently discarded by the phone\cite{TS23.040}. 

\subsubsection*{Assessment}

Attacks based on unauthenticated uplink traffic or on passively exploitable downlink traffic are vastly outnumbered by active radio attacks based on pre-authentication traffic. 
While potentially having a very severe impact, an active radio attacker is limited to his/her radio vicinity. Most of these attacks undermine the victim's data or location privacy. Many commercially available products exploit unsecured pre-authentication traffic \cite{pki_imsi_catcher2017, Biddle2016}, hence making it a high priority to be addressed.

\newCountermeasure{tmsichange}{Frequent TMSI Reallocation}{Arapinis2014, TR33.821, Hong2018}{core,enb*}{I}{?}{2} 

\sokentrystart\sokVector{Air Interface} \sokAim{Privacy} \sokAttack{TMSI Deanonymization (Paging Attack)} \sokCapabilitiesList{airpassive,pstn*,inet*,ss7*} \sokAttackerM{Passiv Radio.}\sokTarget{U} \sokReference{Shaik2015, Kune2012, Hong2018} \sokTech{2}{2}{0} \sokRange{0}{2}{2}{0}{0} \sokCausesList{pat} \sokAddCountermeasure{pubkey} 
\sokAddCountermeasure{tmsichange*,pubkey}
\sokentryend

\sokentrystart\sokVector{Air Interface} \sokAim{Service}\sokAttack{GPS Receiver Denial of Service} \sokAttackerM{Act. Radio} \sokTarget{U} \sokCapabilitiesList{airactive} \sokRange{0}{2}{0}{0}{0} \sokReference{Farivar2008, nokiagps} \sokCause{pat} \sokAddCountermeasure{pubkey}  \sokTech{2}{0}{0} \sokentryend

\sokentrystart\sokVector{Air Interface} \sokAim{Privacy}\sokAttack{Location/Tracking Area not Allowed (Downgrade)} \sokAttackerM{Act. Radio} \sokTarget{U} \sokCapabilitiesList{airpassive,airactive} \sokRange{0}{2}{0}{0}{0} \sokReference{Zhang2016,Shaik2015,TS24.301} \sokCause{pat} \sokAddCountermeasure{pubkey}  \sokTech{0}{2}{2} \sokentryend

\sokentrystart\sokVector{Air Interface} \sokAim{Privacy}\sokAttack{Measurement Reports Localization} \sokAttackerM{Act. Radio} \sokTarget{U} \sokCapabilitiesList{airpassive,airactive} \sokRange{0}{2}{0}{0}{0} \sokReference{Shaik2015, Forsberg2007} \sokCause{pat} \sokAddCountermeasure{pubkey}  \sokTech{0}{0}{2} \sokentryend

\sokentrystart\sokVector{Air Interface}\sokAim{Privacy}\sokAttack{AKA Protocol Linkability Attack} \sokTarget{U}\sokCapabilitiesList{airpassive,airactive}\sokReference{Arapinis2012,Borgaonkar2017}\sokTech{0}{2}{2}\sokRange{0}{2}{0}{0}{0} \sokCause{pat} \sokAddCountermeasure{pubkey} 
\sokentryend

\sokentrystart\sokVector{Air Interface}\sokAim{Privacy}\sokAttack{IMSI Paging Attack} \sokTarget{U}\sokCapabilitiesList{airpassive,airactive}\sokReference{Arapinis2012}\sokTech{2}{2}{2}\sokRange{0}{2}{0}{0}{0} \sokCause{pat} \sokAddCountermeasure{pubkey} 
\sokentryend

\newCountermeasure{pmsi}{Pseudonymous IMSI and Non-Static Identifiers}{vandenbroek2015,TR33.821,Khan2015,Ateniese1999,Choudhury2012}{sim,enb,core,hss,pgw,sgw,inter}{S}{0}{0}
\newCountermeasure{pubkey}{Asymmetric Cryptography with PKI}{TR33.821,Horn1998,Kambourakis2004}{sim,hss,inter}{S}{1}{0}
\newCountermeasure{btsfingerprint}{Base Station Fingerprinting}{Li2017,Dabrowski2014,Malette2015,Ney2017,AIMSICD,Snoopsnitch,Park2017}{ue*}{I}{2}{2} 
\newCountermeasure{btsgeodb}{Geographic Mapping of Network Infrastructure}{Dabrowski2014,Malette2015, AIMSICD, Snoopsnitch, Steig2016}{ue*}{I}{2}{2} 
\newCountermeasure{btsbehave}{Passive Behavioral Analysis of Control Traffic}{Snoopsnitch,GSMKcryptophone2017}{ue}{I}{2}{2}
\newCountermeasure{MNOlocationInconsistency}{Location Updates Inconsistency Check}{Dabrowski2016}{core*}{I}{1}{?}
\newCountermeasure{MNOroundtriptime}{Round-Trip Time Analysis}{Dabrowski2016}{core*}{I}{1}{?}
\newCountermeasure{MNOdisappearance}{Check for Disappearing Phones}{vanDo2015}{core*}{I}{0}{?}
\newCountermeasure{phoneradioreports}{Collect Neighbor Cells IDs by Phones and BTSs}{Steig2016}{core}{I}{1}{?}
\newCountermeasure{dedicatedmonitoring}{Monitor Radio Vicinity for Changing BTSs w/ dedicated HW}{GSMKOverwatch2017, Dabrowski2014, Ney2017, NetworkGuard}{ue}{I}{2}{2}

\sokentrystart\sokVector{Air Interface}\sokAim{Privacy}\sokAttack{Unauthenticated IMSI Request (IMSI Catcher)}\sokTarget{U}\sokAttackerM{Active Radio.}\sokCapabilitiesList{airpassive,airactive}\sokReference{Dabrowski2014 ,Dabrowski2016, Nohl2014, Malette2015, AIMSICD, Snoopsnitch, Paget2010,Mjolsnes2017}\sokTech{2}{2}{2}\sokRange{0}{2}{0}{0}{0}\sokMitigation{}
\sokAddCountermeasure{pmsi,pubkey,btsgeodb*,btsfingerprint*,btsbehave*,phoneradioreports*,dedicatedmonitoring*}
\sokCause{pat}
\sokentryend

\sokentrystart\sokVector{Air Interface}\sokAim{Privacy}\sokAttack{Unauthenticated IMEI Request}\sokTarget{U}\sokAttackerM{Active Radio.}\sokCapabilitiesList{airpassive,airactive}\sokReference{}\sokTech{2}{2}{0}\sokRange{0}{2}{0}{0}{0}
\sokCause{pat}
\sokAddCountermeasure{btsgeodb*,btsfingerprint*,btsbehave*}
\sokReference{Dabrowski2014, Nohl2014, Malette2015, AIMSICD, Snoopsnitch, Paget2010}
\sokentryend

\newCountermeasure{pagingstatistics}{Paging Repeat Statistics}{Dabrowski2014}{core*}{I}{2}{?}

\sokentrystart\sokVector{Air Interface} \sokAim{Service} \sokAttackerM{Radio Act.} \sokTarget{P} \sokAttack{Paging Response Race DoS} \sokCapabilitiesList{airpassive,airactive} \sokRange{0}{2}{0}{0}{0} \sokReference{Golde2013} \sokTech{2}{?}{?} \sokCausesList{pat} \sokAddCountermeasure{machinereadablespec,pagingstatistics*}\sokentryend

\subsubsection{Defenses}
\label{sec:PATdefenses}
The research community proposed and implemented multiple detection and mitigation techniques against fake base stations. 
Detection schemes against the client include geographical mapping (\eg, via GPS) of the network structure to detect new---and possible fraudulent---base stations, finding unusual frequency or cell parameter configurations, and behavioral analysis of base stations. 
Some are implemented as smartphone apps \cite{Dabrowski2014,AIMSICD,Snoopsnitch,Li2017} others use dedicated smartphones with baseband firewalls \cite{GSMKcryptophone2017}. Furthermore, a sensor network can detect such changes \cite{Dabrowski2014, GSMKsensorNetwork2017,Ney2017}. Recently, operator-based detection schemes were proposed \cite{Dabrowski2016, vanDo2015, Steig2016}. Some of these approaches also detect large-scale paging race attacks~\cite{Dabrowski2014}.

Besides detection schemes, mitigations and fixes based on protocol changes have been proposed. An ephemeral identifier, \eg, dynamic IMSI or pseudonymic IMSI (P-IMSI) instead of the static IMSI has been proposed~\cite{vandenbroek2015,Khan2015,Hassidim2016}. All of them protect against an unauthorized IMSI request by providing a new, seemingly unrelated number as the identifier for each request. However, the ephemeral identifier require severe changes to the network structure as the IMSI is used as a primary key to link all the subscriber-related information in a network.  
Additionally, protocol changes have been proposed against the paging response race attack and the \ac{AKA} linkability attacks \cite{Golde2016, Arapinis2012}.

Besides these specialized solutions, securing the paging and other broadcast traffic would eliminate the cause for all these attacks, \eg, using a source origin authentication scheme. Some proposed options are based upon asymmetric cryptography with a public key infrastructure~\cite{Horn1998,Kambourakis2004,TR33.899} or upon broadcast authentication schemes~\cite{TR33.821}. 

Attacks that map a public identifier to the temporary identifier (TMSI de-anonymization) are currently countered by frequently changing TMSIs~\cite{Arapinis2014, TR33.821,Shaik2015} even after unsuccessful calls. Hong~\etal pointed out that TMSI reallocation schemes lack randomness in the real world and remain attackable~\cite{Hong2018}. A proper reallocation scheme must be eimplemented by the operators for encountering the threat of identity mapping attacks. The GSMMap project~\cite{GSMmap} provides a rough estimation on the deployment of this workaround.

\subsubsection{Research Questions and Challenges}

The detection of fake base stations via the handset (e.g., behavioral analysis) or with externally deployed sensors only benefits a small number of people or a certain geographical area.

Countermeasures based on protocol changes often have a hard time to get through specifications for current network generations, as they introduce non-backward-compatible changes. For example, dynamic identifiers such as PMSI (as a replacement for TMSI and IMSI) require deep changes in many systems at once (VLR, HLR, Packet Gateways). The importance of such measures influenced the 5G design process and some of them might be included in the new specification~\cite[§5.7.4.3 ff.]{TR33.899}. However, they introduce the risk of a permanent SIM card lock if the SIM and home network HLR/HSS should get  out of sync. New solutions that address this need to be sought. 
In contrast, moving measurement reports from an unauthenticated to an authenticated protocol state is possible in current network generations, as the functionality of currently deployed handsets is unaffected (but stay vulnerable). However, in recent years, new vulnerabilities based on pre-authentication traffic have been revealed~\cite{Shaik2015}. In conclusion, none of the proposed protocol changes abandons non-protected pre-authenticated traffic in its entirety.

Thus, a more general approach is based on abandoning pre-authentication traffic in particular -- or unprotected signaling in general -- completely.
Two main ideas have been proposed: Schemes based on asymmetric cryptography and broadcast-authentication schemes such as the TESLA protocol \cite{Horn1998,TR33.821,Kambourakis2004,Perrig2002}. For example, 5G currently explores ways to protect base station identity signatures using asymmetric cryptography \cite[§5.4.4.8]{TR33.899}.
However, both were not extensively researched in the context of mobile networks with its limited bandwidth, power consumption restrictions, and low computational USIM card. Another related question is, if the protection of signaling uplink traffic from the \ac{UE} to the network would increase the over-all security of the system. Such a general solutions would be desirable for future network generations as to remove the entire threat class.

Nevertheless, formal analysis of proposed protocols should prove the authenticity, confidentiality, privacy, and availability requirements~\cite{Alt2016,Arapinis2012,Hussain2018}. This can be either accomplished by manually proving the protocols or the use of automated tools.

\begin{mdframed}[backgroundcolor=blue!5,nobreak] 
\textbf{Summary:} Attacks based on pre-authentication traffic affect the privacy and aim to downgrade the service. Such attacks are possible across all three mobile generations. Defenses include either attack detection or protocol changes that aim to mitigate specific attacks or abandon the entire pre-authentication traffic. Future research must focus on completely abandoning pre-authentication traffic, \eg, with asymmetric cryptography. Automated and manual inspection aid the goal of provable security by identifying shortcomings.
\end{mdframed}

\subsection{Cause: Non-Existing Mutual Authentication}
The original specification of \acs{GSM} does not include network authentication and, thus, allows a \ac{MitM} attack. While 3G \ac{AKA} can be used in GSM if supported by all parties, no downgrade prevention exists \cite{Meyer2004}.

Although the fact of non-existing mutual authentication originally exploits a specific vulnerability of \acs{GSM} networks, they are still a relevant threat in today's networks as the weakest-link-principle applies. Downgrade attacks via unsecured pre-authentication traffic on UMTS or LTE (Section~\ref{sec:UnsecuredPreAuthTraffic}) still allow to exploit this GSM vulnerability on modern phones.
The difference to pre-authentication traffic (Section~\ref{sec:UnsecuredPreAuthTraffic}) is the lack of mutual authentication. In that sense, the non-existing mutual authentication is an extension of the unsecured pre-authentication traffic issue and has similarities in attacks and defenses with the former cause. 

\subsubsection{Attacks}
\label{sec:MITMimsicatcher}
If the phone cannot verify the authenticity of the network, an unconditional trust of the phone to the network and, thus, to a potential attacker is established. In a network-centric architecture, where most decisions are made by the network, an attacker faking a base station gains excessive power over the handset.

Fake base stations often employ additional techniques to keep a victim in the fake cell, such as not supplying information on neighboring cells or manipulating cell reselection thresholds~\cite{Dabrowski2014}. 
The phone behaves inconspicuously and is able to make phone calls as well as send text messages and data to the fake network. However, without any further exploit, the attacker can not gain the possession of the cryptographic keys. Still, the attacker can downgrade the communication to the null-cipher or an easily attackable cipher (see Section~\ref{sec:GSMciphers}) for passing it to the real network. In this case, the phone remains reachable for the genuine network. Alternatively, calls, SMS, and data could be forwarded with additional modems or SIP, in which case the original caller-ID is lost, and the phone is not reachable from the outside. The impact of the attacks can be increased by an attacker with SS7 capabilities, \eg, she/he can directly inject the traffic into the phone network. 

Similar to measurement reports on LTE, the GSM radio resource location service protocol enables the network to request GPS coordinates from the phone~\cite{TS04.31}. Developed for emergency services, most phones will answer the request even though it is not an emergency call~\cite{Welte2009}. Additionally, the non-existing mutual authentication has been a rich source for location-based SMS spam---mainly in China~\cite{Li2017,Muncaster2014}. Commercially available fake base stations with ready-to-use exploits are a reminder of the urgency with which this threat should be addressed.

\sokentrystart\sokVector{Air Interface}\sokAim{Secrecy}\sokAttack{MitM IMSI Catcher}\sokTarget{U}\sokReference{Dabrowski2014, Dabrowski2016, Nohl2014, Malette2015, AIMSICD, Snoopsnitch} \sokCausesList{nonmutal} 
\sokCapabilitiesList{airpassive,airactive}
\sokTech{2}{0}{0}\sokRange{0}{2}{0}{0}{0}\sokMitigation{}\sokAttackerM{Act. Radio}
\sokAddCountermeasure{pmsi,pubkey,btsgeodb*,btsfingerprint*,btsbehave*,phoneradioreports*,dedicatedmonitoring*}
\sokAddCountermeasure{retrofitgsmmutual,MNOdisappearance,MNOroundtriptime,MNOlocationInconsistency,disableA50ue*}
\sokentryend

\sokentrystart\sokAim{Fraud}\sokAttack{Fake Base Station SMS Spam} \sokTarget{U} \sokCapabilitiesList{airactive}\sokReference{Li2017,Hassidim2016}\sokTech{2}{1}{0}\sokRange{0}{2}{0}{0}{0} \sokCause{nonmutal}\sokAddCountermeasure{btsfingerprint,retrofitgsmmutual} \sokentryend

\sokentrystart\sokVector{Air Interface} \sokAim{Secrecy} \sokAttack{Encryption Downgrade} \sokTarget{U} \sokCausesList{nonmutal} \sokTech{2}{0}{0} \sokRange{0}{2}{0}{0}{0} \sokCapabilitiesList{airactive} \sokAddCountermeasure{btsfingerprint*,btsbehave*,retrofitgsmmutual} \sokReference{Dabrowski2014, Dabrowski2016, Nohl2014, Malette2015, AIMSICD, Snoopsnitch} \sokentryend

\subsubsection{Defenses}
Similar to unsecured pre-authentication traffic, fake base stations that exploit the non-existing mutual authentication can be detected with mobile apps, baseband firewalls, sensors, or network-based sensors~\cite{Dabrowski2014,AIMSICD,Snoopsnitch,Dabrowski2016,Nohl2014,Malette2015,Ney2017}. 
Nevertheless, mobile apps do not sufficiently protect against the threat of fake base stations~\cite{Park2017}. Besides that, fake base stations that send out SMS spam can be detected based on the content of the SMS and meta-information (\eg, signal strength, duration of cell contact)~\cite{Li2017}. Many defenses (and caveats) from unsecured pre-authentication (Section \ref{sec:PATdefenses}) traffic also apply here.

While \acs{UMTS} and later network generations introduced mutual authentication, the literature urges retrofitting mutual authentication to GSM. Some of the proposed schemes need major changes in the specification~\cite{Kumar2006,Chang2005}, while others focus on minimizing changes to ensure a fast roll-out~\cite{Khan2016,Lee2011}. Different proposals for UMTS authentication schemes over GSM exist \cite{ericsson2003, ericsson2004}, but the current used one does neither protect against downgrades nor integrity-protect the CipherModeCommand. Thus, giving an MitM-attacker attacker the ability to choose a weak or no cipher.

\newCountermeasure{retrofitgsmmutual}{Retrofitting Mutual Authentication to GSM}{Kumar2006,Chang2005,Khan2016,Lee2011,ericsson2003, ericsson2004}{sim,ue*,enb*,core*,hss}{S}{1}{0}

\subsubsection{Research Questions and Challenges}
Even the phase-out of \acs{GSM} by some network providers does not protect users against \ac{MitM} attacks, as the phone continues to ``speak'' \acs{GSM}. Research can thus proceed into several directions. Modern authentication protocols could be retrofitted into GSM with a downgrade protection that prohibits legacy GSM operations if both the phone and the network can support newer methods. A downgrade protection scheme will also benefit future network generations with their security problems. For this, we refer to the causes \textit{unsecured pre-authentication traffic} and \textit{wireless channel}, as they are responsible for downgrade attacks.

\begin{mdframed}[backgroundcolor=blue!5,nobreak] 
\textbf{Summary:} GSM has no network authentication, which leads to privacy and confidentiality issues. Nevertheless, the GSM specification will not be addressed by any improvements. For future generations downgrade attacks to the insecure GSM standard open old attack vectors. Future research must suggest technologies for prevention downgrade attacks, \eg, by securing the pre-authentication traffic. Furthermore, the retrofitting network authentication to GSM provides protection in case of a downgrade attack. 
\end{mdframed}

\subsection{Cause: Weak Cryptography} 
Cryptography provides the means to achieve data confidentiality. However, weak cryptography can lead to attacks revealing protected information. This can emerge from intentionally weakened algorithms or by evolving attack methods~\cite{Pell2014}. Cryptographic systems should be designed following Kerckhoffs' principle~\cite{Kerckhoffs1883}, which states that a system should only rely on the secrecy of the key, while everything else might be known to the attacker (or the public). In the following, we describe attacks that aim specifically at weak cryptography. 

\begin{table}[b]
	\noindent
	\caption{Cipher Overview}
	\label{tab:ciphering}
	\resizebox{1\linewidth}{!}{
	\renewcommand{\arraystretch}{1.3}
	\begin{tabular}{|l|l|l|m{2cm}|c|}
		\hline
			& \textbf{Cipher} & \textbf{Type} & \textbf{Effective (nom.) key length}  & \textbf{Attackable} \\
		\hline\hline
			2G 	& A5/0 			& Null Cipher 	& -- 			& \sokCircle{2} \\
		\cline{2-5}
		  		& A5/1 + Comp128v1/2	& LFSR-based	 		& 54 (64) bits & \sokCircle{2}  \\
		\cline{2-5}
		   		& A5/1 + Comp128v3	& LFSR-based			& 64 bits   & \sokCircle{2}  \\
		\cline{2-5}
				& A5/2 & LFSR-based		& 40 (64) bits  & \sokCircle{2}   	\\
		\cline{2-5}
				& A5/3 & KASUMI 	& 64 bits   & \sokCircle{1}  \\
		\cline{2-5}
		   		& A5/4 & KASUMI 	& 128 bits    & \sokCircle{0} \\
		\cline{2-5}
		   		& GEA1 & LFSR-based		& 64 bits 	&  \sokCircle{2} \\
		\cline{2-5}
		   		& GEA2 & LFSR-based		& 64 bits &  \sokCircle{2} \\
		\cline{2-5}
		   		& GEA3 & KASUMI	& 64 bits   & \sokCircle{1} \\
		\cline{2-5}
		   		& GEA4 & KASUMI	& 128 bits    & \sokCircle{0}\\
		\hline
			3G	& UEA0 & Null Cipher & --  &  \sokCircle{2} \\
		\cline{2-5}
		   		& UEA1 & KASUMI		& 128 bits & \sokCircle{0}   \\
		\cline{2-5}
		   		& UEA2 & SNOW 3G		& 128 bits	 & \sokCircle{0}  \\
		\hline
			4G   & EEA0 & Null Cipher	& -- 	 & \sokCircle{2}  \\   	
		\cline{2-5}
		    		& EEA1 & SNOW 3G		& 128 bits   & \sokCircle{0} \\   	
		\cline{2-5}
		   		& EEA2 & AES			& 128 bits	 & \sokCircle{0}  \\
		\cline{2-5}
		   		& EEA3 & ZUC			& 128 bits	 & \sokCircle{0}  \\
		\hline   	
	\end{tabular}
	}%
	
	\begin{minipage}{\linewidth}
	\vspace{0.2em}
	\noindent\raggedright
	
	\sokCircle{0} not attackable 
	\hspace{2em}
	\sokCircle{2} attacks with commodity hardware known
	\linebreak
	\hspace{2em}
	\sokCircle{1} attacks known, but not practicable or not demonstrated	
	\end{minipage}
\end{table}

\subsubsection{Attacks}

Cryptography is used for the encryption algorithms on the air interface, for the handover, and initial key derivation. In all these parts, we identify attacks due to the use of weak cryptography. The found attacks undermine the data confidentially requirement, either by breaking the used session key or the shared key on the SIM card.

\paragraph{Encryption Algorithms}\label{sec:GSMciphers}
Table~\ref{tab:ciphering} depicts the air interface encryption algorithms for all three generations, in particular the type of cipher, the effective key length, and if the cipher is attackable. All cipher suites in GSM except for A5/\{3,4\} are breakable within minutes on commodity hardware for different reasons. A5/1 is a 64-bit cipher based on three connected Linear Feedback Shift Registers (LFSR) with major cryptographic vulnerabilities that have led to passive decryption attacks \cite{A51rainbow,Nohl2009,Barkan2008,Golic1997,Biryukov2000,Briceno1999}.  A5/2 was designed as a stripped-down export version of A5/1 with a shorter effective key length, and Goldberg \etal~\cite{Goldberg1999} showed how to break this cipher in near real-time. Rainbow table approaches are further eased by the predictable padding of messages \cite{Nohl2011defend}. Theoretical attacks exist against KASUMI but they are impractical in terms of space requirements, as they require $2^{26}$ captured data frames \cite{Dunkelman2010} and do not directly translate into A5/3. A5/3 rainbow tables similar to A5/1 were proposed but never published~\cite{Papantonakis2013}. GPRS ciphers GEA$x$ basically mirror the weaknesses and development of their A5/$x$ counterparts \cite{Nohl2011gprs}. In \acs{GSM}, the cipher-capabilities of the network and the user device are not integrity-protected and are therefore vulnerable to manipulation. An attacker can change the encryption handshake to block A5/3 and force a downgrade to A5/1. 

In contrast to GSM, all UMTS and LTE ciphers underwent public development and thus followed Kerckhoffs' principle. As a consequence, only one attack against the KASUMI based UEA1 algorithm was revealed, but still requires an unpractically large amount of captured data~\cite{Jia2011,Dunkelman2014}. 

Additionally, each generation has a null cipher that offers no protection. Since the networks select the encryption algorithm, the user is unaware of sending data in clear text. A ciphering indicator should warn the user on the \ac{UE}. However, just few vendors implement such a ciphering indicator~\cite{GSM02.07,androidissue5353,Androulidakis2012,Rupprecht2016}. 

\paragraph{Interoperability of Access Technologies}
Interesting problems arise due the usage of same key material within the same generation or due to the interconnection of network generations. Barkan \etal~\cite{Barkan2008} describe that it is possible to downgrade to a less secure cipher for a short period of time or to reconstruct the key passively from over-the-air communication and later use it for all the other (secure) ciphers. In order to allow a GSM SIM to connect to UMTS, the key is extended to meet the UMTS key length \cite{Meyer2004a,Meyer2004,TR31.900}. Also, a USIM operating on GSM will simply use a shortened key. Thus, an attack on the much weaker A5/$x$ series reveals parts of the key information for other access technologies during handover. Additionally, the \acs{LTE} handover is  vulnerable to the so-called ``desynchronization'' attack~\cite{Han2014}. As shown in simulations, an attacker can desynchronize the used key with the core network and, thus, an old session key is reused.

\paragraph{Key Derivation}
\label{sec:weakSIMcrypto}
Weak cryptography is also used for the initial key derivation algorithms. By reverse-engineering and breaking the COMP128v1 key derivation algorithm of GSM an attacker can reconstruct the shared secret key\cite{Briceno1998, Briceno2002}. In combination with side-channels of some poorly protected implementations, COMP128v1 attacks can be brought down to nearly instant key recovery~\cite{Kaljevic2003} (see Section \ref{sec:leakyImplementation}).
Another attack by Nohl \etal \cite{Nohl2013} on SIM cards can remotely reconstruct the SIM's software update key based on weak DES encryption or poorly implemented 3DES (proper implementations are safe). They leveraged the fact that the error messages concerning ciphering are sent encrypted with a known plaintext. As this attack is delivered via SMS, there are no proximity limitations to the attacker. %
A reconstructed OTA key enables the attacker to install new applications on the SIM card, subsequently accessing secrets stored in other applications (see Section \ref{sec:insecureImplementation}).

\subsubsection*{Assessment} 
All the attacks based on weak cryptography primarily undermine the data confidentially aim of mobile networks. The attacker might also pursue a secondary aim. For example, the shared key obtained through SIM attacks can later be used to decode encrypted transmissions or write the keys on a freely programmable SIM card. Such cards can be used to impersonate a subscriber, redirect calls, change settings, or commit fee fraud. While attacks on the air interface can be executed by an attacker with passive radio capabilities, the attacks on the SIM card require physical access--- thereby are thereby either limited to the radio transmission range or to the physical range. 
The attacks on the session keys are possible using affordable methods such as rainbow tables on an ordinary PC. \acs{GSM} is especially prone to cryptography attacks. In contrast, newer generations rely on secure algorithms following the Kerckhoffs' principle such that these attacks are not known for now.

\sokentrystart \sokVector{User Equipment} \sokAim{Secrecy} \sokAttack{SIM Key Extraction via COMP128v1 Cryptoanalysis} \sokCausesList{weakcrypto} \sokCapabilitiesList{physical} \sokTarget{U} \sokReference{Briceno2002,Briceno1998, Kaljevic2003} \sokTech{2}{0}{0} \sokRange{2}{0}{0}{0}{0} \sokAddCountermeasure{seckdf} \sokentryend

\sokentrystart \sokVector{Air Interface} \sokAim{Privacy} \sokAttack{OTA SIM Card Update Key Reconstruction} \sokTarget{U} \sokAttackerM{Active Radio.} \sokCapabilitiesList{ss7*,airactive} \sokCausesList{weakcrypto} \sokReference{Nohl2013} \sokTech{2}{2}{2} \sokRange{0}{0}{0}{0}{2} \sokAddCountermeasure{smsfilter,smshomerouting*} \sokentryend

\newCountermeasure{seckdf}{Secure Key Derivation Function (e.g., MILENAGE)}{Alt2016}{sim,hss}{S}{2}{2}

\newCountermeasure{distinctkeymaterial}{Distinct Key Material and Key Derivation for Different Generations}{}{sim,ue,bts*,core*,hss}{S}{?}{0}

\sokentrystart\sokVector{Air Interface}\sokAim{Secrecy}\sokAttack{Key Reusage Across Cipher and Network Generations} \sokTarget{U}\sokReference{Barkan2008} \sokCapabilitiesList{airpassive,airactive}
\sokTech{2}{1}{?}\sokRange{0}{2}{0}{0}{0}\sokMitigation{}\sokAttackerM{Act. Radio}
\sokAddCountermeasure{distinctkeymaterial} \sokCausesList{weakcrypto}
\sokentryend

\sokentrystart\sokVector{Air Interface}\sokAim{Secrecy}\sokAttack{Weak Key due to Inter-Technology Handover}\sokTarget{U}\sokReference{Meyer2004a} \sokCapabilitiesList{airpassive,airactive}
\sokTech{0}{2}{?}\sokRange{0}{2}{0}{0}{0}%
\sokAddCountermeasure{distinctkeymaterial} \sokCausesList{weakcrypto}
\sokentryend

\sokentrystart\sokVector{Air Interface}\sokAim{Secrecy}\sokAttack{Inter eNodeB User Plane Key Desynchronization Attack} \sokTarget{U}\sokReference{Han2014} \sokCapabilitiesList{airpassive}\sokTech{0}{0}{2}\sokRange{0}{2}{0}{0}{0}\sokCausesList{weakcrypto}
\sokAttackerM{Act. Radio} 
\sokAddCountermeasure{nodirektbtshandover,increasesessionkeyupdate*}
\sokentryend

\newCountermeasure{increasesessionkeyupdate}{Increase Session Key Update Interval}{Han2014}{bts,core}{I}{2}{?}

\newCountermeasure{nodirektbtshandover}{Disable Direct eNodeB Handover}{Han2014}{bts,core}{S}{2}{?}

\newCountermeasure{a53}{Introduction of Secure Cipher to GSM (e.g., A5/3)}{TS35.202}{ue,bts,core*}{S}{2}{2}
\newCountermeasure{disableA51}{Disable A5/1 on Network}{}{bts*,core*}{I}{?}{?}
\newCountermeasure{disableA52}{Disable A5/0 and A5/2 on Network}{GSMA-A52prohibition}{bts*,core*}{I}{?}{2}
\newCountermeasure{disableA50ue}{Disable A5/0 on UE for Non-Emergency Calls}{}{ue}{S}{?}{0}
\newCountermeasure{disableA51ue}{Disable A5/1 on UE}{}{ue}{S}{?}{0}
\newCountermeasure{disableA52ue}{Disable A5/2 on UE}{GSMA-A52prohibition}{ue}{S}{?}{2}
\newCountermeasure{gsmpadding}{GSM Random Padding}{TS44.006,TS44.006}{core*,bts,ue*}{S}{2}{2}

\sokentrystart\sokVector{Air Interface}\sokAim{Secrecy}
\sokAttack{Passive Over-the-Air Decryption of A5/1 and A5/2}
\sokTarget{U}\sokReference{Briceno1999,A51rainbow,Nohl2009,Barkan2008,Golic1997,Biryukov2000,Briceno1999,Nohl2011defend} \sokCapabilitiesList{airpassive} \sokCausesList{weakcrypto}
\sokTech{2}{0}{0}\sokRange{0}{2}{0}{0}{0}\sokMitigation{}\sokAttackerM{Act. Radio}
\sokAddCountermeasure{a53,disableA51,disableA52,disableA51ue,disableA52ue,gsmpadding}
\sokentryend

\subsubsection{Defenses}
Because the specification follows the best practices in newer generations, most of the defenses concentrate on the weaknesses of GSM. After weaknesses of A5/$\{$1,2$\}$ became apparent and attacks were feasible, two new ciphers were added which are backports of the UMTS KASUMI cipher, whereas A5/3 simply pads the GSM 64-bit key to 128 bits and A5/4 uses the full 128 bits \cite{TS55.226}. The A5/3 usage is increasing~\cite{Dabrowski2016}, but as of 2017, there is no known network supporting A5/4. GSMA finally mandated the removal of A5/2 support from phones\cite{GSMA-A52prohibition}. Besides A5/2, the A5/0 was a useful downgrade target, so some networks disabled both of them~\cite{Dabrowski2016}. Disabling A5/1 is still not a viable option for operators.

The key derivations and authentication algorithms are exchangeable and also follow the best practice. MILENAGE is based on AES and replaces COMP128\cite{TS35.206}. TUAK, based on SHA-3 (Keccak), is another option for the authentication \cite{TS35.231}. It is important the algorithms is provably secure and hold strong security assumptions~\cite{Alt2016}.

The introduction of new encryption algorithms for old access technologies decreases the effectiveness of attacks. However, this introduction takes a long time as software/hardware needs to be updated and new algorithms must be specified.

\subsubsection{Research Questions and Challenges}
Standardization and implementation of cryptographic protocols for handover, initial key derivation, or encryption did not follow best practices at all times for various reasons \cite{Pell2014}. Advances in cryptanalysis have revealed various vulnerabilities in these algorithms. In the future, new algorithms need to be carefully analyzed following Kerckhoffs' principle. Furthermore, as advances in cryptanalysis and computational power need to be factored in, provisions for possible updates of security-relevant algorithms must be build into standards while simultaneously employing anti-downgrade methods in case old and new methods or key lengths need to coexist. 

This is especially the case for the newly introduced device-to-device (D2D) communication. In the case of direct device communication, two devices agree on a common key by using two protocols that provide no forward secrecy and rely on one master key~\cite{TS33.303,RFC_6507,RFC_6508}. Such a scheme has different security implications~\cite{Murdoch2016}. If asymmetric cryptography is added to 5G or future generation, this might introduce new attack surfaces if not designed and implemented carefully. Additionally, the to-be-introduced embedded SIM card comes with a complex security infrastructure and with protocols that have not yet been analyzed with respect to security \cite{eSIM2017}. A vulnerability in the draft of the 5G \ac{AKA} allowing an attacker to impersonate a victim to the network has been found with a formal symbolic analysis~\cite{Wild2018}. The latter example shows that new security schemes must be carefully analyzed. Future research should prove the security of all the used cryptographic techniques in the mobile context to ensure overall security.

\begin{mdframed}[backgroundcolor=blue!05,nobreak] 
\textbf{Summary:} Weak cryptography has led to many attacks against the data confidentiality aim. In future generations, structural changes such as device-to-device communication challenge the use of secure algorithms. The system's security should continue to rely on well-known and proven secure cryptographic algorithms. Future research must match specified cryptographic algorithms with a realistic scope of attacker capabilities. Additionally, measures to protect against downgrade attacks to older and less secure ciphers and protocols must be developed. 
\end{mdframed}

\subsection{Cause: Insecure Inter-Network Protocols}
\label{sec:insecureInterprotocol}
Nowadays the telecommunication industry is deregulated and \ac{SS7} has been ported to an IP-based network. Both developments make \ac{SS7} easily accessible. Thus, an attacker with \ac{SS7} capabilities becomes more likely. However, for interconnectivity with the \ac{SS7} networks, \ac{SS7} messages are translated to Diameter. This makes Diameter also vulnerable to \ac{SS7} attacks, as this inter-working function does not provide authenticity. Even though Diameter was designed with security features based on protocols like TLS and IPsec, researchers found vulnerabilities in the dedicated Diameter protocol that do not rest upon the inter-working function of \ac{SS7}.

\subsubsection{Attacks}
The general idea of \ac{SS7} attacks is to request services on different layers of the home network or the serving network. As \ac{SS7} offers no authentication mechanisms, the network entities cannot decide if the request is legitimate. Thus, the entity replies properly, even though the request might not be legitimate. 

An attacker can determine the user location on different levels of granularity---in the range of cells up to exact GPS coordinates~\cite{Engel2008,Engel2014,Holtmanns2016}. Additionally, an attacker can map the temporary identity (\acs{TMSI}) to the permanent identity (\acs{IMSI}) of a victim by using the \ac{SS7} system. The permanent identity can then be mapped to the public telephone number. Both attacks are not compliant with the identity confidentiality aim. The misuse of \ac{SS7} can also lead to attacks that undermine the confidentiality of calls or of text messages~\cite{Engel2014}. This can be done by rerouting calls or by requesting the over-the-air encryption key. Besides this, the insecurity of \ac{SS7} can also be exploited for fraud attacks by unblocking a stolen device~\cite{Rao2015}. Additionally, an attacker can run a precise \ac{DoS} attack against a distinct user by deleting subscriber data in the VLR~\cite{Engel2014}. Attacks that are possible due to the inter-working function between Diameter and \ac{SS7} are discussed by Holtmanns~\etal~\cite{Holtmanns2016} and Rao \etal \cite{Rao2016a}. Even Diameter has been found vulnerable and allows to intercept text messages~\cite{Holtmanns2017}.

\paragraph*{Assessment}The insecurity of \ac{SS7} leads to a wide range of attacks. Most of them aim to undermine the (location) privacy of the user. Even commercial services were built upon the insecurity of \ac{SS7} allowing to pinpoint and track a victim \cite{Verint2013,Timberg2014,Engel2014}. This shows that the SS7 vulnerabilities are actively used and are thereby a serious threat to users. Most of the attacks require \ac{SS7} capabilities of the attacker. However, some attacks can be accomplished by using passive radio capabilities, \eg, an attacker can decrypt the traffic as soon as the over-the-air encryption key is revealed. 

\sokentrystart\sokVector{Inter Network} \sokAim{Secrecy} \sokAttack{Session Key Retrieval via SS7} \sokTech{2}{2}{?} \sokTarget{U} \sokReference{Engel2014,Nohl2014} \sokAttackerM{SS7 Attacker} \sokCapabilitiesList{ss7,airpassive} \sokCausesList{insecSS7} \sokAddCountermeasure{ss7scan*,ss7firewall,diameterrollout} \sokRange{0}{2}{0}{0}{1} \sokentryend

\sokentrystart\sokVector{Inter Network} \sokAim{Service} \sokAttack{Insert/Delete Subscriber Data into the VLR/MSC} \sokTech{2}{2}{2} \sokTarget{U} \sokReference{Engel2014} \sokAttackerM{SS7 Attacker} \sokCapabilitiesList{ss7} \sokCausesList{insecSS7} \sokAddCountermeasure{ss7scan*,ss7firewall,diameterrollout}\sokRange{0}{0}{0}{0}{2} \sokentryend

\sokentrystart\sokVector{Inter Network} \sokAim{Secrecy} \sokAttack{Intercepting Calls with SS7/CAMEL} \sokTech{2}{2}{?} \sokTarget{U} \sokReference{Engel2014} \sokAttackerM{SS7 Attacker} \sokCapabilitiesList{ss7} \sokAddCountermeasure{ss7scan*,ss7firewall,diameterrollout}\sokCausesList{insecSS7} \sokRange{0}{0}{0}{0}{2} \sokentryend

\sokentrystart\sokVector{Inter Network} \sokAim{Privacy} \sokAttack{Cell-Level Tracking with SS7/MAP} \sokTech{2}{2}{2} \sokTarget{U} \sokReference{Engel2014} \sokAttackerM{SS7 Attacker} \sokCapabilitiesList{ss7} \sokAddCountermeasure{ss7scan*,ss7firewall,diameterrollout,smshomerouting}\sokCausesList{insecSS7} \sokRange{0}{0}{0}{0}{2} \sokentryend

\sokentrystart\sokVector{Inter Network} \sokAim{Privacy} \sokAttack{GPS Location with SS7/LCS} \sokTech{2}{2}{2} \sokTarget{U} \sokReference{Engel2014} \sokAttackerM{SS7 Attacker} \sokCapabilitiesList{ss7} \sokAddCountermeasure{ss7scan*,ss7firewall,diameterrollout}\sokCausesList{insecSS7} \sokRange{0}{0}{0}{0}{2} \sokentryend

\sokentrystart\sokVector{Inter Network} \sokAim{Fraud} \sokAttack{Unblock Stolen Devices} \sokTech{2}{2}{2} \sokTarget{U} \sokAttackerM{SS7 Attacker} \sokCausesList{insecSS7} \sokCapabilitiesList{ss7} \sokAddCountermeasure{ss7scan*,ss7firewall,diameterrollout}\sokRange{0}{0}{0}{0}{2} \sokReference{Rao2015} \sokentryend

\subsubsection{Defenses}
The most sustainable long-term solution is the complete elimination of SS7. With the specification of Diameter in \acs{LTE}, a more secure protocol is used for inter-networking functions. However, even Diameter is not free of flaws~\cite{Rao2016a,Holtmanns2017}. Additionally, the inter-working function between SS7 and Diameter still allows attacks via Diameter based on \ac{SS7} vulnerabilities, as long as not all the network providers migrate to Diameter. 

Therefore, short-term solutions to mitigate the threats of \ac{SS7} and Diameter insecurity have been proposed. Most of them are based on validating the legitimacy of the request and then blocking the request itself or blacklisting certain classes of message types. For example, a request for the over-the-air encryption key is only allowed by a network that proves the user's registration within its range. Furthermore, certain requests are merely of network-internal interest and are discarded at the network border, \eg, the charging of the prepaid credit. The industry provides solutions for the mobile network operators ranging from \ac{SS7} scans~\cite{P1Sec,GSMA2017} to stateful \ac{SS7} firewalls~\cite{ashdown2001ss7,Engel2014PATENT}. Peeters \etal suggest a detection mechanism of intercepted phone calls by an \ac{SS7} redirection attack using distance bounding and timing information~\cite{Peeters2018}. 

\newCountermeasure{ss7scan}{SS7 Penetration Tests}{P1Sec}{core}{I}{1}{1}

\newCountermeasure{ss7firewall}{Stateful SS7 Firewall}{ashdown2001ss7,Engel2014PATENT}{core}{I}{1}{1}

\newCountermeasure{diameterrollout}{Migration to Diameter}{}{core}{S}{1}{?}

\subsubsection{Research Questions and Challenges}
By now, it is known that \ac{SS7} is an insecure protocol and the backward-compatibility of Diameter rendering also newer systems vulnerable to \ac{SS7} attacks. The exclusive use of Diameter in the (inter)-core network communication would be a step forward in terms of security, but it will not entirely solve the security problems.

Thus, the open research question is to design a protocol that is proven secure and that holds the security requirements, especially the privacy requirements in the (inter)-core network while maintaining the functionality of the mobility management. Such a protocol must withstand an exhaustive security analysis. For example, such a protocol should enforce a proof from the remote network that the subscriber is actually present and only authorize such transactions. A solution explored for 5G is to bind keys to a public key identity of the serving network \cite[§5.2.4.6]{TR33.899}. Both would prevent attacks in which an attacker sends unauthorized requests to the home network, \eg, for the session key. The means of a privacy-preserving protocol are open topics for research.

\begin{mdframed}[backgroundcolor=blue!05,nobreak] 
\textbf{Summary:} Insecure inter-network protocols (\eg, \ac{SS7}) allow privacy and fraud attacks, and will not be entirely switched off in the near future. Firewalls constitute only temporary solutions to the problem. Future research is challenged to design privacy-preserving inter-network protocols that keep the maintenance overhead low.
\end{mdframed}

\subsection{Cause: Resource Usage Asymmetry}
\emph{Resource usage asymmetry} occurs when an simple action on one side triggers a computationally or resource-wise expensive reaction on the other side. This---for example---leads to signaling \ac{DoS} attacks, during which an attacker misuses signaling/control messages to trigger an expensive action. Thus, the network allocates the resources within different components and may eventually run out of them after repeated or coordinated requests.

\subsubsection{Attacks}
Unauthenticated messages like those used in the \textit{attach procedure} can be utilized to overload the core network components~\cite{Bassil2013,Yu2012}. Additionally, they can impersonate legitimate subscribers. Similarly, Lee \etal~\cite{Lee2009} have presented signaling attacks for 3G networks and argue that low-volume but well-timed signaling attacks can have a major impact on the network components. By misusing multiple messages for establishment and release of radio connections, the authors caused a significant increase of message load in the network. Traynor \etal \cite{Traynor2009} evaluated network attacks targeting the \ac{HLR}\textsuperscript{2G,3G}. They found an effective method to tear down an \acs{HLR} by frequently switching the call forwarding service on and off. They suggest that a mobile phone botnet can disable the service of an \ac{HLR}.

Similarly, a mobile phone botnet could attack a 911 response center, which would result in an outage of emergency services~\cite{Guri2017}. While this is not exclusively related to mobile-phone networks, the elevated priority of emergency calls makes it a unique mobile network problem: The network will drop other connections in favor of emergency calls if necessary.
Enck \etal~\cite{Enck2005} evaluated attacks considering the to open SMS functionality on the Internet. They analyze an attacker model that uses open SMS centers on the Internet to saturate the wireless link downstream from the base stations, obstructing the service in the whole cell.

\paragraph*{Assessment} All the attacks based on resource usage asymmetry focus on an exhaustive denial-of-service of the network. However, the impact of these attacks vary. While some attacks require active radio attacker capabilities, others already work with Internet capabilities. 
Besides intentional disturbance of the service, similar problems can occur due to misconfigured mobile apps or unexpected user behavior \cite{Yang2011,Ericsson2015}.

\sokentrystart\sokVector{Air Interface} \sokAim{Service} \sokAttack{Attach Request Attack} \sokRange{0}{2}{0}{0}{0} \sokAttackerM{Act. Radio} \sokCapabilitiesList{airpassive,airactive} \sokTarget{P} \sokReference{Yu2012} \sokCausesList{resource}
 \sokTech{2}{2}{2} \sokAddCountermeasure{statisticaldetection} \sokentryend

\sokentrystart\sokVector{Air Interface} \sokAim{Service} \sokAttack{DDoS HLR: Activate Call Forwarding Request} \sokAttackerM{Semi act. Attacker} \sokReference{Traynor2009} \sokCapabilitiesList{airuser} \sokTarget{P} \sokRange{0}{0}{0}{1}{0} \sokCausesList{resource} \sokTech{2}{2}{0} \sokAddCountermeasure{statisticaldetection} \sokentryend

\sokentrystart\sokVector{Air Interface} \sokAim{Service} \sokAttack{Signaling DoS}  \sokReference{Lee2009,Bassil2013,Leong2014,Kambourakis2011} \sokTarget{U,P} \sokCapabilitiesList{airpassive,airactive} \sokCausesList{resource} \sokRange{0}{2}{0}{0}{0}
 \sokTech{2}{2}{2} \sokAddCountermeasure{statisticaldetection} \sokentryend

\sokentrystart\sokVector{Inter Network} \sokAim{Service} \sokAttack{SMS Link Saturation} \sokTarget{U,P} \sokReference{Enck2005,Traynor2009a} \sokTech{2}{2}{2}  \sokAttackerM{Internet Attacker} \sokCapabilitiesList{inet,ss7*} \sokCausesList{resource} \sokRange{0}{0}{0}{0}{2} \sokAddCountermeasure{statisticaldetection} \sokentryend

\subsubsection{Defenses}
So far, most suggested detection and protection methods are statistical approaches~\cite{Lee2009,Traynor2009,Traynor2009a}. Random connection drops might protect the network functionality as a whole, but inevitably they also deny legitimate requests. Even good statistical methods come with a non-negligible false-positive rate. The suggested protocol changes are unrealistic for currently rolled out networks.

\newCountermeasure{statisticaldetection}{Statistical Attack Detection and Protection}{Enck2005,Lee2009,Traynor2009,Traynor2009a,Yu2012}{enb,core}{I}{1}{?}

\subsubsection{Research Questions and Challenges}
All the defenses suggest reactive schemes that come with a certain false-positive rate and do not prevent attacks. Future research should explore how to prevent resource exhaustion in the first place. This could require protocol changes and is, thus, only viable for new network generations.

Possible approaches can be borrowed from similar problems in the context of other computer networks. RFC5013 \cite{rfc6013} proposes a TCP cookie against connection flood attacks. In contrast, Dwork \cite{Dwork1992} and Back \cite{Back2002} suggested a proof-of-work-based method against flooding and email spam. Before the server or network processes a request, the client has to solve a (moderately hard) computational puzzle, proving its commitment. These puzzles have to be easy to generate, easy to check, but parametrizable hard to solve (\eg, finding bits of a hash-collision). Thus, equalizing the computational load on both sides and making flood-based DoS attacks much more resource-intense for the attacker. However, such schemes have to be adopted to and evaluated in the context of mobile networks. Challenges include the limited resources on mobile devices and low-latency requirements on some operations.

\begin{mdframed}[backgroundcolor=blue!05,nobreak] 
\textbf{Summary:} Resource usage asymmetry allows to flood networks with signaling messages and eventually a denial of service. Future research must aim for complete attack prevention, as current state of the art research can only provide probabilistic detection. This is possible through protocol designs with balanced resource usage.

\end{mdframed}

\section{Root Cause: Implementation Issue}
\label{sec:RCImplementation}
Deviations of the implementation from the original specification can open attack vectors and, thus, can have a security impact on otherwise securely defined systems. Such deviations can be introduced on purpose, \eg, for compatibility trade-offs, or result from faulty implementations. In the following, we discuss the implications of insecure and leaky implementations.

\subsection{Cause: Insecure Implementation}
\label{sec:insecureImplementation}

While insecure implementations can open attack vectors in deployed systems, current research mainly focuses on attacks on the baseband and SIM cards. By sending malicious data to vulnerable devices, an adversary can exploit implementation issues. In the following, we discuss how attacks undermine the system integrity, availability, secrecy, and privacy including potential countermeasures. 

\subsubsection{Attacks}

\sokentrystart \sokVector{Air Interface} \sokAim{Integrity} \sokAttack{Binary Baseband Exploit} \sokCapabilitiesList{airactive} \sokTarget{U} \sokCausesList{insecureimpl}\sokReference{Weinmann2012,Golde2016,VandenBroek2014} \sokTech{2}{2}{2} \sokRange{0}{2}{0}{0}{0} \sokAddCountermeasure{memsafelang,machinereadablespec*} \sokentryend %

\sokentrystart \sokVector{Air Interface} \sokAim{Integrity} \sokAttack{SMS Parsing} \sokCapabilitiesList{airactive} \sokTarget{U} \sokCausesList{insecureimpl}\sokReference{Mulliner2011,Court2017} \sokTech{2}{2}{2} \sokRange{0}{2}{0}{0}{2} \sokAddCountermeasure{smsimpltestframework,smsfilter*}\sokAddCountermeasure{memsafelang,machinereadablespec,machinereadablefsm} \sokentryend

\sokentrystart \sokVector{Air Interface} \sokAim{Integrity} \sokAttack{ASN.1 Heap Overflow} \sokCapabilitiesList{airactive,ss7*} \sokTarget{U,P} \sokCausesList{insecureimpl}\sokReference{Ia-sadosky2016} \sokTech{2}{2}{2} \sokRange{0}{2}{2}{2}{2} \sokAddCountermeasure{memsafelang*}\sokentryend

\sokentrystart \sokVector{Air Interface} \sokAim{Secrecy} \sokAttack{Baseband State Machine Exploits} \sokCapabilitiesList{airactive} \sokTarget{U} \sokCausesList{insecureimpl}\sokReference{Osipov2015,Michau2016,Rupprecht2016,Welte2009} \sokTech{2}{2}{2} \sokRange{0}{2}{0}{0}{0} \sokAddCountermeasure{stateimpltestframework*,machinereadablespec} \sokentryend %
 
The lower layers of the protocol stack run on distinct baseband processors in the \ac{UE}. Parser errors within the baseband processor can occur due to faulty implementations of parsing modules or libraries threatening the device's integrity. 
In 2016, a heap overflow in a widely used ASN.1 compiler was discovered~\cite{Goodin2016,Ia-sadosky2016} affecting baseband implementations of multiple manufacturers.
Weinmann \cite{Weinmann2012} and Golde \cite{Golde2016} demonstrated how to use baseband exploits to further target the application processor and its operating system.

Crashing-only flaws in the parsing and decoding stage of text messages~\cite{Mulliner2011,VandenBroek2014,Court2017} make the phone inoperable until the next reboot. Similar flaws on SMS parsing have been found on other processing levels~\cite{Brown2015}.

Apart from attacks on the baseband, Nohl \etal~\cite{Nohl2013} showed that the application isolation on the SIM card is so weak that processes can access foreign data including authentication credentials. Such applications can be remotely installed after reconstructing the over-the-air (OTA) update key (see Section~\ref{sec:weakSIMcrypto}). 

Implementation flaws in the protocol state machines of the baseband result in the acceptance of a fake base station as a genuine network endangering data secrecy and privacy~\cite{Osipov2015,Michau2016,Rupprecht2016,Welte2009}.

\sokentrystart \sokVector{Air Interface} \sokAim{Integrity} \sokAttack{SIM Card Rooting} \sokTarget{U} \sokAttackerM{Active Radio.} \sokCapabilitiesList{ss7*,airactive*,airuser*} \sokReference{Nohl2013} \sokCausesList{insecureimpl} \sokTech{2}{2}{2} \sokRange{2}{2}{0}{2}{2} \sokAddCountermeasure{smsfilter*,smshomerouting*} \sokentryend

\subsubsection*{Assessment} 
On the one hand, we see that attacks can be launched globally and in a targeted manner that makes the impact of these flaws very high. On the other hand, the most dangerous ASN.1 heap overflow and the staged baseband-to-application-processor attacks required a fake base station with active radio capabilities and is thus locally bounded. The danger lies in the potential to take over the device at the lowest level.

\subsubsection{Defenses}
Intermediate workarounds for multiple of the aforementioned attacks are based on operator-side filtering. 
For example, operators filter out messages that might be used to infer the OTA-key of SIM cards. Such filtering can be easily and quickly deployed by the operator.
However, intermediate workarounds are typically only effective against known attacks and, thus, are not very sustainable.
Furthermore, network filtering only prevents attacks coming through the network. An attacker with active radio capabilities operating a fake base station can still deliver these exploits directly to the phone, albeit with reduced range.

More generic defenses in the field of insecure implementation focus on the detection and prevention of insecure implementations. For SMS parsing errors as well as for state machine errors, various security testing frameworks have been proposed~\cite{Mulliner2011, VandenBroek2014, Michau2016, Rupprecht2016}. These frameworks automatically test for known vulnerability patterns based on predefined test cases. While automated approaches were used to find SMS parsing errors and state machine errors, many memory corruption vulnerabilities  were manually found through reverse engineering~\cite{Weinmann2012, Golde2016}. 
\newCountermeasure{smsimpltestframework}{SMS Parsing Test Framework}{Mulliner2011, VandenBroek2014}{}{I}{?}{2}

\newCountermeasure{stateimpltestframework}{State Machine Test Framework}{Michau2016, Rupprecht2016}{}{I}{?}{2}

\newCountermeasure{smsfilter}{Filtering Malicious SMS}{Nohl2013}{core}{I}{?}{2}
\newCountermeasure{smshomerouting}{SMS Home Routing}{TR23.840}{core*,hlr,inter*}{S}{2}{2}

\newCountermeasure{memsafelang}{Memory-Safe Languages and Runtime-Environments}{Burow2017,Szekeres2013,Larsen2014}{}{I}{1}{2}

\newCountermeasure{machinereadablespec}{Fully Machine-Readable Protocol Specifications}{}{}{S}{0}{1}
\newCountermeasure{machinereadablefsm}{Fully Machine-Readable State Machine Specifications}{}{}{S}{0}{1}

\subsubsection{Research Questions and Challenges} 
All the attacks and mitigations stemming from \textit{insecure implementation} have similarities to classic system security. We distinguish research questions between detection and prevention of vulnerabilities. Additionally, we discuss the shortcomings of the existing work which is the current scope.

\paragraph{Detection of Vulnerabilities} 
Although testing frameworks have been proposed \cite{Mulliner2011, VandenBroek2014, Michau2016, Rupprecht2016}, they usually focus on one particular type of flaw, such as SMS parsing errors, state machine failures, or particular memory vulnerabilities. 
Basebands have complex state machines and exhibit a fragile behavior~\cite{Golde2016}, thus, automated testing tools based on fuzz testing have problems achieving higher levels of code and state coverage. However, alternatives such as manual reverse engineering of the baseband scale poorly and are expensive.

Therefore, reliable detection methods for vulnerabilities in the decoding functions and state machines are needed. The decoding functions are important to protect against integrity and availability attacks. This can be supported by data for security testing that would allow better corner case testing, \eg, error states and illegal state machine transitions.

\paragraph{Prevention of Vulnerabilities} 
Control-flow hijacking, memory corruption, and state machine failures are well-known problems in the context of classic system security~\cite{Burow2017,Szekeres2013,Larsen2014}. However, in mobile security, classic system security defenses face certain challenges. Most notably, the real-time capability is a hard requirement for the baseband, as it needs to stay synchronized with the radio transmissions. In addition to the run-time overhead, many modern countermeasures come with a certain overhead, unreasonable for the baseband. Adapting classic system security countermeasures like memory-safe languages, memory address randomization, or control-flow integrity solutions in this constrained environment remain an open challenge~\cite{Burow2017,Szekeres2013,Larsen2014}.

Another way to reduce implementation bugs is to carefully choose the development framework based on their intrinsic security properties \cite{DSilva2015,Jaeger2014}. Additionally, machine-readable protocol specifications and state machines would allow to generate parsers and state machines directly from the specification, cutting out the error-prone human interpretation of the specification. For parsing, part of the 3GPP specification already employs ASN.1. However, the parser libraries and compilers must be thoroughly tested and audited to avoid the fallout an ASN.1 compiler bug caused in 2016~\cite{Goodin2016,Ia-sadosky2016}.

\paragraph{Current Scope} 
Within implementation security, the research community focuses mainly on the user equipment. However, it is very likely that other network components, \eg, the core network or base stations, suffer from similar vulnerabilities. 
For example, ASN.1 parsing is also implemented on the network side. Thus, it is not unlikely that the known ASN.1 vulnerabilities may also be present in network components. We therefore suggest the examination of network components as well. 

\begin{mdframed}[backgroundcolor=blue!5,nobreak] 
\textbf{Summary:} Insecure implementations open attack vectors for adversaries with active radio capabilities or direct network access. Future research must provide more sustainable defenses of the classical system security context, \eg, control-flow integrity protection for basebands.
\end{mdframed}

\subsection{Cause: Leaky Implementation}
\label{sec:leakyImplementation}
Implementations in software and hardware can leak information about internal states in surprising or non-obvious ways. Besides using a provable secure, an implementation might leak enough information to circumvent the strong security measures due to the implementation insufficiencies.
\subsubsection{Attacks}
\label{sec:simcloning}
The SIM card stores the secret key for authentication and key derivation. Gaining access to this information breaks the security concept at its very core enabling decryption and impersonation. 

Rao \etal~\cite{Rao2002} and Zhou \etal~\cite{Zhou2013} have built a key reconstruction attack upon the cryptanalysis of Comp128v1 on \ac{GSM} SIM cards with chosen plaintexts and by using electromagnetic field probes. In 2015, Liu~\etal~\cite{Liu2015} found that the AES-based \textit{MILENAGE} algorithm on USIM implementations is susceptible to power-based side-channel analysis and were thus able to extract the secret key. 

\newCountermeasure{consttimespec}{Constant Time and Power Specification Requirements}{Rao2002,Zhou2013,Liu2015}{sim,ue*}{S}{2}{?}

\sokentrystart \sokVector{User Equipment} \sokAim{Secrecy} \sokAttack{(U)SIM: COMP128v1 and MILENAGE Side-Channels} \sokCapabilitiesList{physical} \sokTarget{U} \sokReference{Rao2002,Zhou2013,Liu2015} \sokCausesList{leakyimpl} \sokTech{2}{2}{2} \sokRange{2}{0}{0}{0}{0} \sokAddCountermeasure{consttimespec} 
\sokentryend

\subsubsection*{Assessment} 
The primary aim of such attacks is gaining access to the secret key and, thereby, undermining the confidentially requirement. However, once the key is known to the attacker, he/she might fulfill secondary attack aims. It may enable him/her to decrypt the radio communication with passive radio capabilities or to impersonate a subscriber by cloning the SIM card. 

Even though the aforementioned attacks reveal one of the most valuable secrets in mobile networks, the attacks require temporary physical access to the SIM card. Thus, forging SIM card clones is more likely to happen through an internal attacker or through the device owners themselves than through external attackers. 

\subsubsection{Defenses}
The proposed defenses for side-channel attacks are implementation-specific~\cite{Rao2002,Zhou2013,Liu2015}. The common ground for all known defenses is to have constant time and power properties, thus not leaking information about the internal state and making it unfeasible to derive the secret key by non-invasive methods.

\subsubsection{Research Questions and Challenges}
The proposed countermeasures need to be adopted to SIM cards by the industry. Clear requirements for constant time and constant power properties in the specification would help to accelerate the process of adoption. Additionally, it could be helpful to require a third party certification regarding attack resistance. If asymmetric cryptography should make it into 5G USIMs, than this will pose additional challenges to side-channel prevention \cite[§5.1.4.19]{TR33.899}. 

An upcoming technology in the field of SIM cards is the \textit{embedded SIM card}~\cite{eSIM2017}. Embedded SIM cards enable the configuration of the users' credentials via the Internet and are permanently soldered into the user device. From a research perspective it is interesting to examine how embedded SIM cards are secured against side-channel attacks.

\begin{mdframed}[backgroundcolor=blue!5,nobreak] 
\textbf{Summary:} Leaky implementations reveal the secret key of the SIM card via unintended side-channel attacks. With the leaked key, an attacker can passively decrypt the communication or impersonate a victim. Future research needs to investigate new technologies with respect to their side-channel resistance, \eg, embedded SIM cards or asymmetric cryptography implemented on SIM cards.
\end{mdframed}

\section{Root Cause: Protocol Context Discrepancy} 
\label{sec:sys_nsci}
This root cause is based on protocol context issues that are due to deploying a protocol that is not originally intended for the mobile network environment. Protocol properties are not harmful in a non-mobile network environment, but may be exploitable in a mobile environment if not adjusted properly. 

\subsection{Cause: Cross-Layer Information Loss}
The layering of network stacks serves multiple important purposes such as implementation transparency (\eg, upper layers do not have to care about details of lower layers) and interoperability (\eg, upper layer applications can span or exchange data over multiple networks). However, such layering also means loss of information that might be needed at higher levels, \eg, at some point, IP addresses or connections need to be mapped to the subscriber identity. 

\subsubsection{Attacks}
The lack of a strong binding between radio-level authentication and IP-service authentication is the source for multiple vulnerabilities. The literature show that the implementation of such mapping is vulnerable and can be tricked with simple IP-based attacks, such as spoofing of IP addresses \cite{Peng2014,Wang2011}. IP address spoofing can be exploited for over- and under-billing attacks and to reverse the isolation of the internals to the Internet network. IPv4 and IPv6 \ac{NAT} middleboxes pose a threat to the users as well as for the mobile network operator~\cite{Leong2014,Hong2017}. Similarly to the \ac{NAT} middleboxes, the \ac{P-GW} rooting configuration seems to be a problem in cases that allow direct communication between two phones~\cite{Li2015,Kim2015}. Another related problem is the lack of security checks within the SIP-protocol. Manipulated SIP headers can be used to fake the caller ID with UE-originated SMS messages \cite{Tu2016a}. 

\subsubsection*{Assessment} 
All these attacks consider an attacker able to initiate user traffic and optional Internet traffic capabilities. Hence, all the attacks can be easily realized. The range of those attacks is network-wide, thus an attacker can be anywhere in the network and exploit the flaw. We see the trend that newer generations---especially LTE---are more prone to attacks that are based on cross-layer information loss. This happens because LTE aims to be a general-purpose network providing normal Internet connectivity, and the layering of stacks is more prominent in those networks.

\sokentrystart\sokVector{Inter Network} \sokAim{Service} \sokAttack{IPv4/IPv6 Middleboxes Misconfiguration} \sokTarget{U,P} \sokReference{Leong2014,Hong2017} \sokTech{0}{0}{2} \sokCausesList{crosslayer} \sokAttackerM{Internet Attacker}\sokCapabilitiesList{airuser,inet} \sokAddCountermeasure{firewall} \sokRange{0}{0}{0}{2}{0} \sokentryend
  
\sokentrystart\sokVector{Air Interface} \sokAim{Fraud} \sokAttack{Uplink IP Header Spoofing/Cloak-and-Dagger Misbilling}  \sokCausesList{crosslayer} \sokTarget{U} \sokTech{0}{0}{2} \sokReference{Peng2014,Wang2011} \sokAttackerM{Semi. Act. Radio.} \sokCapabilitiesList{airuser} \sokAddCountermeasure{securebinding,deauthorization,feedbackmischarge} \sokRange{0}{0}{0}{2}{0} \sokentryend

\sokentrystart\sokVector{Air Interface} \sokAim{Fraud} \sokAttack{LTE IMS-based SMS Spoofing} \sokCausesList{crosslayer} \sokTarget{U} \sokTech{0}{0}{2} \sokAddCountermeasure{securebinding} \sokReference{Tu2016a,Chalakkal2017} \sokRange{0}{0}{0}{0}{2} \sokCapabilitiesList{airuser} \sokentryend

\sokentrystart\sokVector{Air Interface} \sokAim{Privacy} \sokAttack{Location Leak by SIP Message} \sokCausesList{crosslayer} \sokTarget{U} \sokTech{0}{0}{2} \sokAddCountermeasure{securebinding} \sokReference{Chalakkal2017} \sokRange{0}{0}{0}{0}{2} \sokCapabilitiesList{airuser} \sokentryend

\subsubsection{Defenses}
Higher-level services cannot solely rely on the transport layer security measures of the lower layers and their authentication. Since an attacker can access any network communication on the device, no data from the device should be trusted. A \emph{secure binding} between the user's charging ID and the established connection suppresses any possible misuse~\cite{Peng2014}. Such a secure binding operates across the separated layers. Additionally, Peng \etal suggest active \emph{de-authorization} of a connection and a \emph{feedback-based mischarge correction} scheme for misbilling attacks~\cite{Peng2014}. Other mitigations built upon well-configured and maintained stateful firewalls to encounter threats due to misconfigured routers and NAT middleboxes\cite{Li2015,Kim2015,Leong2014,Hong2017}. All defenses must be implemented at the core network by operator. While firewalls and the secure binding can by simply implemented, more advanced misbilling countermeasures, \eg, deauthorization or feedback-based mischarge correction need to be specified.

\newCountermeasure{securebinding}{Secure Binding Across Networks Layers}{Peng2014}{core,pgw}{I}{1}{?}

\newCountermeasure{deauthorization}{Active Connection Deauthorization}{Peng2014}{ue,core,pgw}{S}{1}{?}

\newCountermeasure{feedbackmischarge}{Feedback-Based Mischarge Correction}{Peng2014}{ue,core,pgw}{S}{1}{?}

\newCountermeasure{firewall}{Stateful IP Firewalls}{Li2015,Kim2015,Leong2014,Hong2017}{core,pgw}{I}{1}{?}

\vspace{0.5em plus 0.5em}
\subsubsection{Research Questions and Challenges}
The research question is, how protocols that were not originally designed for the use in mobile networks can be adapted in such a way that they prevent possible information loss across layers. For instance, instead of making it a duty of the higher service to connect the IP identity with the radio identity, some part of the core network could inject the radio identity into the IP stream. For all the countermeasures, it is important that no data from the user should be trusted, as it could be forged. However, such extensions must be carefully evaluated with respect to sustainability and performance. Additionally, currently discussed 5G additions such as software-defined networking and network virtualization, can introduce new ways for cross-layer information loss. Future research should evaluate whether new protocols introduce information losses.

\begin{mdframed}[backgroundcolor=blue!5,nobreak] 
\textbf{Summary:} Cross-layer information loss causes fraud or \acs{DoS} attacks and is especially exploitable within newer generations. Countermeasures propose a secure binding between the separated network layers and firewalls. Future research must carefully observe new technological proposals, \eg, 5G network virtualization to avoid cross-layer information loss in the future generations to come.
\end{mdframed}

\subsection{Cause: Accounting Policy Inconsistency}
\label{sec:netcontext_accountPolicyInconsistency}
Mobile networks come with a variety of billing methods. Some services are charged by time and geographical distance, others by data volume. In earlier networks, the different billing methods were straightforward to distinguish as they were based on different network services. However, data networks ---such as the Internet---were originally not in mind when earlier networks were built. Another problem are transmission artifacts that occur on lower layers without the knowledge or control of higher layers, such as data retransmissions because of bad connectivity or packet loss. For example, some providers charge for TCP retransmissions while others do not. In addition, some providers have special charging policies for extra services such as music streaming. These policy inconsistencies lead to hidden channels that can be exploited for billing attacks (fraud attacks).

\subsubsection{Attacks}
Hidden channels for different protocols have been found, \eg, in the \acs{DNS} protocol~\cite{Peng2012} or in TCP retransmissions~\cite{Go2013,Go2014}, both leading to under billing attacks. Additionally, TCP retransmission can also be exploited for over-billing attacks~\cite{Go2014}. In this case, the attacker uses an existing connection to send unwanted TCP retransmissions to increase the victim's data usage. With the shift from the circuit voice to a packet-based voice switching, \ac{VoLTE} introduced a new attack surface for under-billing attacks using the \acs{RTP} and the \acs{SIP} protocol \cite{Li2015,Kim2015}. As voice is traditionally charged according to call duration, the voice-related channels can be misused as a hidden channel to transport data and thereby circumvent the accounting mechanism. 

\subsubsection*{Assessment} 
Most hidden channels are still exploited by an attacker with user traffic capabilities and optional Internet capabilities. Similar to the cross-layer information loss, these attacks are exploitable in the latest network generations and can be exploited everywhere in the network.

\sokentrystart\sokVector{Air Interface} \sokAim{Fraud} \sokAttack{Misbilling: TCP Retransmission or DNS Tunneling} \sokCausesList{accounting} \sokTarget{P} \sokTech{0}{2}{2} \sokReference{Go2014,Go2013,Go2014} \sokAddCountermeasure{dpi} \sokRange{0}{0}{0}{2}{0} \sokCapabilitiesList{airuser} \sokentryend

\sokentrystart\sokVector{Air Interface} \sokAim{Fraud} \sokAttack{Underbilling using VoLTE Hidden Channels} \sokCapabilitiesList{airuser} \sokCausesList{accounting} \sokTarget{P} \sokTech{0}{0}{2} \sokRange{0}{0}{0}{2}{0} \sokReference{Li2015,Kim2015} \sokAddCountermeasure{dpi} \sokAttackerM{Semi. Act. Radio.} \sokentryend

\subsubsection{Defenses}
To encounter the threat of accounting policy inconsistency, most countermeasures suggest the use of improved filtering at the gateway to detect possible misuses based on technologies like deep packet inspection, stateful protocol monitoring or ratio detection (of DNS packets or TCP retransmission)~\cite{Go2014,Peng2012,Li2015,Kim2015}. These countermeasures need to be installed by the operators in the core network at the packet gateways to protect against revenue losses.

\newCountermeasure{dpi}{Deep Packet Inspection and Ratio Detection}{Go2014,Go2014,Kim2015}{pgw}{I}{1}{?}

\subsubsection{Research Questions and Challenges}
In the future, more applications will utilize the IP connectivity for their service instead of using the special purpose services such as text messages and voice. These special purpose services originally generated a large proportion of the operators revenue. To encounter revenue losses, operators have established new accounting policies, \eg, fixed rates for music or video streaming \cite{telekomStreamOn,tmobileStream}. Future research should evaluate how such new accounting policies lead to inconsistency and thus open hidden channels for billing attacks. Effective countermeasures against these hidden channels and thus a prevention of billing attacks are remaining challenges.

\begin{mdframed}[backgroundcolor=blue!05,nobreak] 
\textbf{Summary:} Inconsistent accounting policies open up the possibility for hidden channels, which allow to consume resources without being charged for the service. Future research must exclude the possibility for hidden channels by an early detection of conspicuous behavior, \eg, through anomaly detection.
\end{mdframed}

\section{Root Cause: Wireless Channel}
\label{sec:sys_wc}
The wireless channel is essential for realizing mobility in mobile networks. However, this versatility makes the channel also easily accessible by unauthorized persons within the range of the radio transmission. Additionally, the wireless channel has limited resources. Over time more effective modulations and transmission schemes have been developed to improve the wireless transmission performance by reducing transmission redundancies. The easy access to the wireless channel makes mobile networks prone to jamming attacks for which an attacker disturbs the communication between two parties in a targeted manner. Jamming attacks are \ac{DoS} attacks and require an active radio attacker. As a result, the wireless channel is prone to several attacks and exhibits fundamental limitations such that we define it as a root cause.

\subsection{Attacks}
Jamming attacks disturb the communication by increasing the noise on the wireless channel. Most prior research has concentrated on the evaluation of different constant jamming strategies and their effectiveness~\cite{Xiao2013,Bhattarai2015,Aziz2015,Philippe2013,Jover2013}. While constant jamming attacks jam the entire communication bandwidth over time, smart jamming attacks are protocol-aware and intentionally jam  certain control information that affect the rest of the communication. In general, smart jamming attacks are more cost-efficient. Lichtman \etal~\cite{Lichtman2013, Lichtman2016, Rao2017} demonstrated that \acs{LTE} is particularly vulnerable to smart jamming. 

\subsubsection*{Assessment} All jamming attacks require an active radio attacker who needs to be aware of the used frequencies and the bandwidth. For smart jamming attacks, the attacker requires knowledge of the protocol and needs to be synchronized with the cell to obtain the position of control information. Nevertheless, the hardware for such attacks is easily available~\cite{Jammerblog2017, Jover2016}, in particular in the form of software defined radios such as USRPs~\cite{USRP}. While jamming attacks disturb the communication of all the victims, smart jamming attacks are more targeted. In all cases, the effective range of the attack is limited by the transmission power and location of the jammer. The motivation for jamming attacks is versatile. Besides simply obstructing the mobile service~\cite{Dalton2016}, jamming attacks can also serve as \textit{downgrade attacks}. 

\sokentrystart\sokVector{Air Interface} \sokAim{Service} \sokAttack{Continuous Wideband Jamming}  \sokRange{0}{2}{0}{0}{0} \sokTarget{U,P} \sokAttackerM{Act. Radio Att.} \sokCapabilitiesList{airactive} \sokReference{Lichtman2013, Xiao2013, Jover2013, Aziz2015} \sokCausesList{channel} \sokTech{2}{2}{2} \sokAddCountermeasure{spreadspecturmjamming*,beamforming*}
\sokentryend

\sokentrystart\sokVector{Air Interface} \sokAim{Service} \sokAttack{Protocol-Aware Selective Jamming} \sokRange{0}{2}{0}{0}{0} \sokTarget{U,P} \sokAttackerM{Act. Radio Att.} \sokCapabilitiesList{airpassive,airactive}\sokReference{Lichtman2013, Xiao2013, Jover2013, Rao2017} \sokCausesList{channel} \sokTech{0}{0}{2} \sokAddCountermeasure{spreadspecturmjamming*,broadcastencryption*,beamforming*} \sokentryend

\subsection{Defenses}
Different countermeasures against jamming have been proposed by the research community ranging from specification changes to smart implementations using different technologies, \eg, beamforming or spread-spectrum techniques \cite{Jover2014}.
So far, little effort has been devoted to implement or to evaluate jamming countermeasures in mobile networks. Furthermore, it is little known about jamming countermeasure implementations within commercial products and their deployment.

\newCountermeasure{spreadspecturmjamming}{Spread-Spectrum Jamming Resiliency}{Jover2014}{ue,enb}{S}{0}{?}

\newCountermeasure{broadcastencryption}{Broadcast Encryption}{Jover2014}{ue,enb,core*}{S}{0}{0}

\newCountermeasure{beamforming}{Beam-Forming Scheme for Selective Jamming Cancellation}{Jover2014}{enb}{I}{0}{?}

\subsection{Research Questions and Challenges}
Even though different defenses are proposed, none of them have been evaluated in detail for mobile networks. Such measures could negatively impact the transmission speed which is an important selling point for future network generations. Future research should explore the methods that were proposed or adopted by related fields~\cite{Khattab2008,Popper2010} and evaluate their fit and benefits to mobile network setups. The challenge is to integrate efficiently jamming countermeasures which typically linked to performance impairments, into the radio layer, still fulfilling quality of service requirements. This can be achieved by a specification that is dedicated for the use in critical networks with efficiency loss. Additional research should also consider new radio technologies like the narrowband \acs{LTE}~\cite{3GPP2016}. Hardening new generations against jamming attacks is especially important for the availability of safety-critical applications. 

\begin{mdframed}[backgroundcolor=blue!5,nobreak] 
\textbf{Summary:} The wireless channel is open and easily accessible and, thus, can be exploited by jamming attacks. As a consequence, the adversary can impact the availability of services. So far, no strong defenses exist. Future research is challenged by the trade-off between  efficient jamming countermeasures and high data rates to ensure the availability of safety-critical applications. 
\end{mdframed}

\immediate\soktableend
\renderCounterTableIfUpdate

\section{Related Surveys}
We finally compare our work to related surveys from a methodological perspective highlighting parallels and differences.

Different survey papers study a wide range of aspects of next-generation mobile networks (5G). For example, an overall survey of the performance requirements and solutions for 5G networks is given by Agiwal \etal~\cite{Agiwal2016}. Whereas Taleb \etal focus on the particular use case of mobile edge computing in 5G networks~\cite{Taleb2017}. These surveys lack the focus on security in the field of next generations mobile networks.

Security surveys in (mobile) phone networks focus either on one particular aspect of the system or consider just one type of attack. For example, Unger~\etal~\cite{Unger2015} focus on messaging systems and compare them based on desired security features and usability aspects. In contrast to our work, their methodology does not include attacks. Tu~\etal~\cite{Tu2016} directly map telephone spam attacks and their countermeasures, without an abstraction into causes and root causes. Acer~\etal~\cite{Acar2016} identify research issues in the area of Android security and use a methodology that directly addresses the stakeholders who might fix the issues. Our approach has the most similarities with the recently published work by Sahin~\etal~\cite{Sahin2017} since they also categorize attacks and defenses into causes and root causes. However, they limit their considerations to telephony fraud. Unique to our approach is that we abstract attacks and defenses into causes and root causes for all the three mobile network generations and use this approach to derive research questions for future generations of mobile networks.

\section{Conclusion}
In this work, we introduced a systematization methodology for attacks and defenses in mobile networks. We derived technical causes and abstract root causes for existing vulnerabilities and discussed the impacts of attacks and defenses. We used this to derive challenges and research questions with respect to shortcomings of existing work and security implications for new 5G technologies. The results of our systematization have implications on future security research in mobile networks. We finally point to the major areas and challenges for future research on this topic.

Vulnerabilities in earlier generations of mobile networks were addressed through improvements in the following generations. However, the backward compatibility of systems and attack vectors for downgrade attacks render such vulnerabilities a continuing problem. Two factors are responsible for downgrade attacks: \textit{unsecured pre-authentication traffic} and \textit{openness of the wireless channel}. While protocol changes and new cryptographic methods (\eg, asymmetric cryptography) can address unsecured pre-authentication, the wireless channel requires more fundamental changes to provide security against jamming attacks. Future research must address the class of downgrade attacks to overcome these issues. 

A related problem are \textit{insecure inter-network protocols} (\eg, \ac{SS7} or Diameter) in such a way that these legacy systems represent a threat to users as well as network providers. While firewalls constitute a temporary solution, research should develop inter-network protocols that keep the misuse potential as low as possible by minimizing the number of trusted entities. 

\textit{Insecure implementations} of network components (\eg, smartphones or core network) are an attack vector that undermines the system's integrity and immediately affects many users. Research should focus on securing those implementations by adopting means of classical system security while considering the requirements of the mobile network. 

\textit{Resource usage asymmetry} led to the so-called signaling denial-of-service attacks. In future, the number of subscribers and thus the threat of such an attack increases (\eg, by a mobile phone botnet). Therefore, research should investigate protocol designs in which the resource usage is more balanced to mitigate the threat of signaling denial-of-service attacks.

\appendices

\section*{Acknowledgment}
This work has been supported by the Franco-German BERCOM Project (FKZ: 13N13741) co-funded by the German Federal Ministry of Education and Research (BMBF) and by the DFG Research Training Group GRK 1817/1.

This work was also sponsored by the \textit{COMET K1} program by the Austrian Research Promoting Agency (FFG) and the Public Employment Service Austria (AMS). %

\bibliographystyle{IEEEtran} %
\bibliography{../src/bib/misc,../src/bib/mob_sec_clean,../src/bib/3gpp,../src/bib/adrian}

\begin{thebibliography}{100}
\providecommand{\url}[1]{#1}
\csname url@samestyle\endcsname
\providecommand{\newblock}{\relax}
\providecommand{\bibinfo}[2]{#2}
\providecommand{\BIBentrySTDinterwordspacing}{\spaceskip=0pt\relax}
\providecommand{\BIBentryALTinterwordstretchfactor}{4}
\providecommand{\BIBentryALTinterwordspacing}{\spaceskip=\fontdimen2\font plus
\BIBentryALTinterwordstretchfactor\fontdimen3\font minus
  \fontdimen4\font\relax}
\providecommand{\BIBforeignlanguage}[2]{{%
\expandafter\ifx\csname l@#1\endcsname\relax
\typeout{** WARNING: IEEEtran.bst: No hyphenation pattern has been}%
\typeout{** loaded for the language `#1'. Using the pattern for}%
\typeout{** the default language instead.}%
\else
\language=\csname l@#1\endcsname
\fi
#2}}
\providecommand{\BIBdecl}{\relax}
\BIBdecl

\bibitem{StatNo2016}
{The Statistics Portal}, ``{Number of Mobile Phone Users Worldwide from 2013 to
  2019},''
  \url{https://www.statista.com/statistics/274774/forecast-of-mobile-phone-users-worldwide/},
  [Online; accessed 22-May-2017].

\bibitem{StatRe2016}
------, ``{Revenue Forecast Mobile Operators Worldwide},''
  \url{https://www.statista.com/statistics/371899/mobile-operator-total-revenue-forecasts/},
  [Online; accessed 22-May-2017].

\bibitem{First2016}
{FirstNet}, ``{FirstNet: First Responder Network Authority},''
  \url{http://www.firstnet.gov/}, [Online; accessed 22-May-2017].

\bibitem{Shaik2015}
A.~Shaik, R.~Borgaonkar, N.~Asokan, V.~Niemi, and J.-P. Seifert, ``{Practical
  Attacks against Privacy and Availability in 4G/LTE Mobile Communication
  Systems},'' in \emph{Symposium on Network and Distributed System Security
  (NDSS)}.\hskip 1em plus 0.5em minus 0.4em\relax The Internet Society, 2016.

\bibitem{Kune2012}
D.~F. Kune, J.~Koelndorfer, N.~Hopper, and Y.~Kim, ``{Location Leaks on the GSM
  Air Interface},'' in \emph{Symposium on Network and Distributed System
  Security (NDSS)}.\hskip 1em plus 0.5em minus 0.4em\relax The Internet
  Society, 2012.

\bibitem{TS24.301}
\BIBentryALTinterwordspacing
3GPP, ``{Non-Access-Stratum (NAS) protocol for Evolved Packet System (EPS);
  Stage 3},'' {3rd Generation Partnership Project (3GPP)}, TS {24.301}, 06
  2011. [Online]. Available:
  \url{http://www.3gpp.org/ftp/Specs/html-info/24301.htm}
\BIBentrySTDinterwordspacing

\bibitem{Arapinis2012}
M.~Arapinis, L.~Mancini, E.~Ritter, M.~Ryan, N.~Golde, K.~K. Redon, and
  R.~Borgaonkar, ``{New Privacy Issues in Mobile Telephony: Fix and
  Verification},'' in \emph{ACM Conference on Computer and Communications
  Security (CCS)}.\hskip 1em plus 0.5em minus 0.4em\relax ACM, 2012, pp.
  205--216.

\bibitem{Paget2010}
{Chris Paget}, ``{Practical Cellphone Spying},'' in \emph{DEFCON 19}, 2010.

\bibitem{Mjolsnes2017}
{Mj{\o}lsnes, Stig F. and Olimid, Ruxandra F.}, ``{Easy 4G/LTE IMSI Catchers
  for Non-Programmers},'' in \emph{Conference on Mathematical Methods, Models,
  and Architectures for Computer Network Security (MMM-ACNS)}.\hskip 1em plus
  0.5em minus 0.4em\relax Springer, 2017, pp. 235--246.

\bibitem{Forsberg2007}
D.~Forsberg, H.~Leping, K.~Tsuyoshi, and S.~Alan{\"{a}}r{\"{a}}, ``{Enhancing
  Security and Privacy in 3GPP E-UTRAN Radio Interface},'' in \emph{IEEE
  International Symposium on Personal, Indoor and Mobile Radio Communications
  (PIMRC)}.\hskip 1em plus 0.5em minus 0.4em\relax IEEE, 2007.

\bibitem{Golde2016}
\BIBentryALTinterwordspacing
N.~Golde and D.~Komaromy, ``{Breaking Band},'' in \emph{Recon}, jun 2016.
  [Online]. Available:
  \url{https://comsecuris.com/slides/recon2016-breaking_band.pdf}
\BIBentrySTDinterwordspacing

\bibitem{Unger2015}
N.~Unger, S.~Dechand, J.~Bonneau, S.~Fahl, H.~Perl, I.~Goldberg, and M.~Smith,
  ``{SoK: Secure Messaging},'' in \emph{IEEE Symposium on Security and Privacy
  (SP)}.\hskip 1em plus 0.5em minus 0.4em\relax IEEE, 2015, pp. 232--249.

\bibitem{Acar2016}
Y.~Acar, M.~Backes, S.~Bugiel, S.~Fahl, P.~Mcdaniel, and M.~Smith, ``{SoK:
  Lessons Learned from Android Security Research for Appified Software
  Platforms},'' in \emph{IEEE Symposium on Security and Privacy (SP)}.\hskip
  1em plus 0.5em minus 0.4em\relax IEEE, 2016, pp. 433--451.

\bibitem{Ullrich2014}
J.~Ullrich, K.~Krombholz, H.~Hobel, A.~Dabrowski, and E.~Weippl, ``{IPv6
  Security: Attacks and Countermeasures in a Nutshell},'' in \emph{USENIX
  Workshop on Offensive Technologies (WOOT)}.\hskip 1em plus 0.5em minus
  0.4em\relax USENIX Association, 2014.

\bibitem{Tu2016}
H.~Tu, Z.~Zhao, and G.-J. Ahn, ``{SoK: Everyone Hates Robocalls: A Survey of
  Techniques against Telephone Spam},'' in \emph{IEEE Symposium on Security and
  Privacy (SP)}.\hskip 1em plus 0.5em minus 0.4em\relax IEEE, 2016, pp.
  320--338.

\bibitem{Gupta2015}
P.~Gupta, B.~Srinivasan, V.~Balasubramaniyan, and M.~Ahamad, ``{Phoneypot :
  Data-driven Understanding of Telephony Threats},'' in \emph{Symposium on
  Network and Distributed System Security (NDSS)}.\hskip 1em plus 0.5em minus
  0.4em\relax The Internet Society, 2015.

\bibitem{Sahin2017}
M.~Sahin, A.~Francillon, P.~Gupta, and M.~Ahamad, ``{SoK: Fraud in Telephony
  Networks},'' in \emph{IEEE European Symposium on Security and Privacy
  (EuroSP)}.\hskip 1em plus 0.5em minus 0.4em\relax IEEE, 2017.

\bibitem{Jover2017}
\BIBentryALTinterwordspacing
R.~P. Jover, ``{Some Key Challenges in Securing 5G Wireless Networks},''
  \emph{Electronic Comment Filing System}, Jan. 2017. [Online]. Available:
  \url{https://www.fcc.gov/ecfs/filing/10130278051628}
\BIBentrySTDinterwordspacing

\bibitem{TS23.272}
\BIBentryALTinterwordspacing
3GPP, ``{Circuit Switched (CS) fallback in Evolved Packet System (EPS); Stage
  2},'' {3rd Generation Partnership Project (3GPP)}, TS {23.272}, 06 2011.
  [Online]. Available: \url{http://www.3gpp.org/ftp/Specs/html-info/23272.htm}
\BIBentrySTDinterwordspacing

\bibitem{Walker2001}
M.~Walker and T.~Wright, \emph{{Security, in GSM and UMTS: The Creation of
  Global Mobile Communication}}, F.~Hillebrand, Ed.\hskip 1em plus 0.5em minus
  0.4em\relax John Wiley \& Sons, Ltd, Chichester, UK, 2001.

\bibitem{Wagenknecht2016}
S.~Wagenknecht and M.~Korn, ``{Hacking As Transgressive Infrastructuring:
  Mobile Phone Networks and the German Chaos Computer Club},'' in \emph{ACM
  Conference on Computer-Supported Cooperative Work \& Social Computing
  (CSCW)}.\hskip 1em plus 0.5em minus 0.4em\relax ACM, 2016, pp. 1104--1117.

\bibitem{Avizienis2004}
A.~Avizienis, J.-C. Laprie, B.~Randell, and C.~Landwehr, ``{Basic Concepts and
  Taxonomy of Dependable and Secure Computing},'' \emph{IEEE Transactions on
  Dependable and Secure Computing (TDSC)}, vol.~1, no.~1, pp. 11--33, 2004.

\bibitem{TS22.278}
\BIBentryALTinterwordspacing
3GPP, ``{Service requirements for the Evolved Packet System (EPS)},'' {3rd
  Generation Partnership Project (3GPP)}, TS {22.278}, 10 2010. [Online].
  Available: \url{http://www.3gpp.org/ftp/Specs/html-info/22278.htm}
\BIBentrySTDinterwordspacing

\bibitem{TS33.401}
\BIBentryALTinterwordspacing
------, ``{3GPP System Architecture Evolution (SAE); Security architecture},''
  {3rd Generation Partnership Project (3GPP)}, TS {33.401}, 06 2011. [Online].
  Available: \url{http://www.3gpp.org/ftp/Specs/html-info/33401.htm}
\BIBentrySTDinterwordspacing

\bibitem{TS33.102}
\BIBentryALTinterwordspacing
------, ``{3G security; Security architecture},'' {3rd Generation Partnership
  Project (3GPP)}, TS {33.102}, 12 2010. [Online]. Available:
  \url{http://www.3gpp.org/ftp/Specs/html-info/33102.htm}
\BIBentrySTDinterwordspacing

\bibitem{Thompson2015}
\BIBentryALTinterwordspacing
E.~Thompson, ``{Army examines feasibility of integrating 4G LTE with tactical
  network},'' September 25, 2015, [Online; accessed 22-May-2017]. [Online].
  Available: \url{http://www.army.mil/article/87875/event16}
\BIBentrySTDinterwordspacing

\bibitem{TS22.115}
\BIBentryALTinterwordspacing
3GPP, ``{Service aspects; Charging and billing},'' {3rd Generation Partnership
  Project (3GPP)}, TS {22.115}, 04 2010. [Online]. Available:
  \url{http://www.3gpp.org/ftp/Specs/html-info/22115.htm}
\BIBentrySTDinterwordspacing

\bibitem{Peng2014}
C.~Peng, C.-Y. Li, H.~Wang, G.-H. Tu, and S.~Lu, ``{Real Threats to Your Data
  Bills: Security Loopholes and Defenses in Mobile Data Charging},'' in
  \emph{ACM Conference on Computer and Communications Security (CCS)}.\hskip
  1em plus 0.5em minus 0.4em\relax ACM, 2014, pp. 727--738.

\bibitem{gomez2016}
I.~Gomez-Miguelez, A.~Garcia-Saavedra, P.~D. Sutton, P.~Serrano, C.~Cano, and
  D.~J. Leith, ``{srsLTE: an open-source platform for LTE evolution and
  experimentation},'' in \emph{ACM International Workshop on Wireless Network
  Testbeds, Experimental Evaluation and Characterization (WiNTECH)}.\hskip 1em
  plus 0.5em minus 0.4em\relax ACM, 2016, pp. 25--32.

\bibitem{osmocombb}
{Osmocom}, ``{OsmocomBB Open Source GSM Baseband Software Implementation},''
  \url{http://bb.osmocom.org}, [Online; accessed 12-July-2016].

\bibitem{openlte}
B.~Wojtowicz, ``{OpenLTE - An open source 3GPP LTE implementation},''
  \url{http://openlte.sourceforge.net/}, {2015}, [Online; accessed
  05-April-2017].

\bibitem{openbts}
RangeNetworks, ``Openbts,'' \url{http://openbts.org/}, [Online; accessed
  05-April-2017].

\bibitem{Nikaein2014}
N.~Nikaein, R.~Knopp, F.~Kaltenberger, L.~Gauthier, C.~Bonnet, D.~Nussbaum, and
  R.~Ghaddab, ``{OpenAirInterface: An Open LTE Network in a PC},'' in \emph{ACM
  International Conference on Mobile Computing and Networking (MobiCom)}.\hskip
  1em plus 0.5em minus 0.4em\relax ACM, 2014, pp. 305--308.

\bibitem{Timberg2014}
C.~Timberg, ``{For Sale: Systems that can Secretly Track where Cellphone Users
  go Around the Globe},''
  \url{https://www.washingtonpost.com/business/technology/for-sale-systems-that-can-secretly-track-where-cellphone-users-go-around-the-globe/2014/08/24/f0700e8a-f003-11e3-bf76-447a5df6411f_story.html},
  August 24, 2014, [Online; accessed 22-May-2017].

\bibitem{gcm2017}
{Google Firebase}, ``Firebase cloud messaging,'' accessed 2017-04-09.

\bibitem{apn2017}
{Apple Inc.}, ``{Apple Push Notification service Overview},''
  \url{https://developer.apple.com/library/content/documentation/NetworkingInternet/Conceptual/RemoteNotificationsPG/APNSOverview.html#//apple_ref/doc/uid/TP40008194-CH8-SW1},
  [Online; accessed 06-April-2017].

\bibitem{Prevelakis2007}
V.~Prevelakis and D.~Spinellis, ``{The Athens Affair},'' \emph{IEEE Spectrum},
  vol.~44, no.~7, pp. 26--33, 2007.

\bibitem{Scahill2015}
\BIBentryALTinterwordspacing
J.~Scahill and J.~Begley, ``{The Great SIM Heist - How Spies Stole the Keys to
  the Encyption Castle},'' jan 2015. [Online]. Available:
  \url{https://theintercept.com/2015/02/19/great-sim-heist/}
\BIBentrySTDinterwordspacing

\bibitem{Lee2009}
P.~P.~C. Lee, T.~Bu, and T.~Woo, ``{On the Detection of Signaling DoS Attacks
  on 3G/WiMax Wireless Networks},'' \emph{Computer Networks}, vol.~53, no.~15,
  pp. 2601--2616, 2009.

\bibitem{Bassil2013}
R.~Bassil, I.~H. Elhajj, A.~Chehab, and A.~Kayssi, ``{Effects of Signaling
  Attacks on LTE Networks},'' in \emph{IEEE Advanced Information Networking and
  Applications Workshops (WAINA)}.\hskip 1em plus 0.5em minus 0.4em\relax IEEE,
  2013, pp. 499--504.

\bibitem{Leong2014}
W.~K. Leong, A.~Kulkarni, Y.~Xu, and B.~Leong, ``{Unveiling the Hidden Dangers
  of Public IP locationes in 4G/LTE Cellular Data Networks},'' in \emph{ACM
  Workshop on Mobile Computing Systems and Applications (HotMobile)}.\hskip 1em
  plus 0.5em minus 0.4em\relax ACM, 2014, pp. 16:1--16:6.

\bibitem{Kambourakis2011}
G.~Kambourakis, C.~Kolias, S.~Gritzalis, and J.~Park, ``{DoS Attacks Exploiting
  Signaling in UMTS and IMS},'' \emph{Computer Communications}, vol.~34, no.~3,
  pp. 226--235, 2011.

\bibitem{Yu2012}
D.~Yu and W.~Wen, ``{Non-access-stratum request attack in E-UTRAN},'' in
  \emph{IEEE Computing, Communications and Applications Conference
  (ComComAp)}.\hskip 1em plus 0.5em minus 0.4em\relax IEEE, 2012, pp. 48--53.

\bibitem{Farivar2008}
C.~Farivar, ``{Apple removes GPS functionality from Egyptian iPhones},'' 2008,
  \url{http://www.macworld.com/article/1137410/Apple_removes_GPS_func.html}.

\bibitem{nokiagps}
``{Egypt tries to control the use of GPS by banning except with individual
  licences},'' 2008,
  \url{http://www.balancingact-africa.com/news/en/issue-no-429/top-story/egypt-tries-to-contr/en}.

\bibitem{Lichtman2013}
M.~Lichtman, J.~H. Reed, T.~C. Clancy, and M.~Norton, ``{Vulnerability of LTE
  to Hostile Interference},'' in \emph{IEEE Global Conference on Signal and
  Information Processing (GlobalSIP)}.\hskip 1em plus 0.5em minus 0.4em\relax
  IEEE, 2013, pp. 285--288.

\bibitem{Xiao2013}
J.~Xiao, X.~Wang, Q.~Guo, H.~Long, and S.~Jin, ``{Analysis and Evaluation of
  Jammer Interference in LTE},'' in \emph{ACM International Conference on
  Innovative Computing and Cloud Computing (ICCC)}.\hskip 1em plus 0.5em minus
  0.4em\relax ACM, 2013, pp. 46--50.

\bibitem{Jover2013}
R.~P. Jover, ``{Security Attacks against the Availability of LTE Mobility
  Networks: Overview and Research Directions},'' in \emph{IEEE Symposium on
  Wireless Personal Multimedia Communications (WPMC)}.\hskip 1em plus 0.5em
  minus 0.4em\relax IEEE, 2013.

\bibitem{Aziz2015}
F.~M. Aziz, J.~S. Shamma, and G.~L. St{\"{u}}ber, ``{Resilience of LTE Networks
  against Smart Jamming Attacks: Wideband Model},'' in \emph{IEEE Symposium on
  Personal, Indoor, and Mobile Radio Communications (PIMRC)}.\hskip 1em plus
  0.5em minus 0.4em\relax IEEE, 2015, pp. 1344--1348.

\bibitem{Rao2017}
R.~M. Rao, S.~Ha, V.~Marojevic, and J.~H. Reed, ``Lte phy layer vulnerability
  analysis and testing using open-source sdr tools,'' \emph{arXiv preprint
  arXiv:1708.05887}, 2017.

\bibitem{Hong2017}
H.~Hong, H.~Choi, D.~Kim, H.~Kim, B.~Hong, J.~Noh, and Y.~Kim, ``{When Cellular
  Networks Met IPv6: Security Problems of Middleboxes in IPv6 Cellular
  Networks},'' in \emph{IEEE European Symposium on Security and Privacy
  (EuroSP)}.\hskip 1em plus 0.5em minus 0.4em\relax IEEE, 2017.

\bibitem{Enck2005}
W.~Enck, P.~Traynor, P.~McDaniel, and T.~{La Porta}, ``{Exploiting Open
  Functionality in SMS-capable Cellular Networks},'' in \emph{ACM Conference on
  Computer and Communications Security (CCS)}.\hskip 1em plus 0.5em minus
  0.4em\relax ACM, 2005, pp. 393--404.

\bibitem{Traynor2009a}
P.~Traynor, W.~Enck, P.~McDaniel, and T.~{La Porta}, ``{Mitigating attacks on
  open functionality in SMS-capable cellular networks},'' \emph{IEEE/ACM
  Transactions on Networking (TON)}, vol.~17, no.~1, pp. 40--53, 2009.

\bibitem{Golde2013}
N.~Golde, K.~Redon, and J.-P. Seifert, ``{Let Me Answer That For You:
  Exploiting Broadcast Information in Cellular Networks},'' in \emph{USENIX
  Security Symposium (SSYM)}.\hskip 1em plus 0.5em minus 0.4em\relax USENIX
  Association, 2013, pp. 33--48.

\bibitem{Traynor2009}
P.~Traynor, M.~Lin, M.~Ongtang, V.~Rao, T.~Jaeger, P.~Mcdaniel, and T.~L.
  Porta, ``{On Cellular Botnets: Measuring the Impact of Malicious Devices on a
  Cellular Network Core},'' in \emph{ACM Conference on Computer and
  Communications Security (CCS)}.\hskip 1em plus 0.5em minus 0.4em\relax ACM,
  2009, pp. 223--234.

\bibitem{Engel2014}
\BIBentryALTinterwordspacing
T.~Engel, ``{SS7: Locate. Track. Manipulate.}'' in \emph{Chaos Communication
  Congress}, ser. 31C3, 2014. [Online]. Available:
  \url{https://berlin.ccc.de/~tobias/31c3-ss7-locate-track-manipulate.pdf}
\BIBentrySTDinterwordspacing

\bibitem{Rao2002}
J.~R. Rao, P.~Rohatgi, H.~Scherzer, and S.~Tinguely, ``{Partitioning Attacks:
  or how to Rapidly Clone some GSM Cards},'' in \emph{IEEE Symposium on
  Security and Privacy (SP)}.\hskip 1em plus 0.5em minus 0.4em\relax IEEE,
  2002, pp. 31--41.

\bibitem{Zhou2013}
Y.~Zhou, Y.~Yu, F.~X. Standaert, and J.~J. Quisquater, ``{On the Need of
  Physical Security for Small Embedded Devices: a Case Study with COMP128-1
  Implementations in SIM Cards},'' in \emph{International Conference on
  Financial Cryptography and Data Security (FC)}.\hskip 1em plus 0.5em minus
  0.4em\relax Springer, 2013, pp. 230--238.

\bibitem{Liu2015}
J.~Liu, Y.~Yu, F.~X. Standaert, Z.~Guo, D.~Gu, W.~Sun, Y.~Ge, and X.~Xie,
  ``{Small Tweaks Do Not Help: Differential Power Analysis of MILENAGE
  Implementations in 3G/4G USIM Cards},'' in \emph{European Symposium on
  Research in Computer Security (ESORICS)}.\hskip 1em plus 0.5em minus
  0.4em\relax Springer, 2015, pp. 468--480.

\bibitem{Osipov2015}
A.~Osipov and A.~Zaitsev, ``{Adventures in Femtoland: 350 Yuan for Invaluable
  Fun},'' Black Hat USA 2015, 08 2015.

\bibitem{Michau2016}
B.~Michau and C.~Devine, ``{How to not Break LTE Crypto},'' in \emph{ANSSI
  Symposium sur la s{\'{e}}curit{\'{e}} des technologies de l'information et
  des communications (SSTIC)}, 2016.

\bibitem{Rupprecht2016}
D.~Rupprecht, K.~Jansen, and C.~P{\"{o}}pper, ``{Putting LTE Security Functions
  to the Test: A Framework to Evaluate Implementation Correctness},'' in
  \emph{USENIX Workshop on Offensive Technologies (WOOT)}.\hskip 1em plus 0.5em
  minus 0.4em\relax USENIX Association, 2016.

\bibitem{Welte2009}
H.~Welte, ``{OpenBSC - Running your own GSM network},'' 08 2009, talk at
  Hacking at Random 2009. Slides:
  \url{https://openbsc.osmocom.org/trac/raw-attachment/wiki/FieldTests/HAR2009/har2009-gsm-report.pdf}.

\bibitem{Dabrowski2014}
A.~Dabrowski, N.~Pianta, T.~Klepp, M.~Mulazzani, and E.~Weippl, ``{IMSI-Catch
  Me If You Can: IMSI-Catcher-Catchers},'' in \emph{ACM Annual Computer
  Security Applications Conference (ACSAC)}.\hskip 1em plus 0.5em minus
  0.4em\relax ACM, 2014, pp. 246--255.

\bibitem{Dabrowski2016}
A.~Dabrowski, G.~Petzl, and E.~R. Weippl, ``{The Messenger Shoots Back: Network
  Operator Based IMSI Catcher Detection},'' in \emph{Symposium on Recent
  Advances in Intrusion Detection (RAID)}.\hskip 1em plus 0.5em minus
  0.4em\relax Springer, 2016, pp. 279--302.

\bibitem{Nohl2014}
\BIBentryALTinterwordspacing
K.~Nohl, ``{Mobile Self-Defense},'' in \emph{Chaos Communication Congress},
  ser. 31C3, 2014. [Online]. Available:
  \url{https://events.ccc.de/congress/2014/Fahrplan/system/attachments/2493/original/Mobile_Self_Defense-Karsten_Nohl-31C3-v1.pdf}
\BIBentrySTDinterwordspacing

\bibitem{Malette2015}
\BIBentryALTinterwordspacing
L.~Malette, ``{Catcher Catcher},'' 2015. [Online]. Available:
  \url{https://opensource.srlabs.de/projects/mobile-network-assessment-tools/wiki/CatcherCatcher}
\BIBentrySTDinterwordspacing

\bibitem{AIMSICD}
{SecUpwN (Pseudonym, Maintainer)}, ``{Android IMSI-Catcher Detector},''
  \url{https://cellularprivacy.github.io/Android-IMSI-Catcher-Detector/},
  retrieved 2017-04-04.

\bibitem{Snoopsnitch}
{SR Labs}, ``Snoopsnitch,'' 12 2014,
  \url{https://opensource.srlabs.de/projects/snoopsnitch}, accessed Nov 12
  2015.

\bibitem{Briceno2002}
\BIBentryALTinterwordspacing
I.~{Briceno, Marc and Goldberg} and W.~David, ``{GSM Cloning},'' 2002.
  [Online]. Available: \url{http://www.isaac.cs.berkeley.edu/isaac/gsm.html}
\BIBentrySTDinterwordspacing

\bibitem{Briceno1998}
M.~Briceno, I.~Goldberg, and D.~Wagner, ``{An implementation of the GSM A3A8
  algorithm. (Specifically, COMP128.)},'' 1998,
  \url{http://www.scard.org/gsm/a3a8.txt}, accessed June 24th 2016.

\bibitem{Kaljevic2003}
D.~Kaljevic, ``{SIMSCAN v2.01},'' \url{http://dejankaljevic.org/}, accessed
  2017-03-09, 2003.

\bibitem{Meyer2004a}
U.~Meyer and S.~Wetzel, ``{On the Impact of GSM Encryption and
  Man-in-the-Middle Attacks on the Security of Interoperating GSM/UMTS
  Networks},'' in \emph{IEEE International Symposium on Personal, Indoor and
  Mobile Radio Communications (PRIMRC)}.\hskip 1em plus 0.5em minus 0.4em\relax
  IEEE, 2004, pp. 2876--2883.

\bibitem{Han2014}
C.-K. Han and H.-K. Choi, ``{Security Analysis of Handover Key Management in 4G
  LTE/SAE Networks},'' \emph{IEEE Transactions on Mobile Computing (TMC)},
  vol.~13, no.~2, pp. 457--468, 2014.

\bibitem{Barkan2008}
E.~Barkan, E.~Biham, and N.~Keller, ``{Instant Ciphertext-only Cryptanalysis of
  GSM Encrypted Communication},'' \emph{Journal of Cryptology}, vol.~21, no.~3,
  pp. 392--429, aug 2008.

\bibitem{Briceno1999}
\BIBentryALTinterwordspacing
M.~Briceno, I.~Goldberg, and D.~Wagner, ``{A Pedagogical Implementation of
  A5/1.}'' 1999. [Online]. Available: \url{http://www.scard.org/gsm/a51.html}
\BIBentrySTDinterwordspacing

\bibitem{A51rainbow}
S.~Labs, ``{Kraken: A5/1 Decryption Rainbow Tables},'' via Bittorent, 2010,
  \url{https://opensource.srlabs.de/projects/a51-decrypt}, accessed Nov 12
  2015.

\bibitem{Nohl2009}
K.~Nohl and C.~Paget, ``{GSM – SRSLY?}'' in \emph{Chaos Communication
  Congress}, ser. 26C3, dec 2009.

\bibitem{Golic1997}
J.~D. Goli{\'{c}}, ``{Cryptanalysis of Alleged A5 Stream Cipher},'' in
  \emph{International Conference on Theory and Application of Cryptographic
  Techniques (EUROCRYPT)}.\hskip 1em plus 0.5em minus 0.4em\relax Springer,
  1997, pp. 239--255.

\bibitem{Biryukov2000}
A.~Biryukov, A.~Shamir, and D.~Wagner, ``{Real Time Cryptanalysis of A5/1 on a
  PC},'' in \emph{International Workshop on Fast Software Encryption
  (FSE)}.\hskip 1em plus 0.5em minus 0.4em\relax Springer, 2000.

\bibitem{Nohl2011defend}
\BIBentryALTinterwordspacing
K.~Nohl and L.~Melette, ``Defending mobile phones,'' Chaos Communications
  Congress (28C3), 2011. [Online]. Available:
  \url{https://events.ccc.de/congress/2011/Fahrplan/events/4736.en.html}
\BIBentrySTDinterwordspacing

\bibitem{Borgaonkar2017}
R.~Borgaonkar, L.~Hirshi, S.~Park, A.~Shaik, A.~Martin, and J.-P. Seifert,
  ``{New Adventures in Spying 3G \& 4G Users: Locate, Track, Monitor},'' in
  \emph{BlackHat}, 2017.

\bibitem{Chalakkal2017}
\BIBentryALTinterwordspacing
S.~Chalakkal, H.~Schmidt, and S.~Park, ``{Practical Attacks on VoLTE and
  VoWIFI},'' ERNW Enno Rey Netzwerke, Tech. Rep., 2017. [Online]. Available:
  \url{https://www.ernw.de/download/newsletter/ERNW_Whitepaper_60_Practical_Attacks_On_VoLTE_And_VoWiFi_v1.0.pdf}
\BIBentrySTDinterwordspacing

\bibitem{Zhang2016}
\BIBentryALTinterwordspacing
W.~Zhang and H.~Shan, ``{LTE Redirection: Forcing Targeted LTE Cellphone into
  Unsafe Network},'' in \emph{Defcon}, 2016. [Online]. Available:
  \url{https://media.defcon.org/DEF CON 24/DEF CON 24
  presentations/DEFCON-24-Zhang-Shan-Forcing-Targeted-Lte-Cellphone-Into-Unsafe-Network.pdf}
\BIBentrySTDinterwordspacing

\bibitem{Nohl2013}
\BIBentryALTinterwordspacing
K.~Nohl, ``{Rooting SIM cards},'' in \emph{Blackhat}, 2013. [Online].
  Available:
  \url{https://media.blackhat.com/us-13/us-13-Nohl-Rooting-SIM-cards-Slides.pdf}
\BIBentrySTDinterwordspacing

\bibitem{Hong2018}
B.~Hong, S.~Bae, and Y.~Kim, ``{GUTI Reallocation Demystified: Cellular
  Location Tracking with Changing Temporary Identifier},'' in \emph{Symposium
  on Network and Distributed System Security (NDSS)}.\hskip 1em plus 0.5em
  minus 0.4em\relax The Internet Society, 2018.

\bibitem{Ia-sadosky2016}
\BIBentryALTinterwordspacing
Ia-sadosky, ``{Heap memory corruption in ASN.1 parsing code generated by
  Objective Systems Inc. ASN1C compiler for C/C++},'' 2016. [Online].
  Available:
  \url{https://github.com/programa-stic/security-advisories/tree/master/ObjSys/CVE-2016-5080}
\BIBentrySTDinterwordspacing

\bibitem{Weinmann2012}
R.-P. Weinmann, ``{Baseband Attacks: Remote Exploitation of Memory Corruptions
  in Cellular Protocol Stacks},'' in \emph{USENIX Workshop on Offensive
  Technologies (WOOT)}.\hskip 1em plus 0.5em minus 0.4em\relax USENIX
  Association, 2012.

\bibitem{VandenBroek2014}
F.~van~den Broek, B.~Hond, and A.~{Cedillo Torres}, ``{Security Testing of GSM
  Implementations},'' \emph{Engineering Secure Software and Systems (ESSoS)},
  vol. 8364, pp. 179--195, 2014.

\bibitem{Mulliner2011}
C.~Mulliner, N.~Golde, and J.-P. Seifert, ``{SMS of Death: from Analyzing to
  Attacking Mobile Phones on a Large Scale},'' in \emph{USENIX Security
  Symposium (SSYM)}.\hskip 1em plus 0.5em minus 0.4em\relax USENIX Association,
  2011, pp. 363--378.

\bibitem{Court2017}
\BIBentryALTinterwordspacing
T.~Court and N.~Biggs, ``{WAP just happened to my Samsung Galaxy?}'' 2017,
  \url{https://www.contextis.com/resources/blog/wap-just-happened-my-samsung-galaxy/},
  accessed 2017-02-05. [Online]. Available:
  \url{https://www.contextis.com/resources/blog/wap-just-happened-my-samsung-galaxy/}
\BIBentrySTDinterwordspacing

\bibitem{Li2017}
Z.~Li, W.~Wang, C.~Wilson, J.~Chen, C.~Qian, T.~Jung, L.~Zhang, K.~Liu, X.~Li,
  and Y.~Liu, ``{FBS-Radar: Uncovering Fake Base Stations at Scale in the
  Wild},'' in \emph{Symposium on Network and Distributed System Security
  (NDSS)}.\hskip 1em plus 0.5em minus 0.4em\relax The Internet Society, 2017.

\bibitem{Hassidim2016}
vinatan Hassidim, Y.~Matias, M.~Yung, and A.~Ziv, ``{Ephemeral Identifiers:
  Mitigating Tracking \& Spoofing Threats to BLE Beacons},'' 2016,
  \url{https://developers.google.com/beacons/eddystone-eid-preprint.pdf}.

\bibitem{Tu2016a}
G.-H. Tu, C.-Y. Li, C.~Peng, Y.~Li, and S.~Lu, ``{New Security Threats Caused
  by IMS-based SMS Service in 4G LTE Networks},'' in \emph{ACM Conference on
  Computer and Communications Security (CCS)}.\hskip 1em plus 0.5em minus
  0.4em\relax ACM, 2016, pp. 1118--1130.

\bibitem{Go2014}
Y.~Go, E.~Jeong, J.~Won, Y.~Kim, D.~F. Kune, and K.~Park, ``{Gaining Control of
  Cellular Traffic Accounting by Spurious TCP Retransmission},'' in
  \emph{Symposium on Network and Distributed System Security (NDSS)}.\hskip 1em
  plus 0.5em minus 0.4em\relax The Internet Society, 2014.

\bibitem{Go2013}
Y.~Go, D.~F. Kune, S.~Woo, K.~Park, and Y.~Kim, ``{Towards Accurate Accounting
  of Cellular Data for TCP Retransmission},'' in \emph{ACM Workshop on Mobile
  Computing Systems and Applications (HotMobile)}.\hskip 1em plus 0.5em minus
  0.4em\relax ACM, 2013.

\bibitem{Li2015}
C.-Y. Li, G.-H. Tu, S.~Lu, X.~Wang, C.~Peng, Z.~Yuan, Y.~Li, S.~Lu, and
  X.~Wang, ``{Insecurity of Voice Solution VoLTE in LTE Mobile Networks},'' in
  \emph{ACM Conference on Computer and Communications Security (CCS)}.\hskip
  1em plus 0.5em minus 0.4em\relax ACM, 2015, pp. 316--327.

\bibitem{Kim2015}
H.~Kim, D.~Kim, M.~Kwon, H.~Han, Y.~Jang, D.~Han, T.~Kim, and Y.~Kim,
  ``{Breaking and Fixing VoLTE : Exploiting Hidden Data Channels and
  Misimplementations},'' in \emph{ACM Conference on Computer and Communications
  Security (CCS)}.\hskip 1em plus 0.5em minus 0.4em\relax ACM, 2015, pp.
  328--339.

\bibitem{Wang2011}
Z.~Wang, Z.~Qian, Q.~Xu, Z.~Mao, and M.~Zhang, ``{An Untold Story of
  Middleboxes in Cellular Networks},'' in \emph{ACM SIGCOMM Computer
  Communication Review}.\hskip 1em plus 0.5em minus 0.4em\relax ACM, 2011, pp.
  374--385.

\bibitem{Rao2015}
S.~P. Rao, S.~Holtmanns, I.~Oliver, and T.~Aura, ``{Unblocking Stolen Mobile
  Devices using SS7-MAP Vulnerabilities: Exploiting the Relationship between
  IMEI and IMSI for EIR access},'' in \emph{IEEE Conference on Trust, Security
  and Privacy in Computing and Communications (TrustCom)}.\hskip 1em plus 0.5em
  minus 0.4em\relax IEEE, 2015, pp. 1171--1176.

\bibitem{TR33.899}
\BIBentryALTinterwordspacing
3GPP, ``{Technical Specification Group Services and System Aspects; Study on
  the security aspects of the next generation system},'' {3rd Generation
  Partnership Project (3GPP)}, TR {33.899}, 08 2017. [Online]. Available:
  \url{http://www.3gpp.org/ftp/specs/archive/33_series/33.899/}
\BIBentrySTDinterwordspacing

\bibitem{Arapinis2014}
M.~Arapinis, L.~I. Mancini, E.~Ritter, and M.~Ryan, ``{Privacy through
  Pseudonymity in Mobile Telephony Systems},'' in \emph{Symposium on Network
  and Distributed System Security (NDSS)}.\hskip 1em plus 0.5em minus
  0.4em\relax The Internet Society, 2014.

\bibitem{TR33.821}
\BIBentryALTinterwordspacing
3GPP, ``{Rationale and track of security decisions in Long Term Evolution (LTE)
  RAN / 3GPP System Architecture Evolution (SAE)},'' {3rd Generation
  Partnership Project (3GPP)}, TR {33.821}, 06 2009. [Online]. Available:
  \url{http://www.3gpp.org/ftp/Specs/html-info/33821.htm}
\BIBentrySTDinterwordspacing

\bibitem{vandenbroek2015}
F.~van~den Broek, R.~Verdult, and J.~de~Ruiter, ``{Defeating IMSI Catchers},''
  in \emph{ACM Conference on Computer and Communications Security (CCS)}.\hskip
  1em plus 0.5em minus 0.4em\relax ACM, 2015, pp. 340--351.

\bibitem{Khan2015}
M.~S.~A. Khan and C.~J. Mitchell, ``{Improving Air Interface User Privacy in
  Mobile Telephony},'' in \emph{International Conference on Research in
  Security Standardisation (SSR)}, vol. 9497.\hskip 1em plus 0.5em minus
  0.4em\relax Springer, 2015, pp. 165--184.

\bibitem{Ateniese1999}
G.~Ateniese, A.~Herzberg, H.~Krawczyk, and G.~Tsudik, ``{Untraceable Mobility
  or How to Travel Incognito},'' \emph{Computer Networks}, vol.~31, no.~8, pp.
  871--884, 1999.

\bibitem{Choudhury2012}
H.~Choudhury, B.~Roychoudhury, and D.~K. Saikia, ``{Enhancing User Identity
  Privacy in LTE},'' in \emph{IEEE International Conference on Trust, Security
  and Privacy in Computing and Communications (TrustCom)}.\hskip 1em plus 0.5em
  minus 0.4em\relax IEEE, 2012, pp. 949--957.

\bibitem{Horn1998}
G.~Horn and B.~Preneel, ``{Authentification and Payment in Future Mobile
  Systems},'' in \emph{European Symposium on Research in Computer Security
  (ESORICS)}.\hskip 1em plus 0.5em minus 0.4em\relax Springer, 1998, pp.
  183--207.

\bibitem{Kambourakis2004}
G.~Kambourakis, A.~Rouskas, and S.~Gritzalis, ``{Performance Evaluation of
  Public Key-Based Authentication in Future Mobile Communication Systems},''
  \emph{EURASIP Journal on Wireless Communications and Networking}, vol. 2004,
  no.~1, pp. 184--197, 2004.

\bibitem{Ney2017}
P.~Ney, I.~Smith, G.~Cadamuro, and K.~Tadayoshi, ``{Sea-Glass: Enabling
  City-wide IMSI-Catcher Detection},'' in \emph{Privacy Enhancing Technologies
  Symposium (PETS)}.\hskip 1em plus 0.5em minus 0.4em\relax Springer, 2017.

\bibitem{Park2017}
S.~Park, A.~Shaik, R.~Borgaonkar, A.~Martin, and J.-P. Seifert,
  ``White-stingray: Evaluating {IMSI} catchers detection applications,'' in
  \emph{USENIX Workshop on Offensive Technologies (WOOT)}.\hskip 1em plus 0.5em
  minus 0.4em\relax USENIX Association, 2017.

\bibitem{Steig2016}
S.~Steig, A.~Aarnes, T.~v.~Do, and H.~T. Nguyen, ``A network based imsi catcher
  detection,'' in \emph{2016 6th International Conference on IT Convergence and
  Security (ICITCS)}, Sept 2016, pp. 1--6.

\bibitem{GSMKcryptophone2017}
{GSMK mbH}, ``Gsmk cryptophone,''
  \url{http://www.cryptophone.de/en/products/mobile/}, {2017}, [Online;
  accessed 03-April-2017].

\bibitem{vanDo2015}
T.~van Do, H.~T. Nguyen, and N.~Momchil, ``{Detecting IMSI-Catcher Using Soft
  Computing},'' in \emph{Springer International Conference on Soft Computing in
  Data Science (SCDS)}.\hskip 1em plus 0.5em minus 0.4em\relax Springer, 2015,
  pp. 129--140.

\bibitem{GSMKOverwatch2017}
{GSMK Gesellschaft für sichere Mobile Kommunikation mbH}, ``Gsmk debuts new
  security systems to protect mobile network operators against eavesdropping
  and fraud,''
  \url{http://www.cryptophone.de/en/company/news/gsmk-debuts-new-security-systems-to-protect-mobile-network-operators-against-eavesdropping-and-fraud/},
  {2017}, [Online; accessed 03-April-2017].

\bibitem{NetworkGuard}
C.~Technologies, ``Protect yourself from cell phone interception,''
  \url{http://www.charontech.com/network_guard.html}, accessed 2017-03-12.

\bibitem{Kumar2006}
K.~P. Kumar, G.~Shailaja, A.~Kavitha, and A.~Saxena, ``{Mutual Authentication
  and Key Agreement for GSM},'' in \emph{IEEE International Conference on
  Mobile Business (ICMB)}.\hskip 1em plus 0.5em minus 0.4em\relax IEEE, 2006.

\bibitem{Chang2005}
C.-C. Chang, J.-S. Lee, and Y.-F. Chang, ``{Efficient Authentication Protocols
  of GSM},'' \emph{Computer Communications}, vol.~28, no.~8, pp. 921--928,
  2005.

\bibitem{Khan2016}
M.~S.~A. Khan and C.~J. Mitchell, ``{Retrofitting Mutual Authentication to GSM
  using RAND Hijacking},'' in \emph{International Workshop on Security and
  Trust Management (STM)}.\hskip 1em plus 0.5em minus 0.4em\relax Springer,
  2016, pp. 17--31.

\bibitem{Lee2011}
C.-C. Lee, I.-E. Liao, and M.-S. Hwang, ``{An Efficient Authentication Protocol
  for Mobile Communications},'' \emph{Telecommunication Systems}, vol.~46,
  no.~1, pp. 31--41, 2011.

\bibitem{ericsson2003}
Ericsson, ``{Enhancements to GSM/UMTS AKA: 3GPP TSG SA WG3 Security,
  S3-030542},''
  \url{http://www.3gpp.org/ftp/tsg_sa/wg3_security/tsgs3_30_povoa/docs/pdf/S3-030542.pdf},
  [Online; accessed 06-April-2017].

\bibitem{ericsson2004}
------, ``{On the introduction and use of UMTS AKA in GSM: 3GPP TSG SA WG3
  Security, S3-040534},''
  \url{ftp://www.3gpp.org/tsg_sa/WG3_Security/TSGS3_34_Acapulco/Docs/PDF/S3-040534.pdf},
  [Online; accessed 06-April-2017].

\bibitem{Alt2016}
S.~Alt, P.-A. Fouque, G.~Macario-rat, C.~Onete, and B.~Richard, ``{A
  Cryptographic Analysis of UMTS/LTE AKA},'' in \emph{International Conference
  on Applied Cryptography and Network Security (ACNS)}.\hskip 1em plus 0.5em
  minus 0.4em\relax Springer, 2016, pp. 18--35.

\bibitem{TS35.202}
\BIBentryALTinterwordspacing
3GPP, ``{3G Security; Specification of the 3GPP confidentiality and integrity
  algorithms; Document 2: Kasumi specification},'' {3rd Generation Partnership
  Project (3GPP)}, TS {35.202}, 12 2009. [Online]. Available:
  \url{http://www.3gpp.org/ftp/Specs/html-info/35202.htm}
\BIBentrySTDinterwordspacing

\bibitem{GSMA-A52prohibition}
\emph{{Prohibiting A5/2 in mobile stations and other clarifications regarding
  A5 algorithm support}}, GSM Association Std. SP-070\,671,
  \url{http://www.3gpp.org/ftp/tsg_sa/TSG_SA/TSGS_37/Docs/SP-070671.zip}.

\bibitem{TS44.006}
\BIBentryALTinterwordspacing
3GPP, ``{Mobile Station - Base Station System (MS - BSS) interface; Data Link
  (DL) layer specification},'' {3rd Generation Partnership Project (3GPP)}, TS
  {44.006}, 03 2010. [Online]. Available:
  \url{http://www.3gpp.org/ftp/Specs/html-info/44006.htm}
\BIBentrySTDinterwordspacing

\bibitem{P1Sec}
\BIBentryALTinterwordspacing
P1Sec, ``{SS7 Security Map}.'' [Online]. Available:
  \url{http://ss7map.p1sec.com/}
\BIBentrySTDinterwordspacing

\bibitem{ashdown2001ss7}
\BIBentryALTinterwordspacing
M.~Ashdown and S.~Lynchard, ``{SS7 Firewall System},'' 2001. [Online].
  Available: \url{https://www.google.com/patents/US6308276}
\BIBentrySTDinterwordspacing

\bibitem{Engel2014PATENT}
T.~Engel and H.~Freyther, ``\BIBforeignlanguage{German}{Verfahren und eine
  vorrichtung zur sicherung einer signalisierungssystem- nr.
  7-schnittstelle},'' DE Patent 102\,014\,117\,713, 2016, also filed as
  US20160156647.

\bibitem{TR23.840}
\BIBentryALTinterwordspacing
3GPP, ``{Study into routeing of MT-SMs via the HPLMN},'' {3rd Generation
  Partnership Project (3GPP)}, TR {23.840}, 03 2007. [Online]. Available:
  \url{http://www.3gpp.org/ftp/Specs/html-info/23840.htm}
\BIBentrySTDinterwordspacing

\bibitem{Burow2017}
N.~Burow, S.~A. Carr, J.~Nash, P.~Larsen, M.~Franz, S.~Brunthaler, and
  M.~Payer, ``{Control-Flow Integrity: Precision, Security, and Performance},''
  \emph{ACM Computing Surveys (CSUR)}, vol.~50, no.~1, pp. 16:1--16:33, Apr.
  2017.

\bibitem{Szekeres2013}
L.~Szekeres, M.~Payer, T.~Wei, and D.~Song, ``{SoK: Eternal War in Memory},''
  in \emph{IEEE Symposium on Security and Privacy (SP)}.\hskip 1em plus 0.5em
  minus 0.4em\relax IEEE, 2013, pp. 48--62.

\bibitem{Larsen2014}
P.~Larsen, A.~Homescu, S.~Brunthaler, and M.~Franz, ``{SoK: Automated Software
  Diversity},'' in \emph{IEEE Symposium on Security and Privacy (SP)}.\hskip
  1em plus 0.5em minus 0.4em\relax IEEE, 2014, pp. 276--291.

\bibitem{Jover2014}
R.~P. Jover, J.~Lackey, and A.~Raghavan, ``{Enhancing the Security of LTE
  Networks against Jamming Attacks},'' \emph{Journal on Information Security
  (EURASIP)}, vol. 2014, no.~1, p.~7, 2014.

\bibitem{Welte2008}
H.~Welte and D.~Spaar, ``{Running your own GSM network},'' 12 2008, talk at
  25th Chaos Communication Congress, Berlin, Germany.
  \url{https://events.ccc.de/congress/2008/Fahrplan/events/3007.en.html},
  Slides:
  \url{http://events.ccc.de/congress/2008/Fahrplan/attachments/1259_25C3-OpenBSC.pdf},
  Video:
  \url{https://media.ccc.de/v/25c3-3007-en-running_your_own_gsm_network}.

\bibitem{TS23.040}
\BIBentryALTinterwordspacing
3GPP, ``{Technical realization of the Short Message Service (SMS)},'' {3rd
  Generation Partnership Project (3GPP)}, TS {23.040}, 09 2010. [Online].
  Available: \url{http://www.3gpp.org/ftp/Specs/html-info/23040.htm}
\BIBentrySTDinterwordspacing

\bibitem{pki_imsi_catcher2017}
{PKI Electronic Intelligence GmbH Germany}, ``{3G UMTS IMSI Catcher},''
  \url{http://www.pki-electronic.com/products/interception-and-monitoring-systems/3g-umts-imsi-catcher/},
  {2017}, [Online; accessed 03-April-2017].

\bibitem{Biddle2016}
\BIBentryALTinterwordspacing
S.~Biddle, ``{Long-Secret Stingray Manuals Detail how Police an Spy on
  Phones},'' pp. 1--10, sep 2016. [Online]. Available:
  \url{https://theintercept.com/2016/09/12/long-secret-stingray-manuals-detail-how-police-can-spy-on-phones/}
\BIBentrySTDinterwordspacing

\bibitem{GSMKsensorNetwork2017}
{GSMK mbH}, ``Gsmk debuts new security systems to protect mobile network
  operators against eavesdropping and fraud,''
  \url{http://www.cryptophone.de/en/company/news/gsmk-debuts-new-security-systems-to-protect-mobile-network-operators-against-eavesdropping-and-fraud/},
  {2017}, [Online; accessed 03-April-2017].

\bibitem{GSMmap}
``{GSM} security map,'' \url{http://gsmmap.org/}.

\bibitem{Perrig2002}
A.~Perrig, R.~Canetti, J.~D. Tygar, and D.~Song, ``{The TESLA Broadcast
  Authentication Protocol},'' \emph{CryptoBytes}, vol.~5, no.~2, pp. 2--13,
  2005.

\bibitem{Hussain2018}
S.~R. Hussain, O.~Chowdhury, S.~Mehnaz, and E.~Bertino, ``{LTEInspector: A
  Systematic Approach for Adversarial Testing of 4G LTE},'' in \emph{Symposium
  on Network and Distributed System Security (NDSS)}.\hskip 1em plus 0.5em
  minus 0.4em\relax The Internet Society, 2018.

\bibitem{Meyer2004}
U.~Meyer and S.~Wetzel, ``{A Man-in-the-Middle Attack on UMTS},'' in \emph{ACM
  Workshop on Wireless Security (WiSe)}.\hskip 1em plus 0.5em minus 0.4em\relax
  ACM, 2004, pp. 90--97.

\bibitem{TS04.31}
\BIBentryALTinterwordspacing
3GPP, ``{Location Services (LCS); Mobile Station (MS) - Serving Mobile Location
  Centre (SMLC) Radio Resource LCS Protocol (RRLP)},'' {3rd Generation
  Partnership Project (3GPP)}, TS {04.31}, 06 2007. [Online]. Available:
  \url{http://www.3gpp.org/ftp/Specs/html-info/0431.htm}
\BIBentrySTDinterwordspacing

\bibitem{Muncaster2014}
P.~Muncaster, ``Chinese cops cuff 1,500 in fake base station spam raid,'' The
  Register, 26 Mar 2014,
  \url{http://www.theregister.co.uk/2014/03/26/spam_text_china_clampdown_police/}.

\bibitem{Pell2014}
S.~K. Pell and C.~Soghoian, ``{Your secret stingray's no secret anymore: The
  vanishing government monopoly over cell phone surveillance and its impact on
  national security and consumer privacy.}'' \emph{Harvard Journal of Law \&
  Technology}, vol.~28, no.~1, pp. 1--76, 2014.

\bibitem{Kerckhoffs1883}
A.~Kerckhoffs, ``La cryptographie militaire,'' \emph{Journal des Sciences
  Militaires}, pp. 5--83, 1883.

\bibitem{Goldberg1999}
I.~Goldberg, D.~Wagner, and L.~Green, ``{The (Real-Time) Cryptanalysis of
  A5/2},'' in \emph{Rump session of Crypto'99}, 1999.

\bibitem{Dunkelman2010}
\BIBentryALTinterwordspacing
O.~Dunkelman, N.~Keller, and A.~Shamir, ``{A Practical-Time Attack on the A5/3
  Cryptosystem Used in Third Generation GSM Telephony},'' 2010. [Online].
  Available: \url{http://eprint.iacr.org/2010/013}
\BIBentrySTDinterwordspacing

\bibitem{Papantonakis2013}
P.~Papantonakis, D.~Pnevmatikatos, and I.~Papaefstathiou, ``{Fast, FPGA-based
  Rainbow Table creation for attacking encrypted mobile communications},'' in
  \emph{Field Programmable Logic and Applications (FPL)}.\hskip 1em plus 0.5em
  minus 0.4em\relax IEEE, 2013.

\bibitem{Nohl2011gprs}
K.~Nohl and L.~Melette, ``{GPRS Intercept: Wardriving your country},'' Chaos
  Communications Camp 2011, 2011.

\bibitem{Jia2011}
K.~Jia, C.~Rechberger, and X.~Wang, ``Green cryptanalysis: Meet-in-the-middle
  key-recovery for the full kasumi cipher,'' Cryptology ePrint Archive, Report
  2011/466, 2011, \url{http://eprint.iacr.org/2011/466}.

\bibitem{Dunkelman2014}
\BIBentryALTinterwordspacing
O.~Dunkelman, N.~Keller, and A.~Shamir, ``A practical-time related-key attack
  on the kasumi cryptosystem used in gsm and 3g telephony,'' \emph{Journal of
  Cryptology}, vol.~27, no.~4, pp. 824--849, 2014. [Online]. Available:
  \url{http://dx.doi.org/10.1007/s00145-013-9154-9}
\BIBentrySTDinterwordspacing

\bibitem{GSM02.07}
\emph{Digital cellular telecommunications system (Phase 2+); Mobile Stations
  (MS) features}, European Telecommunications Standards Institute Std. GSM
  02.07, Rev. version 8.0.0 Release 1999, 1999.

\bibitem{androidissue5353}
``{Android Issue 5353: Ciphering Indicator},'' 2009,
  \url{https://code.google.com/p/android/issues/detail?id=5353}, accessed July
  14th 2013.

\bibitem{Androulidakis2012}
I.~Androulidakis, D.~Pylarinos, and G.~Kandus, ``{Ciphering Indicator
  Approaches and User Awareness},'' \emph{Maejo International Journal of
  Science and Technology}, vol.~6, no.~3, pp. 514--527, 2012.

\bibitem{TR31.900}
\BIBentryALTinterwordspacing
3GPP, ``{SIM/USIM internal and external interworking aspects},'' {3rd
  Generation Partnership Project (3GPP)}, TR {31.900}, 12 2009. [Online].
  Available: \url{http://www.3gpp.org/ftp/Specs/html-info/31900.htm}
\BIBentrySTDinterwordspacing

\bibitem{TS55.226}
\BIBentryALTinterwordspacing
------, ``{3G Security; Specification of the A5/4 Encryption Algorithms for GSM
  and ECSD, and the GEA4 Encryption Algorithm for GPRS},'' {3rd Generation
  Partnership Project (3GPP)}, TS {55.226}, 02 2011. [Online]. Available:
  \url{http://www.3gpp.org/ftp/Specs/html-info/55226.htm}
\BIBentrySTDinterwordspacing

\bibitem{TS35.206}
\BIBentryALTinterwordspacing
------, ``{3G Security; Specification of the MILENAGE algorithm set: An example
  algorithm set for the 3GPP authentication and key generation functions f1,
  f1*, f2, f3, f4, f5 and f5*; Document 2: Algorithm specification},'' {3rd
  Generation Partnership Project (3GPP)}, TS {35.206}, 12 2009. [Online].
  Available: \url{http://www.3gpp.org/ftp/Specs/html-info/35206.htm}
\BIBentrySTDinterwordspacing

\bibitem{TS35.231}
\BIBentryALTinterwordspacing
------, ``{Specification of the TUAK algorithm set: A second example algorithm
  set for the 3GPP authentication and key generation functions f1, f1*, f2, f3,
  f4, f5 and f5*;},'' {3rd Generation Partnership Project (3GPP)}, TS {35.231},
  12 2015. [Online]. Available:
  \url{http://www.3gpp.org/ftp/Specs/html-info/35231.htm}
\BIBentrySTDinterwordspacing

\bibitem{TS33.303}
\BIBentryALTinterwordspacing
------, ``{Proximity-based Services (ProSe); Security aspects},'' {3rd
  Generation Partnership Project (3GPP)}, TS {33.303}, 12 2009. [Online].
  Available: \url{http://www.3gpp.org/ftp/Specs/html-info/33303.htm}
\BIBentrySTDinterwordspacing

\bibitem{RFC_6507}
{M. Groves}, ``{Elliptic Curve-Based Certificateless Signatures for
  Identity-Based Encryption},'' \url{https://tools.ietf.org/html/rfc6507},
  [Online; accessed 12-July-2016].

\bibitem{RFC_6508}
------, ``{Sakai-Kasahara Key Encryption (SAKKE)},''
  \url{https://tools.ietf.org/html/rfc6508}, [Online; accessed 12-July-2016].

\bibitem{Murdoch2016}
S.~J. Murdoch, ``{Insecure by Design: Protocols for Encrypted Phone Calls},''
  \emph{IEEE Computer}, vol.~49, no.~3, pp. 25--33, 2016.

\bibitem{eSIM2017}
{GSM Association}, ``{Remote Provisioning Architecture for Embedded UICC
  Technical Specification (27 May 2016)},''
  \url{http://www.gsma.com/iot/wp-content/uploads/2016/07/SGP.02_v3.1.pdf},
  [Online; accessed 06-April-2017].

\bibitem{Wild2018}
M.~Dehnel-Wild and C.~Cremers, ``{Security Vulnerability in 5G-AKA Draft},''
  {Department of Computer Science, University of Oxford}, Tech. Rep., 2018,
  \url{https://www.cs.ox.ac.uk/5G-analysis/5G-AKA-draft-vulnerability.pdf}.

\bibitem{Engel2008}
\BIBentryALTinterwordspacing
T.~Engel, ``{Locating Mobile Phones using Signalling System 7},'' in
  \emph{Chaos Communication Congress}, ser. 25C3, dec 2008. [Online].
  Available:
  \url{http://events.ccc.de/congress/2008/Fahrplan/attachments/1262_25c3-locating-mobile-phones.pdf}
\BIBentrySTDinterwordspacing

\bibitem{Holtmanns2016}
S.~Holtmanns, S.~P. Rao, and I.~Oliver, ``{User Location Tracking Attacks for
  LTE Networks using the Interworking Functionality},'' in \emph{IFIP
  Networking Conference and Workshops}, 2016, pp. 315--322.

\bibitem{Rao2016a}
S.~P. Rao, B.~T. Kotte, and S.~Holtmanns, ``{Privacy in LTE networks},'' in
  \emph{ICST International Conference on Mobile Multimedia Communications},
  2016, pp. 176--183.

\bibitem{Holtmanns2017}
S.~Holtmanns and I.~Oliver, ``Sms and one-time-password interception in lte
  networks,'' in \emph{IEEE International Conference on Communications
  (ICC)}.\hskip 1em plus 0.5em minus 0.4em\relax IEEE, 2017.

\bibitem{Verint2013}
\BIBentryALTinterwordspacing
Verint, ``{Skylock Product Description 2013},'' 2013. [Online]. Available:
  \url{http://apps.washingtonpost.com/g/page/business/skylock-product-description-2013/1276/}
\BIBentrySTDinterwordspacing

\bibitem{GSMA2017}
\BIBentryALTinterwordspacing
D.~Maxwell, ``Guidelines for independent remote interconnect security
  testing,'' GSM Association, Tech. Rep., 2017. [Online]. Available:
  \url{https://www.gsma.com/aboutus/wp-content/uploads/2017/11/FS.26_v1.0.pdf}
\BIBentrySTDinterwordspacing

\bibitem{Peeters2018}
C.~Peeters, H.~Abdullah, N.~Scaife, J.~Bowers, P.~Traynor, B.~Reaves, and
  K.~Butler, ``{Sonar: Detecting SS7 Redirection Attacks With Audio-Based
  Distance Bounding},'' in \emph{IEEE Symposium on Security and Privacy
  (SP)}.\hskip 1em plus 0.5em minus 0.4em\relax IEEE, 2018, pp. 86--101.

\bibitem{Guri2017}
M.~Guri, Y.~Mirsky, and Y.~Elovici, ``{9-1-1 DDoS via Malicious Baseband
  Firmware},'' in \emph{IEEE European Symposium on Security and Privacy
  (EuroSP)}.\hskip 1em plus 0.5em minus 0.4em\relax IEEE, 2017.

\bibitem{Yang2011}
\BIBentryALTinterwordspacing
C.~Yang, ``{Weather the signaling storm},'' Huawei, Tech. Rep.~61, 2011.
  [Online]. Available: \url{http://www1.huawei.com/en/static/HW-094153.pdf}
\BIBentrySTDinterwordspacing

\bibitem{Ericsson2015}
\BIBentryALTinterwordspacing
Ericsson, ``{Handling of Signaling Storms in Mobile Networks: The Role of the
  User Data Management System},'' Ericsson, Tech. Rep., 2015. [Online].
  Available:
  \url{https://www.ericsson.com/res/docs/2015/handling-of-signaling-storms-in-mobile-networks-august.pdf}
\BIBentrySTDinterwordspacing

\bibitem{rfc6013}
\BIBentryALTinterwordspacing
W.~Simpson, ``{RFC 6013: TCP Cookie Transactions (TCPCT)},'' 2011, [Online;
  accessed 22-May-2017]. [Online]. Available:
  \url{https://www.rfc-editor.org/rfc/rfc6013.txt}
\BIBentrySTDinterwordspacing

\bibitem{Dwork1992}
C.~Dwork and M.~Naor, ``Pricing via processing or combatting junk mail,'' in
  \emph{Annual International Cryptology Conference (Crypto)}.\hskip 1em plus
  0.5em minus 0.4em\relax Springer, 1992, pp. 139--147.

\bibitem{Back2002}
A.~Back, ``Hashcash-a denial of service counter-measure,''
  \url{http://www.hashcash.org/papers/hashcash.pdf}, 2002, [Online; accessed
  22-May-2017].

\bibitem{Goodin2016}
\BIBentryALTinterwordspacing
D.~Goodin, ``{Software Flaw puts Mobile Phones and Networks at Risk of Complete
  Takeover},'' July 2016. [Online]. Available:
  \url{https://arstechnica.com/security/2016/07/software-flaw-puts-mobile-phones-and-networks-at-risk-of-complete-takeover/}
\BIBentrySTDinterwordspacing

\bibitem{Brown2015}
{Aaron Brown}, ``{WARNING: This text message will CRASH and reboot YOUR
  iPhone},''
  \url{http://www.express.co.uk/life-style/science-technology/580211/iPhone-Messages-iMessage-Bug-Text-Reboot-Crash},
  June 2015.

\bibitem{DSilva2015}
\BIBentryALTinterwordspacing
V.~D’Silva, M.~Payer, and D.~Song, ``{The Correctness-Security Gap in
  Compiler Optimization},'' in \emph{IEEE CS Security and Privacy
  Workshops}.\hskip 1em plus 0.5em minus 0.4em\relax IEEE, 2015. [Online].
  Available: \url{http://spw15.langsec.org/papers/dsilva-gap.pdf}
\BIBentrySTDinterwordspacing

\bibitem{Jaeger2014}
E.~Jaeger and O.~Levillain, ``Mind your language (s): A discussion about
  languages and security,'' in \emph{IEEE Security and Privacy Workshops
  (SPW)}.\hskip 1em plus 0.5em minus 0.4em\relax IEEE, 2014, pp. 140--151.

\bibitem{Peng2012}
C.~Peng, C.-y. Li, G.-H. Tu, S.~Lu, and L.~Zhang, ``{Mobile Data Charging: New
  Attacks and Countermeasures},'' in \emph{ACM Conference on Computer and
  Communications Security (CCS)}.\hskip 1em plus 0.5em minus 0.4em\relax ACM,
  2012, pp. 195--204.

\bibitem{telekomStreamOn}
{Telekom}, ``{StreamOn Optionen},''
  \url{https://www.telekom.de/unterwegs/tarife-und-optionen/streamon}, [Online;
  accessed 06-April-2017].

\bibitem{tmobileStream}
{T-Mobile}, ``{Binge On},''
  \url{https://www.t-mobile.com/offer/binge-on-streaming-video.html}, [Online;
  accessed 06-April-2017].

\bibitem{Bhattarai2015}
S.~Bhattarai, S.~Wei, S.~Rook, W.~Yu, R.~F. Erbacher, and H.~Cam, ``{On
  Simulation Studies of Jamming Threats Against LTE Networks},'' in \emph{IEEE
  International Conference on Computing, Networking and Communications
  (ICNC)}.\hskip 1em plus 0.5em minus 0.4em\relax IEEE, 2015, pp. 99--103.

\bibitem{Philippe2013}
G.~Philippe, F.~Montaigne, J.-c. Schiel, E.~Georgeaux, C.~Gruet, P.-y. Roy,
  P.~Force, and P.~M{\`{e}}ge, ``{LTE Resistance to Jamming Capability},'' in
  \emph{Military Communications and Information Systems Conference (MCC)},
  2013, pp. 7--9.

\bibitem{Lichtman2016}
M.~Lichtman, R.~P. Jover, M.~Labib, R.~Rao, V.~Marojevic, and J.~H. Reed,
  ``{LTE/LTE-a Jamming, Spoofing, and Sniffing: Threat Assessment and
  Mitigation},'' \emph{IEEE Communications Magazine}, vol.~54, no.~4, pp.
  54--61, 2016.

\bibitem{Jammerblog2017}
{electronicsforu}, ``{How to Build: Cell Phone Jammer},''
  \url{https://electronicsforu.com/electronics-projects/build-cell-phone-jammer},
  October 2016.

\bibitem{Jover2016}
\BIBentryALTinterwordspacing
R.~P. Jover, ``{LTE security, protocol exploits and location tracking
  experimentation with low-cost software radio},'' \emph{arXiv}, 2016.
  [Online]. Available: \url{http://arxiv.org/abs/1607.05171}
\BIBentrySTDinterwordspacing

\bibitem{USRP}
{Ettus Research}, ``{Universal Software Radio Peripheral},''
  \url{https://www.ettus.com/product}, [Online; accessed 22-May-2017].

\bibitem{Dalton2016}
{Andrew Dalton}, ``{Florida man fined 48k for jamming cellphones while
  driving},''
  \url{https://www.engadget.com/2016/05/25/florida-man-fined-48k-fcc-jamming-cellphones/},
  May 2016, [Online; accessed 22-May-2017].

\bibitem{Khattab2008}
S.~Khattab, D.~Mosse, and R.~Melhem, ``{Jamming Mitigation in Multi-Radio
  Wireless Networks: Reactive or Proactive?}'' in \emph{ACM International
  Conference on Security and Privacy In Communication Netowrks
  (SecureComm)}.\hskip 1em plus 0.5em minus 0.4em\relax ACM, 2008.

\bibitem{Popper2010}
C.~P{\"{o}}pper, M.~Strasser, and S.~{\v{C}}apkun, ``{Anti-jamming broadcast
  communication using uncoordinated spread spectrum techniques},'' \emph{IEEE
  Journal on Selected Areas in Communications}, vol.~28, no.~5, pp. 703--715,
  Jun 2010.

\bibitem{3GPP2016}
\BIBentryALTinterwordspacing
3GPP, ``{Standardization of NB-IOT completed},'' 2016, [Online; accessed
  06-April-2017]. [Online]. Available:
  \url{http://www.3gpp.org/news-events/3gpp-news/1785-nb_iot_complete}
\BIBentrySTDinterwordspacing

\bibitem{Agiwal2016}
M.~Agiwal, A.~Roy, and N.~Saxena, ``{Next Generation 5G Wireless Networks: A
  Comprehensive Survey},'' \emph{IEEE Communications Surveys Tutorials},
  vol.~18, no.~3, pp. 1617--1655, 2016.

\bibitem{Taleb2017}
T.~Taleb, K.~Samdanis, B.~Mada, H.~Flinck, S.~Dutta, and D.~Sabella, ``{On
  Multi-Access Edge Computing: A Survey of the Emerging 5G Network Edge Cloud
  Architecture and Orchestration},'' \emph{IEEE Communications Surveys
  Tutorials}, vol.~19, no.~3, pp. 1657--1681, 2017.

\end{thebibliography}

\vfill\null %

\begingroup
\footnotesize \section*{Mobile Communications Acronyms}
\renewcommand{\IEEEiedlistdecl}{\IEEEsetlabelwidth{MSISDN}}
\begin{acronym}
\acro{3GPP}{3rd Generation Partnership Project}
\acro{AKA}{Authentication and Key Agreement}
\acro{AuC}{Authentication Center}
\acro{DNS}{Domain Name System}
\acro{DoS}{Denial of Service}
\acro{DPI}{Deep Packet Inspection}
\acro{EDGE}{Enhanced Data Rates for \ac{GSM} Evolution}
\acro{eNodeB}{Evolved NodeB}
\acro{EPC}{Evolved Packet Core} 
\acro{ETSI}{European Telecommunications Standards Institute}
\acro{GCM}{Google Cloud Messaging}
\acro{GPRS}{General Packet Radio Service} 
\acro{GSM}{Global System for Mobile Communications}
\acro{HLR}{Home Location Register} 
\acro{HSS}{Home Subscriber Server}
\acro{IMEI}{International Mobile Station Equipment Identity}
\acro{IMSI}{International Mobile Subscriber Identity}
\acro{IMS}{IP Multimedia Subsystem}
\acro{LA}{Location Area}
\acro{LTE}{Long Term Evolution}
\acro{MAC}{Message Authentication Code}
\acro{MitM}{Man-in-the-Middle}
\acro{MSISDN}{Mobile Station Integrated Services Digital Network Number}
\acro{NAT}{Network Address Translation}
\acro{OTA}{Over-the-Air}
\acro{P-GW}{Packet Data Network Gateway}
\acro{PSTN}{Public Switched Telephone Network}
\acro{PKI}{Public Key Infrastructure}
\acro{QoS}{Quality of Service}
\acro{RA}{Routing Area}
\acro{RAN}{Radio Access Network}
\acro{RTP}{Real-Time Transport Protocol}
\acro{SDR}{Software Defined Radio}
\acro{SIP}{Session Initiation Protocol}
\acro{SS7}{Signalling System \#7}
\acro{TA}{Tracking Area} 
\acro{TMSI}{Temporary Mobile Subscriber Identity}
\acro{TLS}{Transport Layer Security}
\acro{UE}{User Equipment} 
\acro{UMTS}{Universal Mobile Telecommunications System}
\acro{USIM}{Universal Subscriber Identity Module}
\acro{VoLTE}{Voice over LTE} 
\end{acronym}
\renewcommand{\IEEEiedlistdecl}{\relax}%
\endgroup

\begin{figure*}[!p]
\centering
\parbox{0.8\linewidth}{
	\footnotesize \slightlybold  Aim \hfill Attack \hfill Cause \hfill Root Cause \hspace{5mm}
	\vspace{3mm}
}
\includegraphics[height=0.9\textheight]{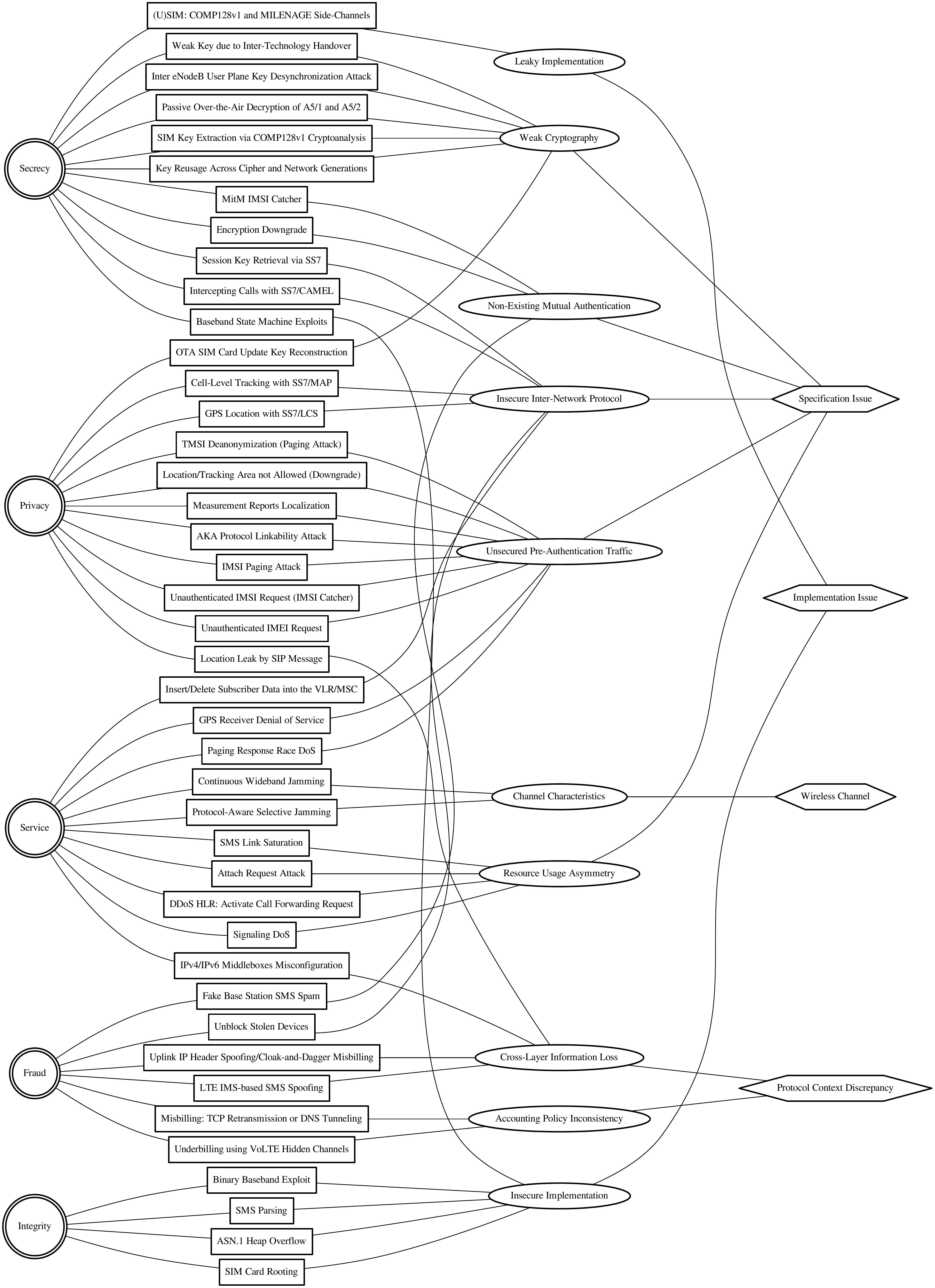} %
\caption{Visualization of Systematization including Attack Aims, Attacks, Causes, and Root Causes.}
\label{fig:graphviz_diagram_all}
\end{figure*}

\end{document}